\documentclass[aps,prb,twocolumn,superscriptaddress,showpacks]{revtex4-2}

\usepackage{comment}
\usepackage[colorlinks, citecolor=red]{hyperref}
\usepackage{hyperref}

\usepackage[utf8]{inputenc}
\usepackage[T1]{fontenc}    %
\usepackage[english]{babel}
\usepackage[pdftex]{graphicx}
\usepackage{amsmath,amssymb,amsthm,bbm, mathtools} %
\usepackage{mathrsfs}
\usepackage{tikz}   %
\usepackage[normalem]{ulem}

\renewcommand{\vec}[1]{\boldsymbol{#1}}

\newcommand{\CR}[1]{{\color{red} #1}}

\begin{document}

\title{Fabry-Perot superconducting diode}

\author{Xian-Peng Zhang}
\affiliation{Centre for Quantum Physics, Key Laboratory of Advanced Optoelectronic Quantum Architecture and Measurement (MOE), School of Physics, Beijing Institute of Technology, Beijing, 100081, China}

\begin{abstract}
Superconducting diode effects (SDEs) occur in systems with asymmetric critical supercurrents $|I^c_+|\neq |I^c_-|$ yielding   
dissipationless flow in one direction $(e.g., +)$, while dissipative transport in the opposite direction $(-)$. 
Here we investigate the SDE in a phase-biased $\phi$ Josephson junction with a 
double-barrier resonant-tunneling InAs nanowire nested between proximitized InAs/Al 
leads with finite momentum $\hbar q$ Cooper pairing. Within the Bogoliubov-de Gennes (BdG) approach, we  
obtain the exact BCS ground state energy $\mathcal{E}_G(q,\phi)$ and $I^{c}_{+} \neq |I^{c}_{-}|$  
from the current-phase relation $I_G(q,\phi) \sim \partial_{\phi}\mathcal{E}_G(q,\phi)$. The SDE arises 
from the accrued Andreev phase shifts $\delta \phi_{L,R}(q,\phi)$ leading to asymmetric BdG spectra for $q\neq 0$. 
Remarkably, the diode efficiency $\gamma=(I^{c}_{+} - |I^{c}_{-}|)/(I^{c}_{+} + |I^{c}_{-}|)$ shows 
multiple Fabry-Perot resonances $\gamma \simeq 26\%$ at the double-barrier 
Andreev bound states as the well depth $V_g$ is varied. Our $\gamma$ also features sign reversals for increasing $q$ and high sensitiveness 
to fermion-parity transitions. The latter enables $I^{c}_{+} (\phi_+)\rightleftarrows I^{c}_{-}(\phi_-)$ switchings 
over narrow phase windows, i.e., $\phi_+, \phi_- \in \Delta \phi\ll\pi$, possibly relevant for future superconducting electronics.

\end{abstract}

\maketitle

\section{Introduction}
Nonreciprocity in superconducting materials \cite{ando2020observation,bauriedl2022supercurrent,shin2021magnetic,bocquillon2017gapless,pal2022josephson,baumgartner2022supercurrent,wu2022field,turini2022josephson,lin2022zero,scammell2022theory,davydova2022universal,bauriedl2022supercurrent,narita2022field,jiang2022field,daido2022intrinsic,yuan2022supercurrent,ilic2022theory,he2022phenomenological,legg2022superconducting,zhang2022general,tanaka2022theory,jiang2022superconducting} is currently a
subject of particular interest. It refers to the asymmetry between the forward $I^{c}_{+}>0$ and reverse 
$I^{c}_{-}<0$ critical supercurrents  such that $I^{c}_{+}\neq |I^{c}_{-}|$.  This has been observed 
in superconducting films \cite{ando2020observation,bauriedl2022supercurrent,shin2021magnetic}, 
Josephson junctions \cite{bocquillon2017gapless,pal2022josephson,baumgartner2022supercurrent,wu2022field,turini2022josephson},  superconductor/ferromagnet multilayers \cite{narita2022field,jiang2022field}, as well as twisted trilayer 
graphene \cite{lin2022zero,scammell2022theory}. Currents $I$ in the range $| I^{c}_{-}|< I < I^{c}_{+}$, assuming $I^c_+ > |I^c_-|$, flow dissipationlessly in 
one direction (zero resistance), but dissipatively in the opposite direction (non-zero resistance). This is the 
superconducting diode effect (SDE) 
\cite{daido2022intrinsic,yuan2022supercurrent,ilic2022theory,he2022phenomenological,legg2022superconducting,zhang2022general,tanaka2022theory}, conceptually similar to the p-n junction semiconducting diode.

A non-zero SDE generally requires breaking both time-reversal and spatial inversion 
symmetries \cite{yuan2022supercurrent,he2022phenomenological}. Time-reversal breaking can be achieved 
via, e.g.,  exchange fields which displace the Fermi surfaces in a spin-resolved fashion, 
thus making the Cooper pairs acquire a finite center-of-mass momentum $\hbar \vec{q}$ \cite{fulde1964superconductivity,larkin1965nonuniform,casalbuoni2004inhomogeneous,bowers2002crystallography,kenzelmann2008coupled,mayaffre2014evidence,kinjo2022superconducting}. Inversion asymmetry allows for a non-zero spin-orbit coupling
and can result in a Fulde-Ferrell type phase-modulated pairing potential $\Delta(\vec{r})=\Delta e^{i \vec{q} \cdot \vec{r}}$ \cite{matsuda2007fulde,kaur2005helical,sigrist2007superconductivity,agterberg2007magnetic}. 
The momentum $\hbar \vec{q}$ can be controlled, e.g., by an external 
in-plane magnetic field and the spin-orbit interaction, which combined can change the Fermi surface  \cite{chen2018finite,wu2013unconventional,hart2017controlled,zhang2013topological,chan2014pairing,dimitrova2007theory,yuan2021topological,yuan2018zeeman},
by externally injecting currents \cite{levine1965dependence,hansen1969observation,bardeen1962critical} 
into the system and via the intrinsic screening supercurrents through the Meissner 
effect \cite{davydova2022universal,zhu2021discovery}. Bulk superconductors and Josephson junctions alike 
can exhibit SDE. As we show next, superconductor-semiconductor hybrids enable proximity superconductivity 
in a semiconducting matrix, thus providing a unique setting to exploit SDE.

\begin{figure*}[t!]
\begin{center}
\includegraphics[width=0.85\linewidth]{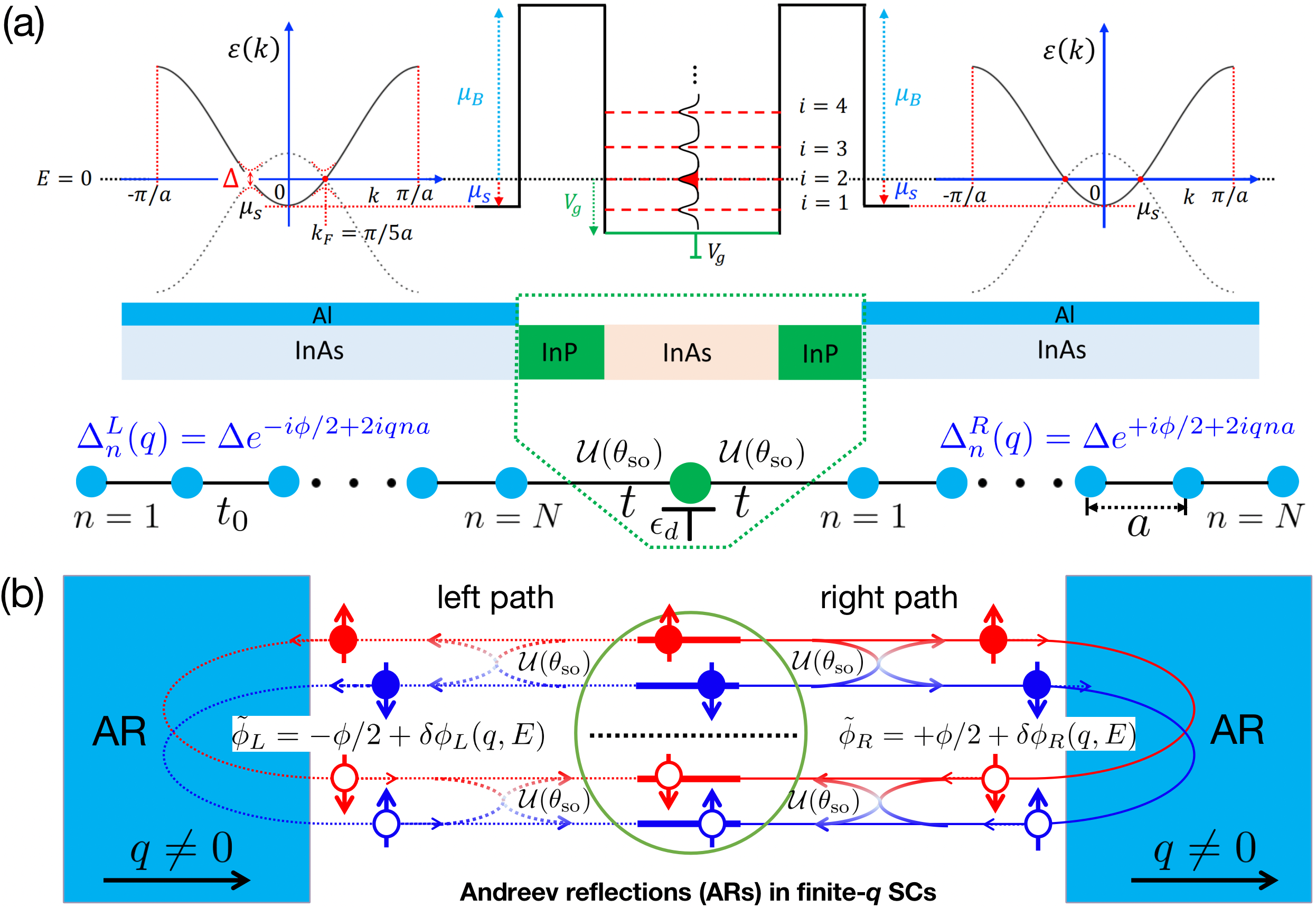}
\end{center}
\caption{(a) InAs-based resonant-tunneling Josephson junction 
and its single-site quantum dot counterpart. (b) Multiply-reflected 
Andreev paths between finite-momentum $\hbar q$ superconductors and corresponding 
accrued phase shifts $\delta \phi_{L,R}[q,E(q,\phi)]$. 
Electrons ($\bullet$) and holes ($\circ$) can be further coupled via the SO-induced 
spin rotation $\mathcal{U}(\theta_{so})$. }
\label{STORY}
\end{figure*} 

Here we consider a Josephson junction formed by a 1D resonant-tunneling double-barrier InAs semiconducting 
nanowire placed between two adjacent proximitized InAs/Al superconducting leads \cite{dartiailh2021phase} with finite momentum $\hbar q$ 
Cooper pairing, combining the extraordinary tunability of semiconductors, the remarkable scalability of superconducting circuits, and the compact footprint of quantum dots. Within the Bogoliubov-de Gennes (BdG) formalism, we determine the exact 
ground-state energy $\mathcal{E}_G(q,\phi)$ and the accumulated phase shifts 
$\delta \phi_{L,R}[q,E(q,\phi)]$ Fig.~\ref{FIGDB}(a), due to multiple Andreev reflections, for our phase-biased $\phi$ 
junction. We also obtain the supercurrent phase relation 
$I_G(q,\phi)=(2e/\hbar)\partial_{\phi}\mathcal{E}_G(q,\phi)~vs.~\phi$, from which we extract  
the depairing supercurrents $I^c_+ \neq |I^c_-|$  signaling SDE, Fig.~\ref{FIGDB}(b). 
The asymmetry about $\phi=\pi$ of the phase shifts $\delta \phi_{L,R}[q,E(q,\phi)]$ Fig.~\ref{FIGDB}(a)
results in asymmetric dispersions for $q\neq 0$[inset of Fig.~\ref{FIGDB}(a)] known to originate SDE. 

For our resonant-tunneling Josephson junction, 
we find a sizable SDE with the diode efficiency 
$\gamma=(I^{c}_{+} - |I^{c}_{-}|)/(I^{c}_{+} + |I^{c}_{-}|)$ exhibiting sharp gate-tunable ($V_g$) 
Fabry-Perot type resonances, peaking at about $\gamma \simeq 26\%$, 
Fig.~\ref{FIGDB}(c). These arise from the many zero-energy Andreev bound states [Fig.~\ref{FIGDB}(d), red lines lines] stemming from the ordinary quasi-bound states of the double barrier  
 [Fig.~\ref{FIGDB}(d), black solid lines]. We can further investigate SDE within the simpler 
single-site quantum dot model for a Josephson junction, Fig.~\ref{STORY}(a) (lower part). We find that the SO in the dot-lead 
tunnel coupling, finite momentum $\hbar q$, and Zeeman fields (dot and leads) can substantially affect SDE, thus 
producing multiple sign reversals in the diode efficiency $\gamma$, Figs.~\ref{SDE} and \ref{GFSDE}. The 
Zeeman fields, in particular, can induce fermion-parity (even/odd) transitions of the ground state, which 
can greatly affect the current phase relation thus significantly 
changing the SDE, Fig.~\ref{SDE}(b). The sharp jumps in $I_G(q,\phi)$ vs $\phi$ not only offer signatures 
for parity changes, but also yield unique $I^{c}_{+} (\phi_+)\rightleftarrows I^{c}_{-}(\phi_-)$ switchings within  
a very narrow phase range, $\phi_+, \phi_- \in \Delta \phi \ll \pi$, Fig.~\ref{SDE}(b) (green curve). This  
enables high sensitivity in the SDE tuning, possibly being a resource for superconducting electronics. 
Furthermore, the dot model yields approximate expressions (via Green functions) for the Andreev bound states, thus providing insight into the crucial role of the phase shifts 
$\delta \phi_{L,R}[q,E(q,\phi)]$ (see Sec.~\ref{singlesiteQDJJ}).

The paper is organized as follows. In Sec.~\ref{modelandtheory}, we present our model and theory, including the continuum and tight-bind models of double-barrier Josephson junction (Sec.~\ref{continuutotightbinding}), ground-state supercurrent (Sec.~\ref{groundstatesupercurrent}), additional phase shifts in the Andreev reflection of a finite-momentum superconductors (Sec.~\ref{Additionalphaseshifts}),  as well as  single-site quantum dot Josephson junction \label{singlesiteQDJJ} (Sec.~\ref{singlesiteQDJJ}).  Section~\ref{resultsanddiscussions} presents our results and discussions, such as control of SDE via fermion-parity change (Sec.~\ref{CSDEFPC}) and  tuning SDE via the spin-orbit angle (Sec.~\ref{tuningSEDSOC}). Our paper ends with conclusion and acknowledgement in Sec.~\ref{conclusion} and Sec.~\ref{acknowledgement}. The appendices contain detailed information and derivations.  Appendix \ref{fvndkkaa} presents the detailed evolution of the effective Hamiltonian from the continuum (Sec.~\ref{continuumtotignht}) to the tight-binding (Sec.~\ref{tightbindingmodel}) models of the double-barrier Josephson junction, which is then simplified into the single-site quantum dot Josephson junction (Secs.~\ref{gatetunableAL} and \ref{fvandkfk}). Appendix \ref{sec1} and Appendix \ref{supercurrent} provide the detailed derivations of the diagonal Hamiltonian [Eq. \eqref{fdbldfl}] and the ground-state supercurrent [Eq. \eqref{fdjvkdfr}], respectively. Appendix \ref{AndreevrReflection} presents the additional phase shifts of Andreev reflection in the finite-momentum superconductor, which leads to the microscopic origin of the asymmetric Andreev dispersions discussed in Appendix \ref{sec-asym}. Finally, the investigation of ground state fermion parity changes is presented in Appendix \ref{sec-parity}.

\section{Model and theory}  \label{modelandtheory}

\subsection{From continuum to tight-bind models of double-barrier Josephson junction}   \label{continuutotightbinding}

Here, we start with a double-barrier Josephson junction, plotted in Fig.~\ref{STORY}(a). The lower panel shows a possible layered structure 
comprising InAs/Al proximitized superconducting leads with InP barriers and a InAs well \cite{baumgartner2022supercurrent}.  The corresponding BdG Hamiltonian is 
\begin{align} \label{fvgbvfglbm}
    H=\int dx \psi^{\dagger}(x)\begin{bmatrix}
        -\frac{\hbar^2\partial^2_x}{2m^{*}} +\mu(x) &-\Delta(x) \\
         -\Delta^{*}(x) & +\frac{\hbar^2\partial^2_x}{2m^{*}} -\mu(x)
    \end{bmatrix}\psi(x),
\end{align}
where $m^*$ is the electron mass and  $\psi^{\dagger}(x)=\begin{bmatrix}
        \psi^{\dagger}_{\uparrow} (x) &-\psi^{}_{\downarrow}(x)
    \end{bmatrix}$ is the spinor field. The pair potential $  \Delta(x)$ is zero within semiconductor (green layer) and nonzero within the 
finite-momentum $\hbar q$ superconductors (grey layers), i.e.,
\begin{align}
    \Delta(x)=\left\{\begin{matrix*}
        \Delta e^{+2iqx+\phi_L}, & x <-L/2\\
        0, & -L_M/2<x <+L_M/2\\
        \Delta e^{+2iqx+\phi_R}, & x >L/2
    \end{matrix*}\right.,
\end{align}
where $L$ in width of the semiconducting layer.
The full chemical potential profile $\mu(x)$, including the double-barrier structure, is described by 
\begin{align} \label{fvvdldlvm}
    \mu(x)=\left\{\begin{matrix*}
       \mu_S,& x <-L/2\\
          \mu_B=\mu_S+V_B, & -L/2<x <-L_W/2\\
         \mu_S+V_g, & -L_W/2<x <+L_W/2\\
         \mu_B= \mu_S+V_B, & +L_W/2<x <+L/2\\
       \mu_S, & x >L/2
    \end{matrix*}\right.,
\end{align}
where $L_W$ denotes the well width, $\mu_S$ the chemical potential of superconductor, $\mu_B$ the chemical potential of semiconductor,  $V_B$ 
the height of the double barriers, and $V_g$ the electrostatic gate controlling the quantum well depth. For simplicity, we do not consider Zeeman fields and SO coupling in the 
continuum model; these are included in the simpler single-site quantum dot Josephson junction in Sec.~\ref{singlesiteQDJJ}.

\begin{figure*}[t!]
\begin{center}
\includegraphics[width=0.7\linewidth]{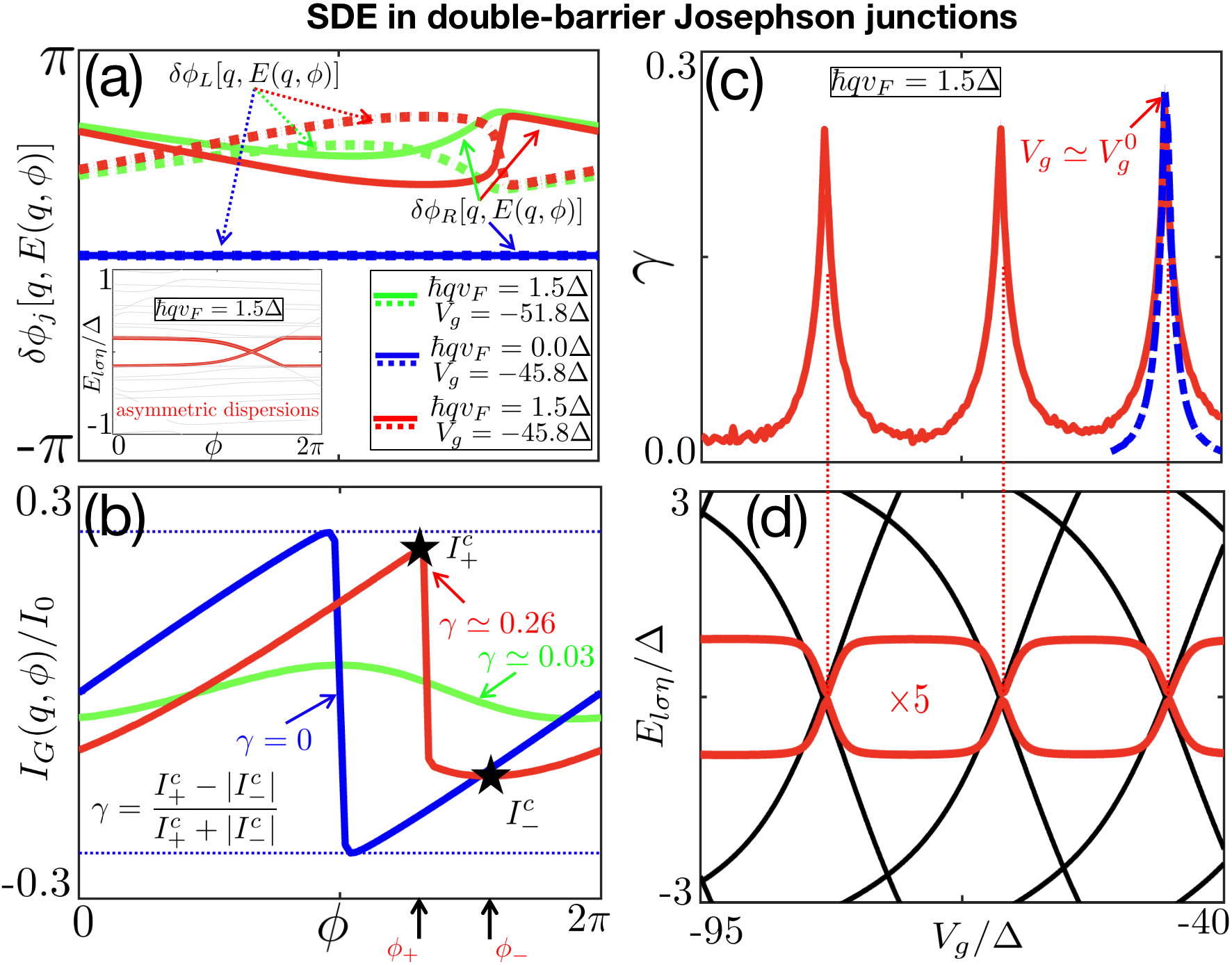}
\end{center}
\caption{(a) Additional phase shifts of the Andreev reflection, defined in Eq.~\eqref{fdavkkflw}. When $q\neq 0$, the asymmetry of $\delta \phi_{L,R}[q,E(q,\phi)]$ around $\phi=\pi$ leads to an asymmetric dispersion [inset of panel (a)] and supercurrent [panel (b)], resulting in a significant SDE [red line in panel (b)]. 
Remarkably, the SDE efficiency, denoted as $\gamma$, exhibits multiple Fabry-Perot resonances as a function of the well depth $V_g$ [panel (c)]. Peaks
at $\gamma \simeq 26\%$ correspond to zero-energy Andreev levels [red lines of panel (d)] stemming from 
the (non-superconducting) double-barrier quasi-bound states [black lines of (d)]. 
The resonance of $\gamma$ at $V_g^0\simeq -45.8 \Delta$ is well captured by the single-site quantum dot model, as indicated by the blue dashed line in panel (c).
Parameters: $\Delta=0.5$ meV, $m^*=0.03m_0$, $\mu_S=-19$ meV, and $\mu_B= 9$ meV.} 

\label{FIGDB}
\end{figure*}

By considering the three-point (second derivative) finite difference method with $N_T$ equally spaced 
discretized points $x_n$ between the end points $x_i$ and $x_f$, with $x_n=x_i + n\delta x$ for $n=0, 1,2,\cdots, N_T$, 
$\delta x=(x_f-x_i)/N_T$ being the (sufficiently small) discretization step, we can approximate 
the BdG equation of our Hamiltonian \eqref{fvgbvfglbm} by the coupled set of equations 
\begin{align} \label{fvmkfk0m}
    &-\frac{\hbar^2}{2m^{*}} \frac{\psi_{\uparrow} (x_n+\delta x)-2\psi_{\uparrow}(x_n)+\psi_{\uparrow}(x_n-\delta x)}{(\delta x_n)^2} \\
    &+\mu(x_n)\psi_{\uparrow}(x_n)+\Delta(x_n)\psi^{\dagger}_{\downarrow}(x_n)=E\psi_{\uparrow}(x_n), \notag
\end{align}
\begin{align} \label{fvmkfk1m}
    &-\frac{\hbar^2}{2m^{*}} \frac{\psi^{\dagger}_{\downarrow} (x_n+\delta x)-2\psi^{\dagger}_{\downarrow}(x_n)+\psi^{\dagger}_{\downarrow}(x_n-\delta x)}{(\delta x_n)^2} \\
    &+\mu(x_n)\psi^{\dagger}_{\downarrow}(x_n)-\Delta^{*}(x_n)\psi^{}_{\uparrow}(x_n)=E\psi^{\dagger}_{\downarrow}(x_n).\notag
\end{align}
As well known, the three-point second derivative approximation for numerical discretization, e.g., Eqs.~(\ref{fvmkfk0m}) and (\ref{fvmkfk1m}), 
can be formally mapped onto a 1D tight-binding model with nearest-neighbor hopping. More specifically, by making the replacements $\delta x\rightarrow a$, $\psi_s(x_n)\rightarrow c_{jns}$,  as well as $-\frac{\hbar^2}{2m^{*}(\delta x)^2} \rightarrow t_0$, we can immediately write down the corresponding 1D tight-biding model Hamiltonian
\begin{align} \label{0maindvmdlmain}
    H&=\sum_{j=L,R,C}\sum^{N_j}_{n=1}\sum_{s=\uparrow,\downarrow}(-2t_0+\mu_{jn})c^{\dagger}_{jns}c^{}_{jns}   \notag \\
   &+\sum_{j=L,R,C}\sum^{N_j-1}_{n=1}\sum_{s=\uparrow,\downarrow}\left(t_{0}c^{\dagger}_{jns}c^{}_{jn+1s}+h.c.\right)\notag \\
&+t_0\sum_{s=\uparrow,\downarrow}\left(c^{\dagger}_{LNs} c^{}_{C1s}+c^{\dagger}_{CMs}c^{}_{R1s} +h.c. \right)\notag\\
&+\sum_{j=L,R}\sum^{N_j}_{n=1}\left(\Delta^{j}_{n}c^{\dagger}_{jn\uparrow}c^{\dagger}_{jn\downarrow}+h.c. \right).
\end{align}
The subscript $j=C,L,R$ of electron operator $c_{jns}$ includes the semiconducting (Central) 
region $C$ in addition to the (Left and Right) superconducting leads $L,R$. The chemical potential $\mu_{jn}$ define the relevant offsets 
between the several layers in terms of $\mu_S$, $V_B$, and $V_g$, as shown by  Eq.~\eqref{fvvdldlvm}. Here, $N_j$ 
denotes the respective number of points in the $j=C,L,R$ layers, and  the total number of sites is given by $N_T=2N+M$ with $N=L_S/a$ and $M=L/a$, where $a$ is the spacing of the tight-binding mode.  The first (last) site $c_{C1s}$ ($c_{CMs}$) of the semiconductor layer is 
tunnel coupled to the last (first) site $c_{LNs}$ ($c_{R1s}$) of the left (right) superconducting lead. The tunnel 
coupling $t_0$ denotes the nearest-neighbor hopping amplitude in all regions. 
 For the left (right) lead, we 
assume a Fulde-Ferrell type \textit{proximitized} order 
parameter $\Delta^{j}_n=\Delta e^{i\phi_j+2iqna}$, where  $\phi_j$ and $\Delta$ are, respectively, the phase and absolute value of the proximitized gap of the superconducting lead $j$, $\hbar q$ is the momentum of the Cooper pairs, 
 and $a$ is the lattice constant. Below we assume $\phi_R=-\phi_L=\phi/2$, where $\phi$ is a global flux-tunable phase difference. The finite-momentum Cooper pairs can be realized by, e.g., direct current injection \cite{levine1965dependence,hansen1969observation,bardeen1962critical}, screening currents 
via the Meissner effect \cite{davydova2022universal,zhu2021discovery}, SO interaction + Zeeman field  \cite{dartiailh2021phase},  and exchange-mediated 
Fulde-Ferrel type mechanism \cite{fulde1964superconductivity}. In Appendix \ref{Qestimate}, we present a detailed explanation and estimation of the mechanism behind the finite-momentum Cooper pairs in our proximitized InAs/Al layer. These pairs can attain significant values of $\hbar qv_F\sim \Delta $, where $v_F$ represents the Fermi velocity, for experimentally accessible parameters. It is important to note that the momentum of the Cooper pairs should be smaller than the critical dispersing momentum ($q_c$) determined by the closure of the gap in the energy spectra of the continuum quasiparticles. Numerical analysis confirms that $q<q_c$, as evidenced by the presence of a clear gap in the energy spectra of the continuum quasiparticles [Fig.~\ref{SDE}(a) and the inset of Fig.~\ref{FIGDB}(a)].

\subsection{Ground-state supercurrent} \label{groundstatesupercurrent}

The tight-binding Hamiltonian \eqref{0maindvmdlmain} is exactly solvable via a (real space) Bogoliubov transformation 
\begin{align} \label{fdbldfl}
    H&=\frac{1}{2}\sum^{2N+M}_{l=1}\sum_{\sigma=\Uparrow,\Downarrow}\sum_{\eta=+,-} E^{}_{l\sigma\eta}(q,\phi)\gamma_{l\sigma\eta}^{\dagger} \gamma_{l\sigma\eta }^{}+\mathcal{E},
\end{align} 
where $l \in \mathbb{N}$ enumerates the dot-lead orbital states and $\eta=+,-$ labels the particle-hole degrees of freedom. Here, $\sigma=\Uparrow,\Downarrow$ denotes (pseudo) spin components (if we include spin-orbit coupling in Sec.~\ref{singlesiteQDJJ}). The prefactor $1/2$ in Eq.~\eqref{fdbldfl}  arises from the artificial doubling  \cite{bernevig2013topological} of the BdG formalism. The quasiparticle eigenenergies $E_{l\sigma\eta}$ and operators $\gamma^{}_{l\sigma\eta}$ are obtained via numerical diagonalization of the BdG Hamiltonian in the $(8N+4M)$-component Nambu spin space ($\{c_{n\uparrow},c^{}_{n\downarrow}, -c^{\dagger}_{n\downarrow},+c^{\dagger}_{n\uparrow}\}$ for all $n$) and obey particle-hole symmetry $E_{l\sigma\eta}(q,\phi)=-E_{l-\sigma-\eta}(q,\phi)$ and conjugate relation $\gamma^{\dagger}_{l\sigma\eta}= \gamma^{}_{l-\sigma-\eta}$. Hamiltonian \eqref{fdbldfl} is written in the  two-quasiparticle representation $\{\gamma_{l\sigma+}, \gamma_{l\sigma-}\}$, with $\mathcal{E}=\sum_{jn}\epsilon_{jn}$ being $\phi$-independent but dependent  on $\mu_S$.

Next we obtain the ground state of our 
system as defined by 
$\gamma^{}_{l\sigma\eta} \vert G\rangle=0$ for all $E_{l\sigma\eta}>0$, implying  that all positive-energy quasiparticle states are empty in the ground state, i.e., $\langle G\vert\gamma^{\dagger}_{l\sigma\eta}\gamma^{}_{l\sigma\eta} \vert G\rangle=0$ for all $E_{l\sigma\eta}>0$. Calculating 
the ground state of an entangled system is by itself challenging.  
Further difficulty arises from fermion-parity changes with system parameters. For 
convenience, we determine the ground 
state from the effective vacuum state $\vert V\rangle_{+}$ defined by $\gamma_{l\sigma+}\vert V\rangle_{+}=0$ 
in the orthogonal basis set $\{\gamma_{l\sigma+}\}$ for all $l$ and $\sigma$, 
with the vacuum energy   
$\mathcal{E}_{+}(q,\phi)=\mathcal{E}+\sum_{l\sigma}\frac{1}{2}E^{}_{l\sigma-}(q,\phi)$. This choice of basis suitably guarantees the effective vacuum state $\vert V\rangle_{+}$ has even parity so that the ground state is even for 
$E_{1\Downarrow+}(q,\phi)>0$ and odd for $E_{1\Downarrow+}(q,\phi)<0$, [Fig.~\ref{SDE}(a)]. Adding all negative-energy quasiparticles 
[$E_{l\sigma+}(q,\phi)<0$] to the vacuum 
state $\vert V\rangle_{+}$,  we obtain \textit{by construction} the ground-state wave function 
\begin{align}
    \vert G \rangle=\left(\prod_{E_{l\sigma+}<0}\gamma^{\dagger}_{l\sigma+}\right)\vert V\rangle_{+}, 
\end{align}
and, from $H \vert G \rangle = \mathcal{E}_G\vert G \rangle$, the ground-state energy $  \mathcal{E}_G(q,\phi)=\mathcal{E}_{+}(q,\phi)+ \sum_{E_{l\sigma+}<0}E_{l\sigma+}(q,\phi).$ We recast the above as 
\begin{align} \label{fvkdfmvk}
    \mathcal{E}_{G}(q,\phi)=\mathcal{E}+ \frac{1}{2}\sum_{E_{l\sigma\eta}<0}E_{l\sigma\eta}(q,\phi),
\end{align}
where we have used $E_{l\sigma\eta}(q,\phi)=-E_{l-\sigma-\eta}(q,\phi)$. As shown in Appendix~\ref{supercurrent}, the 
ensemble-averaged supercurrent
$I_G(T,q,\phi)\sim \partial_\phi \Omega(T,q,\phi)$, 
with $\Omega(T,q,\phi)$ being the grand potential function and $T$ the temperature 
\cite{anderson1964many,bagwell1992suppression,riedel1998low,beenakker1991universal,beenakker2013fermion}. At $T=0$, $\Omega(0,q,\phi)=\mathcal{E}_G(q,\phi)$ [Eq.~\eqref{fvkdfmvk}] and $I_G(q,\phi)=I_G(0,q,\phi)$ is
\begin{align} \label{fdjvkdfr}
I_G(q,\phi)=\frac{I_0}{2}\sum_{E_{l\sigma\eta}<0}\partial_{\phi}E_{l\sigma\eta}(q,\phi),
\end{align}
where $I_0=2e/\hbar$, $e<0$ is the electron charge, $\hbar=h/2\pi$, and $h$ is the Planck constant. Equation~\eqref{fdjvkdfr} has the great advantage as it enables one to calculate $I_G(q,\phi)$ including all negative-only eigenenergies without worrying about parity changes and/or double counting \cite{bardeen1969structure,chtchelkatchev2003andreev,beenakker2013fermion}.

\subsection{Additional phase shifts in the Andreev reflection of a finite-momentum superconductors}\label{Additionalphaseshifts}

Following the same way as Ref.~\cite{zhang2022logic}, directly diagonalizing the total Hamiltonian \eqref{maindvmdl} causes the following reduced determinant equation whose solutions contain exact Andreev levels
\begin{align} \label{mfdvkdfmv}
   \det\left\{\mathcal{H}_C-E-\Sigma(E)\right\}
    =0.
\end{align}
The spin degeneracy in Hamiltonian \eqref{0maindvmdlmain} halves the dimension of the Hamiltonian matrix and thus $\mathcal{H}_{j}$ is the $2N_j\times 2N_j$ Hamiltonian matrix of the layer $j$ in the Nambu space $\Psi^{\dagger}_j=\prod^{N_j}_{n=1}[c^{\dagger}_{jn\uparrow}, -c^{}_{jn\downarrow}] 
   $. The self-energy arises from  integrating out the superconducting leads~\cite{silvester2000determinants} 
\begin{widetext}
\begin{align} \label{mselfenergy}
   \Sigma(q,E,\phi_j) &=t^2_0\begin{bmatrix}
   \tau_z[G_{L}(q,E,\phi_L)]_{2\times 2} \tau_z & 0 & \cdots & 0 & 0\\
    0 & 0 & \cdots & 0 &0\\
     \vdots & \vdots & \ddots & 0& 0\\
       0 & 0 & 0 & 0 &0\\
    0 & 0 &0& 0 & \tau_z[G_{R}(q,E,\phi_R)]_{2\times 2} \tau_z
    \end{bmatrix}_{2M\times 2M}.
\end{align}
\end{widetext}
Here,  $[G_{L}(q,E,\phi_L)]_{2\times 2}$ and $[G_{R}(q,E,\phi_R)]_{2\times 2}$ are the lowermost and uppermost $2\times 2$ diagonal blocks of $G_{L}=1/(\mathcal{H}_{L}-E)$ and $G_{R}=1/(\mathcal{H}_{R}-E)$, respectively, 
\begin{align} \label{fdvmdllLm}
   G_{L}(q,E,\phi_L)\equiv
    \begin{bmatrix}
        \square & \square & \square & \square \\
        \square & \ddots & \vdots & \vdots \\
        \square & \cdots & \square &  \square\\
        \square & \cdots & \square & [G_{L}(q,E,\phi_L)]_{2\times 2}
 \end{bmatrix}_{2N\times 2N},
\end{align}
\begin{align} \label{fdvmdllRm}
       G_{R}(q,E,\phi_R)\equiv
    \begin{bmatrix}
       [G_{R}(q,E,\phi_R)]_{2\times 2} & \square & \cdots & \square \\
        \square & \square & \cdots & \square  \\
        \vdots & \vdots & \ddots & \square  \\
        \square & \square & \square  & \square  
 \end{bmatrix}_{2N\times 2N}.
\end{align}
Each empty square in Eqs.~\eqref{fdvmdllLm} and \eqref{fdvmdllRm} is a two-by-two matrix, irrelevant to the Andreev levels. Next, we write out $[G_{j}(q,E,\phi_j)]_{2\times 2}$, explicitly using their Hermiticity
\begin{align} \label{fdvmdkfmkvM}
    [G_{j}(q,E,\phi_j)]_{2\times 2}\equiv\begin{bmatrix}
    \mathcal{G}^{1,1}_{j}(q,E,\phi_j) & -\mathcal{F}_{j}(q,E,\phi_j)  \\
     -\mathcal{F}^{*}_{j}(q,E,\phi_j) & \mathcal{G}^{2,2}_{j}(q,E,\phi_j) 
    \end{bmatrix}.
\end{align}
Hermiticity requires that diagonal Green functions
$\mathcal{G}^{1,1}_{j}(q,E,\phi_j)$ and $\mathcal{G}^{2,2}_{j}(q,E,\phi_j)$ are real number, which can have different values for the proximitized InAs/Al 
leads (see Appendix~\ref{asymmetricdispersions}). We  extract the moduli and phases of 
the anomalous Green functions $\mathcal{F}_{j}(q,E,\phi_j)$ by using their polar form
\begin{align} \label{fdavkkflw}
    \mathcal{F}_{j}(q,E,\phi_j) \equiv \left| \mathcal{F}_{j}(q,E)\right| e^{i\phi_j+i\delta\phi_j(q,E)}.
\end{align}
Notably, $\delta\phi_j(q,E)$ are additional phase shifts due to the Andreev reflections at the finite-momentum superconducting leads (see detailed explanation in Appendix \ref{AndreevrReflection}). The minus sign in front of $\mathcal{F}_{j}(q,E,\phi_j)$ in Eq.~(\ref{fdvmdkfmkvM}) is added such that $\delta\phi_j(q,E)=0$ when $q=0$ [blue lines in Fig.~\ref{FIGDB}(a)]. Remarkably, the form of the off-diagonal component of the self 
energy [i.e., Eq. \eqref{fdavkkflw}] is quite general. Irrespective of the details of the 
superconducting leads, their effect on the Andreev reflections are  captured by the additional phase shifts 
$\delta\phi_j(q,E)$ that can be numerically calculated.

Note that the anomalous Green functions \eqref{fdavkkflw} are responsible for the Andreev reflections which couple electrons with holes of opposite spins. 
As shown in Fig.~\ref{STORY}(b),  the interference between left- and right-lead Andreev reflections generates phase-tunable Andreev levels and ground-state supercurrent.  For finite-$q$ leads [red and blue curves in Fig.~\ref{FIGDB}(a)], holes acquire additional phase shifts after the multiply reflections $\delta\phi_{j}[q,E(q,\phi)]$, where the $\phi$-modulation of the additional phase shifts is introduced by the phase-tunable Andreev levels, e.g., $E_{1+}(q,\phi)$. Notably, these additional phase shifts are asymmetric with respect to $\phi$, resulting in an asymmetric dispersion [inset of Fig.~\ref{FIGDB}(a)] and a finite SDE [Fig. \ref{FIGDB}(b)]. Figure \ref{FIGDB}(b) illustrates $I_G(q,\phi)$ vs. $\phi$ 
\CR{[Eq.~\eqref{fdjvkdfr}]} for our resonant-tunneling Josephson junction, showing no effect for $q=0$ (blue line) and a sizable SDE for $q \neq 0$ (red and green lines)   \cite{ando2020observation,daido2022intrinsic,jiang2022superconducting,lin2022zero}, and reveals that more asymmetric additional phase shifts contribute to larger SDE. 
Interestingly, the diode efficiency $\gamma=( I^{c}_{+} - \vert I^{c}_{-}\vert)/(I^{c}_{+} + \vert I^{c}_{-}\vert)$ 
features [Fig. \ref{FIGDB}(c)] multiple Fabry-Perot resonances due to the 
many $V_g$-tunable Andreev bound states [vertical dotted lines in Figs. \ref{FIGDB}(c) and (d)]. 
The resonant peaks at $\gamma \simeq 26\%$ in Fig. \ref{FIGDB}(c) stem from the ordinary 
zero-energy quasi-bound states of the double-barrier potential 
[Fig.~\ref{FIGDB}(d), black lines] morphing into Andreev bound states [Fig.~\ref{FIGDB}(d), red lines] due to the superconducting leads.
The troughs in Fig.~\ref{FIGDB}(c) have $\gamma \simeq 3\%$, see green solid line in Fig.~\ref{FIGDB}(b) for $V_g =-51.8\Delta$. Therefore, we microscopically explain the SDE from the additional phase shifts acquired during the Andreev reflections of finite-momentum superconductors, which exhibits multiple Fabry-Perot resonance.

\subsection{Single-site quantum dot Josephson junction} \label{singlesiteQDJJ}

\begin{figure*}[t!]
\begin{center}
\includegraphics[width=0.7\linewidth]{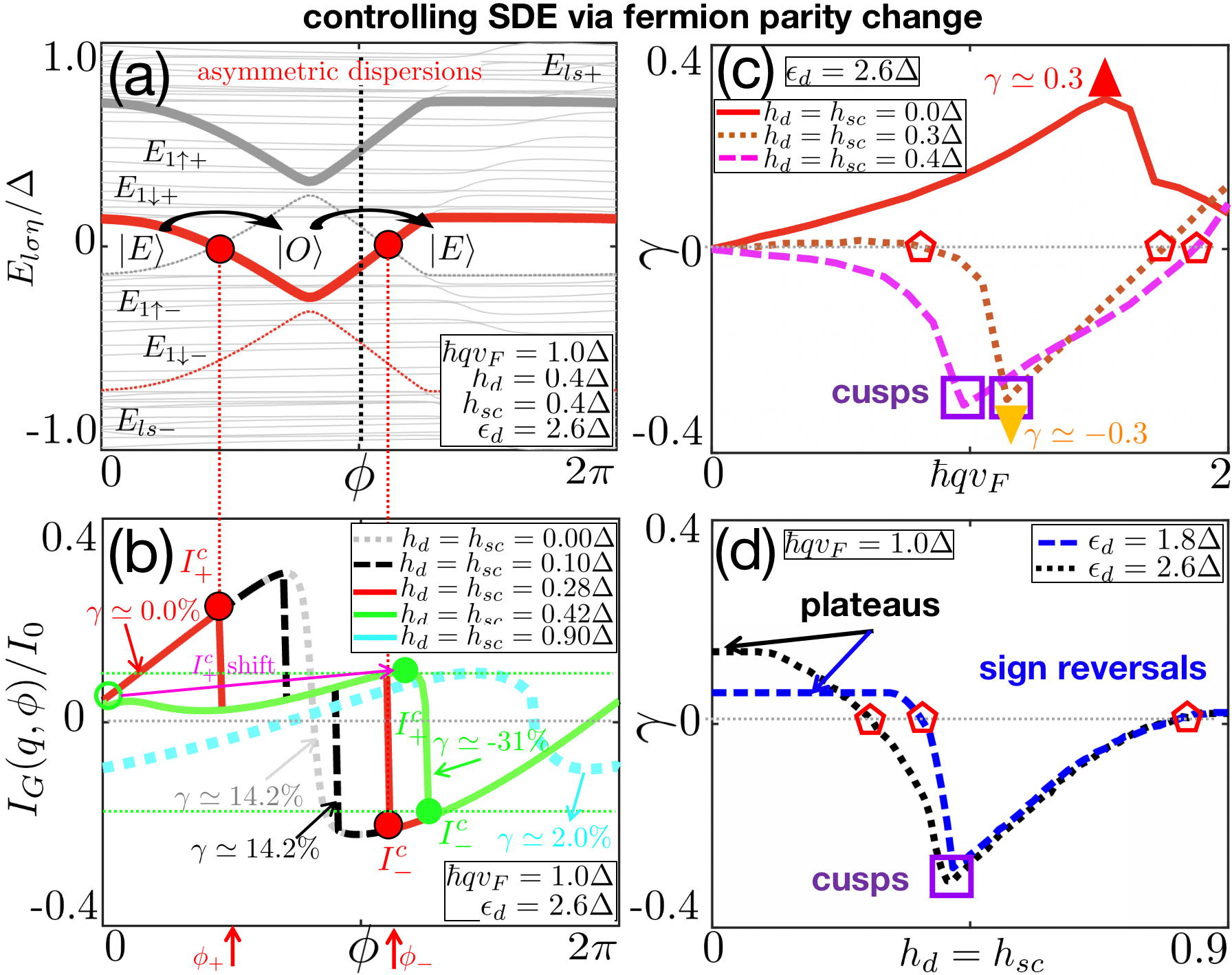}
\end{center}
\caption{(a) Full BdG spectrum $E_{ls\eta}$ of our single-site dot Josephson junction showing its asymmetry about $\phi=\pi$, i.e., 
$E_{l\sigma\eta} (q,\pi - \phi) \neq E_{l\sigma\eta}(q,\pi+\phi)$ for $q\neq 0$. Here we chose $\epsilon_d=-64\Delta$ and 
adjust the dot-lead coupling $t=-13\Delta$ to obtain the resonance at $V_g^0$ of the continuum model in Fig.~\ref{FIGDB}(d). 
Red and thin grey lines indicate the Zeeman-split Andreev levels crossing at zero 
energy and inducing even-odd-even fermion parity changes in the 
ground state as $0<\phi<2\pi$. These transitions induce discontinuities 
in $I_G(q,\phi)$ vs $\phi$, see solid red circles in (b), and can strongly affect the SDE. The diode efficiency $\gamma$ undergoes 
multiple sign reversals as a function of (c) $q$ (for several $h_d=h_{sc}$) and (d) the Zeeman field,  
for a fixed $q$ and $\epsilon_d=1.8\Delta$, $2.6\Delta$; it also exhibits cusps and plateaus. Cusps arise when 
$I_c^+$ shifts significantly to the right [see empty and filled green circles in (b)] due to a parity transition. 
After a cusp $\max[I_G(q,\phi)]$ and $\min[I_G(q,\phi)]$ can yield $I^c_+$ and $I^c_-$ occurring within a very narrow 
range of $\phi$, see, e.g., green curve in (b) thus enabling high sensitiveness of $\gamma$. 
Plateaus occur when the magnetic fields, though inducing significant changes in $I_G(q,\phi)$ vs $\phi$ due to parity changes
[see grey and dashed lines in (b)], do not affect the critical currents $I^c_+$, $I^c_-$. 
Parameters: $N=2000$, $\Delta=0.5$ meV, $t_0=-100\Delta$, and $\theta_{\text{so}}=0$.}
\label{SDE}
\end{figure*}

It is important to know that the single-site quantum-dot Josephson-junction model describes well the resonant behavior of $\gamma$; 
see, e.g., the peak at $V_g^0=-45.8\Delta$ [see blue dashed lines in Fig.~\ref{FIGDB}(c)].  In the following analysis, we focus on this simplified single-site quantum dot model, which offers several advantages. Firstly, it allows us to consider the influence of additional parameters such as Zeeman magnetic field and spin-orbit coupling. Moreover, the simplicity of this model enables us to perform analytical calculations, facilitating a better understanding of the underlying physical mechanisms. Below, for simplicity, we only present the 1D tight binding 
model \cite{davydova2022universal}  
for a single-site quantum dot coupled to superconducting leads, plotted in bottom panel of Fig.~\ref{STORY}(a) 
\begin{align} \label{maindvmdl}
    H&=\sum^{N}_{n=1}\sum_{j=L,R}\sum_{s=\uparrow,\downarrow}\epsilon_{jns}c^{\dagger}_{jns}c^{}_{jns}+\sum_{s=\uparrow,\downarrow}(\epsilon_d+sh_d)d^{\dagger}_{s}d^{}_{s}\notag  \\
   &+\sum^{N-1}_{n=1}\sum_{j=L,R}\sum_{s=\uparrow,\downarrow}\left(t_{0}c^{\dagger}_{jns}c^{}_{jn+1s}+h.c.\right)\notag \\
&+\sum^{N}_{n=1}\sum_{j=L,R}\left(\Delta^{j}_{n}c^{\dagger}_{jn\uparrow}c^{\dagger}_{jn\downarrow}+h.c. \right)\\
 &+t\sum_{ss'}\left[c^{\dagger}_{LNs} \mathcal{U}_{ss'}(\theta_{so})d_{s'}+d^{\dagger}_{s} \mathcal{U}_{ss'}(\theta_{so})c_{R1s'} +h.c. \right].\notag
\end{align}
In the presence of the magnetic field, the electron energy in lead $j$ is given by $\epsilon_{jns}=-2t_0+\mu_{j}+sh_{sc}$, with $h_{sc}$ being the Zeeman energy. 
The operator $d_{s}^{}$  annihilates an electron state in the dot having 
 spin component $s$ and energy $\epsilon_d+sh_d$, with 
 $\epsilon_d$ and $h_d$ being the dot and Zeeman energies, respectively.  The dot energy $\epsilon_d$ is assumed 
 to be one of the gate-tunable quasi-bound states $\epsilon^{i}_{w}(V_g)$, $i=1,2,\cdots$, of the double-barrier potential well. In Figs. \ref{SDE} 
 and \ref{GFSDE}, for instance, we take $\epsilon_d=\epsilon_{w}(-45.8)=2.6\Delta$, corresponding to the resonance at $V^0_g=-45.8\Delta$ 
 in Fig.~\ref{FIGDB}(d). The spin rotation matrix $\mathcal{U}
 (\theta_{\text{so}})=e^{i\theta_{\text{so}} s^{y}/2}$ accounts for the spin-orbit induced 
spin rotation $\theta_{so}$ in the tunnelling 
 between the dot and the last (first) site $c_{LNs}$ ($c_{R1s}$) of the left (right) lead with  coupling strength 
 $t$~\cite{dell2007josephson,Hoffman2017,padurariu2010theoretical,GonzalezRosado2021,spethmann2022coupled}. Noting that $t$ is related to the the double-barrier 
potential, we treat $t$ as a fitting parameter to reproduce one of the 
resonances of the double-barrier Josephson junction.

The natural starting point for studying single-site quantum dot Josephson junction is a superconducting quantum dot described by an effective dot Hamiltonian where the leads are integrated out to generate an order parameter on the dot \cite{meng2009self,kurilovich2021microwave,rozhkov1999josephson,bauer2007spectral,meden2019anderson,fatemi2021microwave}. Again, we obtain a reduced determinant equation for exact Andreev levels (see detailed derivations in Appendix \ref{fvdkvoav})
\begin{align} \label{mfdvkdfmv}
   \det\left\{\epsilon_D\tau_z+h_Ds^{z}-E-\Sigma(E)\right\}
    =0,
\end{align}
with
\begin{align} \label{mselfenergy}
    \Sigma(E)&= t^2U_{}(\theta_{so})[G_{L}(q,E,\phi_L)]_{4\times 4}U^{+}(\theta_{so})\\
    &+t^2U^{+}(\theta_{so})[G_{R}(q,E,\phi_R)]_{4\times 4}U^{}(\theta_{so}),\notag 
\end{align}
where $U(\theta_{so})=\text{diag}[\mathcal{U}(\theta_{so}); -\mathcal{U}(\theta_{so})]$ is the spin rotation of the spin-orbit-coupled tunneling in Nambu space. In the absence of the spin degeneracy, $[G_{L}(q,E,\phi_L)]_{4\times 4}$  and $[G_{R}(q,E,\phi_R)]_{4\times 4}$ are given by the last and first four-by-four matrix of $G_{L}=1/(\mathcal{H}_L-E)$  and $G_{R}=1/(\mathcal{H}_R-E)$, respectively, which are tunnel coupled to the quantum dot  
\begin{align} \label{fdvadavfv}
   [G_{j}&(q,E,\phi_j)]_{4\times 4}\equiv\begin{bmatrix}
    \mathcal{G}^{1,1}_{j\uparrow} & 0 & -\mathcal{F}^{}_{j\uparrow} & 0 \\
    0 & \mathcal{G}^{1,1}_{j\downarrow} & 0 & -\mathcal{F}^{}_{j\downarrow}\\
     -\mathcal{F}^{*}_{j\uparrow} & 0 & \mathcal{G}^{2,2}_{j\uparrow} & 0\\
    0 & -\mathcal{F}^{*}_{j\downarrow} & 0 & \mathcal{G}^{2,2}_{j\downarrow}
    \end{bmatrix}.
\end{align}
Again, the anomalous Green function is expressed by the polar form, i.e.,  $\mathcal{F}_{js}(q,E,\phi_j) \equiv \left| \mathcal{F}_{js}(q,E)\right| e^{i\phi_j+i\delta\phi_j(q,E)}$. In the absence of spin-orbit coupling ($\theta_{so}$), the interference between the left-lead and right-lead Andreev reflections can be captured by 
\begin{align} \label{fvjdfadjv}
    \sum_j\mathcal{F}_{js}(q,E,\phi_j)=\left| \mathcal{F}_{s}(q,E)\right|\cos\left[\frac{\phi+\delta\phi(q,E)}{2}\right]e^{i\phi_{\text{eff}}},
\end{align}
with $\phi_{\text{eff}}=[\delta\phi_R(q,E)+\delta\phi_L(q,E)]/2$, where  
$\delta\phi(q,E)=\delta\phi_R(q,E)-\delta\phi_L(q,E)$ is the additional \textit{phase-shift difference}  from the Andreev reflections in the left and right finite-momentum 
superconductors. Hereafter, we assume the identical left and right superconducting leads except for the phase of order parameter. 

In the absence of Zeeman fields, we have also derived the approximate implicit solution for the Andreev levels of this single-site quantum dot model
$E_{1\sigma+}(q,\phi) \simeq \sqrt{(\epsilon^{}_d-\epsilon^S_d)^2+4\Gamma^2\cos^2\left[\frac{\phi+\delta\phi[q,E_{1\sigma +}(q,\phi)]}{2}\right]}$, where the additional dot energy $\epsilon^S_d$ arises from the renormalization of the lead-dot
tunneling coupling (Appendix~\ref{asymmetricdispersions}). This clearly shows that the 
Andreev dispersions are asymmetric about $\phi=\pi$ [Fig.~\ref{SDE}(a)], due to the asymmetry in 
$\delta\phi[q,E_{1\sigma+}(q,\phi)]$ [Fig.~\ref{FIGDB}(a)], thus leading to the SDE; 
here $\Gamma/\hbar$ denotes the total dot-lead tunnel rate. Note that $E_{1\sigma+}(q,\phi)$ above
reduces to the known results for $q=0$. In the presence of Zeeman fields, Fig.~\ref{SDE}(a) shows the full BdG spectrum $E_{l\sigma\eta}(q,\phi)$ vs. $\phi$ with finite-$q$ leads and no spin-orbit coupling in the dot-lead coupling ($\theta_{so}=0$). The Zeeman fields in the dot $h_d$ 
and in the leads $h_{sc}$ can cause fermion-parity changes of the ground state due to Zeeman-split negative-energy 
levels $E_{l\sigma+}(q,\phi)$. We find even-odd-even parity transitions at $\phi= \phi_{\pm}$ 
as the subgap levels $E_{1\Uparrow-}(q,\phi)$, $E_{1\Downarrow+}(q,\phi)$ cross each at zero energy [see filled red
circles in Fig.~\ref{SDE}(a)]. In Appendix \ref{zeroSO}, we show that   
$\phi_{\pm}=-\delta\phi(q,h_{sc})\pm2\arccos\left[\frac{[h_d+2 t^2h_{sc}\mathcal{K}(q,h_{sc})]^2- [\epsilon_d-2(\mu_S-2t_0)t^2\mathcal{D}(q,h_{sc}\tilde{\mu}_S)]^2}{4t^4\mathcal{R}^2(q,h_{sc})}\right]^{1/2}$, with the functions $\mathcal{K}(q,h_{sc})$,  $\mathcal{D}(q,h_{sc})$, $\mathcal{R}(q,h_{sc})$ 
defined following Eq.~\eqref{fvndkdvk}. Parity changes can affect $I_G(q,\phi)$ and the SDE as we discuss \CR{in Sec.~\ref{CSDEFPC}}.

\section{Results and discussions} \label{resultsanddiscussions}

\subsection{Controlling SDE via fermion-parity change} \label{CSDEFPC}

\begin{figure*}[t]
\begin{center}
\includegraphics[width=0.7\linewidth]{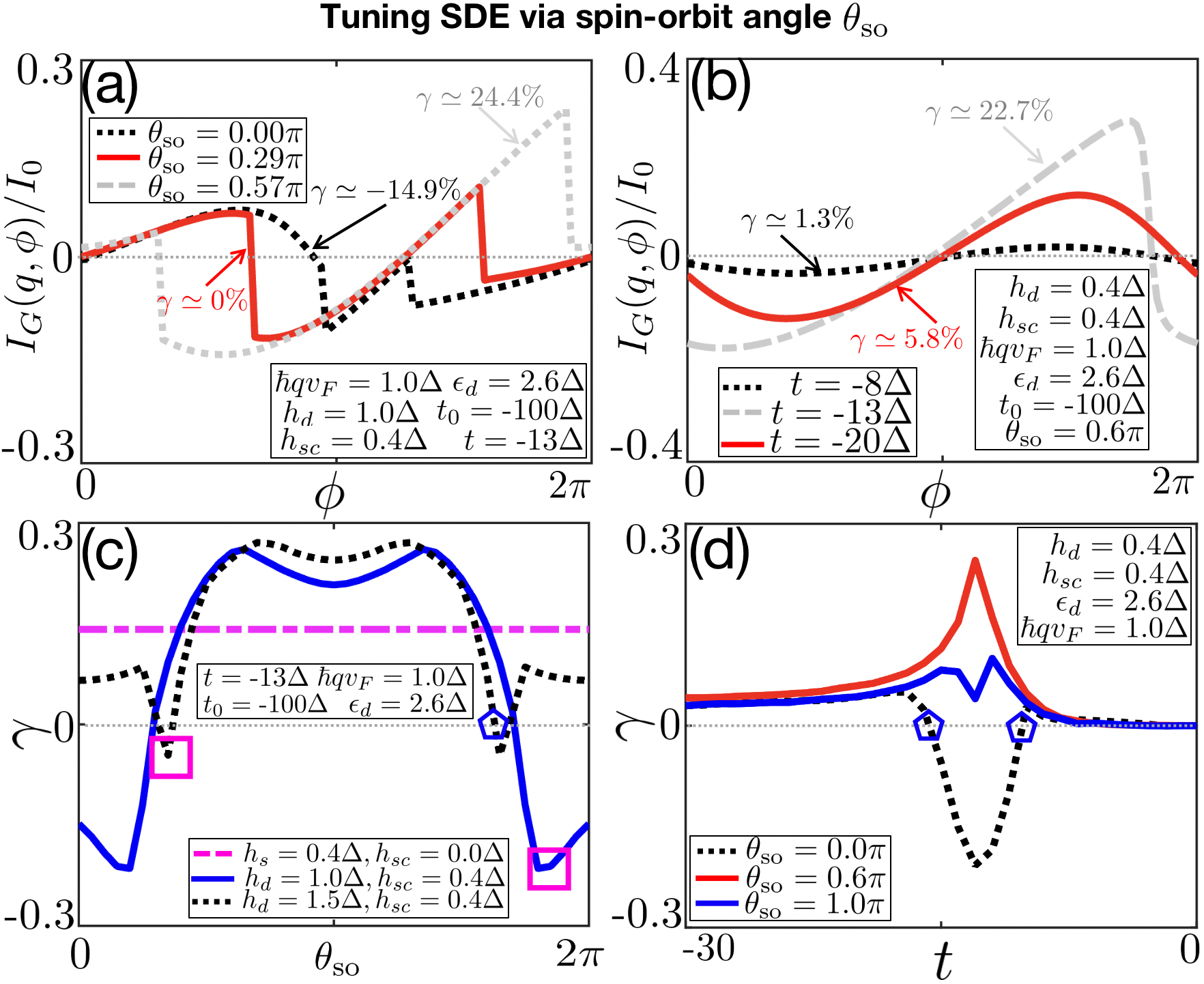}
\end{center}
\caption{Supercurrent $I_G(q,\phi)$ vs $\phi$ for various (a) angles $\theta_{\text{so}}$ describing the 
SO-induced spin rotation in the dot-lead tunneling process and (b) tunnel-coupling 
strengths $t$. We use $\hbar qv_F= 0.1  \Delta$ in all panels. Varying  
$\theta_{\text{so}}$ affects the fermionic parity transitions manifest in $I_G(q,\phi)$ vs $\phi$ 
such: that $I^c_+ < I^c_-$ for $0< \theta_{so}<  0.29\pi$ 
($\gamma<0$), $I^c_+ =I^c_-$ at $\theta_{\text{so}}\simeq 0.29\pi$ ($\gamma\simeq 0$), 
and $I^c_+ > I^c_-$ for $0.29\pi<\theta_{so}<0.56\pi$ ($\gamma>0$). This is more clearly seen in (c) that shows $\gamma$
being \textit{periodically modulated}  by $\theta_{so}$ for a variety of Zeeman parameters 
$h_{sc}$ and $h_d$; the blue curve in (c) corresponds to the 
parameters in panel (a). Note also its four $\gamma$-sign reversals [see pentagon symbols in (c)]. 
Other curves in (c) also shows sign reversals and, additionally, cusps.  
For $h_{sc}=0$ [cyan curve in (c)], the spin basis of superconducting leads can be rotated so as 
to remove the effect of the spin-orbit coupling, thus leading to a diode efficiency independent of $\theta_{\text{so}}$. The diode 
efficiency is also strongly modulated by tuning the tunneling strength $t$ as shown in (d), 
which shows resonant behaviour around $t=-13\Delta$. The effective dot energy induced by the tunneling coupling with superconducting leads cancels with $\epsilon_d=2.6\Delta$.  Similarly to (c) and (a), the blue curve in (d) corresponds to the parameters in (b). }
\label{GFSDE}
\end{figure*}

The discontinuities in $I_G(q,\phi)$ vs. $\phi$, see $I_G(q,\phi_+)$ 
and  $I_G(q,\phi_-)$ [red filled circles in Fig.~\ref{SDE}(b)], correspond to Andreev levels crossing zero [Fig.~\ref{SDE}(a)] 
and thus signal fermion parity changes. For $q\neq 0$, these transitions {\it can} significantly alter the forward and reverse 
critical supercurrents $I_+^c$, $I_-^c$, affecting in turn the SDE [Fig.~\ref{SDE}(b)]. For $\hbar qv_F\simeq 1\Delta$ 
[Fig.~\ref{SDE}(b)], we have a finite diode efficiency $\gamma \simeq 14.2\%$ (gray line). 
Interestingly, the SDE can be suppressed (red line, $\gamma\simeq 0$) or enhanced -- and even sign 
reversed (green line, $\gamma\simeq -30\%$) --  by Zeeman tuning the fermion parity. 
Note that parity transitions do not affect $I_-^c$, $I_+^c$ when 
they take place away from the extrema of the current-phase relation $I_G(q,\phi)$; cf dotted grey line and long-short dashed black line in Fig.~\ref{SDE}(b): both have $\gamma \simeq 14.2\%$.  

Figures \ref{SDE} (c) and (d) show the diode efficiency $\gamma$ as a function of $q$ and 
$h_{d}=h_{sc}$, respectively. For $h_d=h_{sc}=0$ [red line of Fig.~\ref{SDE} (c)], the maximum efficiency $\gamma\simeq 30\%$ 
is attained at $\hbar qv_F\simeq 1.5\Delta$, which 
describes a large asymmetry between forward and reverse critical supercurrents 
$I^{c}_{+}/I^{c}_{-}\simeq 186\%$. For nonzero fields, $\gamma$ shows sign reversals and cusps
as $q$ increases [Fig.~\ref{SDE}(c), see dotted and dashed lines]. Interestingly, for a larger field $h_d=h_{sc}=0.4\Delta$, $\gamma$ becomes fully 
negative with a cusp ($\gamma\simeq -30\%$) at $\hbar qv_F \simeq \Delta$.  
Figure~\ref{SDE}(d) shows that $\gamma$ is initially independent of $h_d$, $h_{sc}$ 
because the fermion parity transition does not affect $I^c_+$ and $I^c_-$ 
[see long-short dashed black curve in Fig.~\ref{SDE} (b)]. It then undergoes two sign reversals and exhibits a cusp. 
The cusp at $h_{sc}=h_d\sim 0.4 \Delta$ is due to the $I^{c}_{+}$ 
shifting closer to $I^c_-$ as shown in Fig.~\ref{SDE}(b) (see 
arrow connecting the empty and filled green circles). Interestingly, due to this shift, $I^c_+$ and $I^c_-$  
occur at essentially the same critical phases, i.e., $\phi_{+}\simeq\phi_{-}$, as shown by the green 
curve, Fig.~\ref{SDE}(b). This should enable phase-tunable SDE devices with high sensitivity 
over a narrow phase range. It also allows for short switching times between forward and reverse 
critical currents. The diode efficiency is linear for  
small $q$ [Fig.~\ref{SDE}(c)] and its slope can change sign with field. 
Distinct fields ($h_d \neq h_{sc}$) do not qualitatively change the above results.

\subsection{Tuning SDE via the spin-orbit angle} \label{tuningSEDSOC}

In the presence of spin-orbit interaction in the tunneling between the dot and the leads, spin rotation mixes 
the spin-dependent Andreev-reflection paths and hence can affect 
$I_G(q,\phi)$. This couples electrons and holes with opposite \textit{and} same spins, see gradient arrows in Fig.~\ref{STORY} (b). 
The spin-rotation unitary matrix $\mathcal{U}(\theta_{\text{so}})$ accounts for the SO induced rotation in the dot-lead 
tunneling. Figures~\ref{GFSDE}(a) and (b) show the supercurrent $I_G(q,\phi)$ as a function of $\phi$ for different 
$\theta_{\text{so}}$ and tunnel coupling amplitude $t$, respectively. For $\hbar qv_F= 1.0 \Delta$ and a vanishing dot Zeeman 
field $h_d=0$ and non-zero $h_{sc}$, Fig.~\ref{GFSDE} (a), we obtain a sizable SDE with diode efficiency 
$\gamma\simeq -15 \%$ for $\theta_{\text{so}}=0$ (black dashed line).
As the spin-orbit angle $\theta_{\text{so}}$ increases, $\gamma$ is first suppressed 
$\theta_{\text{so}}< 0.29\pi$ (grey dashed line, $\gamma\simeq 0$) 
and then enhanced for $\theta_{\text{so}}<0.57\pi$ (green line, $\gamma\simeq 24.4\%$). 
Note that varying $\theta_{so}$ affects $I_G(q,\phi)$ vs. $\phi$ as it significantly alters the fermion 
parity transition and in turn the  forward and reverse critical supercurrents.
A similar effect happens when we vary $t$, possibly via an electrostatic gate, as shown in Fig.~\ref{GFSDE} (b).

The above features are more systematically seen in Figs.~\ref{GFSDE}(c) and (d) showing the diode efficiency 
$\gamma$ as a functions of $\theta_{\text{so}}$ and $t$, respectively. 
Interestingly, for $h_{sc}=0$, we can rotate the spin basis of the superconducting 
leads to remove the effect of spin-orbit coupling in the dot-lead tunnel couplings so that the diode efficiency $\gamma$ becomes independent of $\theta_{\text{so}}$. 
For finite $h_{sc}$ the spin-orbit significantly modulates the SDE, see the blue curve in Fig.~\ref{GFSDE}(c) with $h_{sc}=0.4$ and $h_d=0$. 
For this blue curve we see four $\gamma$ sign reversals with increasing $\theta_{so}$, while for a 
higher dot Zeeman field [red line in Fig.~\ref{GFSDE} (c)], the diode efficiency is always positive but does show a 
highly non-linear behavior. Similarly, Fig.~\ref{GFSDE}(d) shows $\gamma$ sign reversals and cusps as $t$ is varied [cf. Fig.~\ref{GFSDE}(b)].


\section{Conclusion} \label{conclusion}
We investigated the SDE in InAs-based resonant-tunneling Josephson 
junctions with proximitized InAs/Al leads. We find sizable SDE with diode efficiencies $\gamma$ showing 
tunable Fabry-Perot resonances, peaking at $\sim 26\%$. 
We identify the asymmetry of the phase shifts $\delta \phi_{L,R}$, 
due to the multiple Andreev reflections, as the mechanism inducing asymmetric Andreev dispersions and unequal 
depairing currents $I^c_+ \neq |I^c_-|$. Within a simpler single-site dot model for a 
Joseph junction, which captures resonant tunneling, we have additionally found that 
our SDE is highly tunable with $\gamma$ exhibiting multiple sign reversals as a function of the SO coupling in 
the dot-lead tunneling process, Zeeman fields (dot and leads), and finite momentum 
$\hbar q$. Fermion parity transitions can also significantly change SDE. 
The tunability afforded by our proposed Fabry-Perot superconducting diode should allow the 
design of new multifunctional devices for future superconducting electronics.

\section{Acknowledgements} \label{acknowledgement}
We sincerely thank J. Carlos Egues and Yugui Yao for helpful and sightful discussions. This work was supported by the National Key R\&D Program of China (Grant No. 2020YFA0308800).

\appendix

\section{Effective Hamiltonians} \label{fvndkkaa}

In this section discuss the Hamiltonians used to obtain the results in the main text. We first present the 
continuum model consisting of a semiconducting double-barrier potential sandwiched between two 
proximitized superconducting leads within the BdG formalism. We then discuss the equivalent 
lattice tight-binding model describing such a structure in the limit of vanishing lattice parameter $a$ taken to be the 
discretization step $\delta x$ (three-point finite differences) of the continuum model. 
Finally we describe the tight binding model for a Josephson junction with just one resonant level, i.e., a single-site dot 
coupled to two superconductors. This simpler model captures the essential feature 
of the continuum description of the double-barrier Josephson junction, i.e., resonant tunneling. It allows the 
inclusion of additional features in the model (e.g., Zeeman fields in the dot and leads and SO interaction) to 
more easily investigate the SDE in terms of the corresponding parameters.

\begin{figure*}[t!]
\begin{center}
\includegraphics[width=0.85\linewidth]{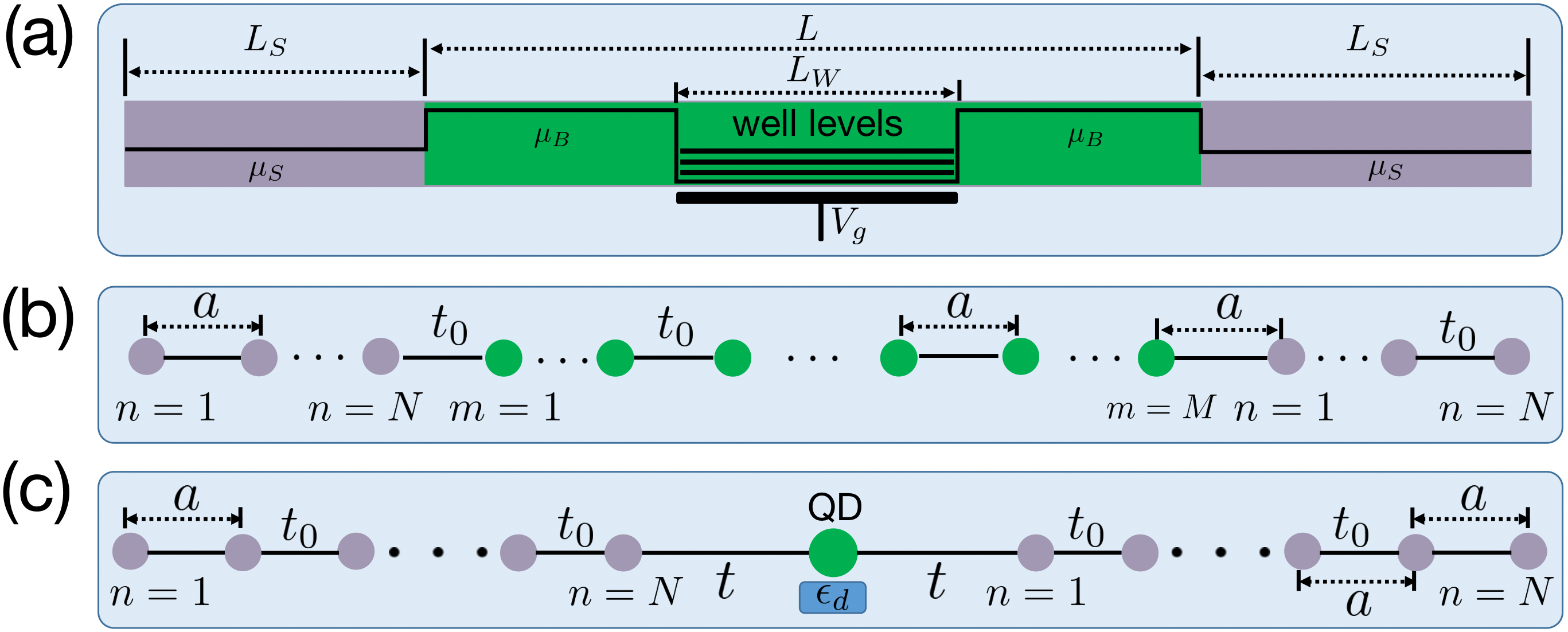}
\end{center}
\caption{(a) Sketch of a generic 1D  superconductor-semiconductor-superconductor Josephson junction. The semiconductor layer contains a double-barrier potential with a gate-tunable $V_g$ well depth hosting many quasi-bound states. In absence of superconductivity (i.e., normal semiconducting leads), this system represents the usual resonant-tunneling double-barrier semiconducting diode. (b) Schematic of a 1D tight-binding model for the hybrid superconductor-semiconductor resonant-tunneling Josephson junction in (a). In the limit $a\rightarrow 0$, this model reproduces the continuum limit of a discretized (finite difference) 1D model. (c) 1D tight binding model for a single-site quantum dot Josephson junction.} 
\label{Effectivemodel}
\end{figure*}  

\subsection{Continuum BdG model for the resonant-tunneling Josephson junction} \label{continuumtotignht}

Figure~\ref{Effectivemodel}(a) shows the hybrid proximitized-superconductor/semiconductor resonant-tunneling Josephson 
junction investigated here. The corresponding BdG Hamiltonian is 
\begin{align} \label{fvgbvfglb}
    H=\int dx \psi^{\dagger}(x)\begin{bmatrix}
        -\frac{\hbar^2\partial^2_x}{2m^{*}} +\mu(x) &-\Delta(x) \\
         -\Delta^*(x) & +\frac{\hbar^2\partial^2_x}{2m^{*}} -\mu(x)
    \end{bmatrix}\psi(x),
\end{align}
wher $m^*$ is the electron mass and the spinor field
\begin{align}
    \psi(x)=\begin{bmatrix}
        \psi^{}_{\uparrow} (x)\\-\psi^{\dagger}_{\downarrow}(x)
    \end{bmatrix}.
\end{align}
The pair potential $  \Delta(x)$ is zero within semiconductor (green layer) and nonzero within the 
finite-momentum $\hbar q$ superconductors (grey layers), i.e.,
\begin{align}
    \Delta(x)=\left\{\begin{matrix*}
        \Delta e^{+2iqx+\phi_L}, & x <-L/2\\
        0, & -L_M/2<x <+L_M/2\\
        \Delta e^{+2iqx+\phi_R}, & x >L/2
    \end{matrix*}\right.,
\end{align}
where $L$ in width of the semiconducting layer, [Fig.~\ref{Effectivemodel} (a)].
The full chemical potential profile $\mu(x)$, including the double-barrier structure, is described by 
\begin{align} \label{fvvdldlv}
    \mu(x)=\left\{\begin{matrix*}
       \mu_S,& x <-L/2\\
          \mu_B=\mu_S+V_B, & -L/2<x <-L_W/2\\
         \mu_S+V_g, & -L_W/2<x <+L_W/2\\
         \mu_B= \mu_S+V_B, & +L_W/2<x <+L/2\\
       \mu_S, & x >L/2
    \end{matrix*}\right.,
\end{align}
where $L_W$ denotes the well width, $\mu_S$ the chemical potential of superconductor, $\mu_B$ the chemical potential of semiconductor,  $V_B$ 
the height of the double barriers, and $V_g$ the electrostatic gate controlling the quantum well depth. The origin $x=0$ is 
defined at the center of the double barrier potential, Fig.~\ref{Effectivemodel}(a). 
For simplicity, we do not consider Zeeman fields and SO coupling in the 
continuum model; these are included in the simpler single-site model described in the main text.  Here, we consider the  system size of $L_S=3.75$ $\mu$m and $L=450$ nm, and hence the total number of sites is given by $N_T=2N+M$, where $N=L_S/a$ and $M=L/a$, where $a$ is the spacing of the tight-binding mode (Fig. \ref{Effectivemodel}).

\subsection{Discretized continuum model as a 1D nearest-neighbor tight-binding model} \label{tightbindingmodel}

By considering the three-point (second derivative) finite difference method with $N_T$ equally spaced 
discretized points $x_n$ between the end points $x_i$ and $x_f$, with $x_n=x_i + n\delta x$ for $n=0, 1,2,\cdots, N_T$, 
$\delta x=(x_f-x_i)/N_T$ being the (sufficiently small) discretization step, we can approximate 
the BdG equation of our Hamiltonian \eqref{fvgbvfglb} by the coupled set of equations 
\begin{align} \label{fvmkfk0}
    &-\frac{\hbar^2}{2m^{*}} \frac{\psi_s (x_n+\delta x)-2\psi_s(x_n)+\psi_s(x_n-\delta x)}{(\delta x_n)^2} \notag\\
    &+\mu(x_n)\psi_s(x_n)+s\Delta(x_n)\psi^{\dagger}_{\bar{s}}(x_n)=E\psi_s(x_n).
\end{align}
The pair potential $\Delta(x_n)$ couples the spin-dependent components of the field 
spinors $\psi_s(x_n)$ and $\psi_{\bar{s}}(x_n)$. A more familiar form of Eq.~(\ref{fvmkfk0}) is
\begin{align} \label{fvmkfk}
    -\frac{\hbar^2}{2m^{*}} \frac{\psi_s^{n+1} -2\psi_s^n+\psi_s^{n-1}}{(\delta x)^2} +\mu^n\psi_s^n+s\Delta^n(\psi_{\bar{s}}^n)^{\dagger}=E\psi_s^n,
\end{align}
where we have used the discretized versions of the relevant quantities: $\psi(x_n) \rightarrow \psi_s^n$, $\mu_s(x_n) \rightarrow \mu_S^n$ and $\Delta(x_n) \rightarrow \Delta^n$.
As well known, the three-point second derivative approximation for numerical discretization, e.g., Eq.~(\ref{fvmkfk}), 
can be formally mapped onto a 1D tight-binding model with nearest-neighbor hopping, similar to the one introduced in the main text for the single-site dot. More specifically, by making the replacements
\begin{align}  \label{fvmfdkk}
    \psi_s^n&\rightarrow c_{jns}, \\
    \delta x&\rightarrow a, \\
    -\frac{\hbar^2}{2m^{*}(\delta x)^2} &\rightarrow t_0,
    \end{align}
we can immediately write down the 1D tight-biding model corresponding to Fig.~\ref{Effectivemodel}(b)
\begin{align} 
    H&=\sum_{j=L,R,C}\sum^{N_j}_{n=1}\sum_{s=\uparrow,\downarrow}(-2t_0+\mu_j)c^{\dagger}_{jns}c^{}_{jns}   \notag \\
   &+\sum_{j=L,R,C}\sum^{N_j-1}_{n=1}\sum_{s=\uparrow,\downarrow}\left(t_{0}c^{\dagger}_{jns}c^{}_{jn+1s}+h.c.\right)\notag \\
&+t_0\sum_{s=\uparrow,\downarrow}\left(c^{\dagger}_{LNs} c^{}_{C1s}+c^{\dagger}_{CMs}c^{}_{R1s} +h.c. \right)\notag\\
&+\sum_{j=L,R}\sum^{N_j}_{n=1}\left(\Delta^{j}_{n}c^{\dagger}_{jn\uparrow}c^{\dagger}_{jn\downarrow}+h.c. \right).
\label{0maindvmdl}
\end{align}
Note that the subscript $j=C,L,R$ of fermionic electron operator $c_{jns}$ includes the semiconducting (Central) 
region $C$ in addition to the (Left and Right) superconducting leads $L,R$. 
The field operator $c_{jns}$  annihilates a spin-$s$ electron at position $x_n$ in region $j$. Here $N_j$ 
denotes the respective number of points in the $j=C,L,R$ layers, i.e., $N_C=M$ and $N_{R,L}=N$. For 
definiteness, we index the superconducting leads with (similar) indices $n=1..N$ and the central semiconducting 
region with index $m=1, 2\cdots M$. The first (last) site $c_{C1s}$ ($c_{CMs}$) of the semiconductor layer is 
tunnel coupled to the last (first) site $c_{LNs}$ ($c_{R1s}$) of the left (right) superconducting lead. The tunnel 
coupling $t_0$ denotes the nearest-neighbor hopping amplitude in all regions. For the left (right) superconducting 
lead, we consider the Fulde-Ferrell type order parameter 
\begin{equation} \label{phaseys}
\Delta^{j}_n=\Delta e^{i\phi_j+2iqna},
\end{equation}
where $\hbar q$ is the Cooper-pair momentum of the superconducting 
lead $j$ with global superconducting phase $\phi_j$, and $a$ is the lattice constant.

\begin{figure*}[t!]
\begin{center}
\includegraphics[width=0.9\linewidth]{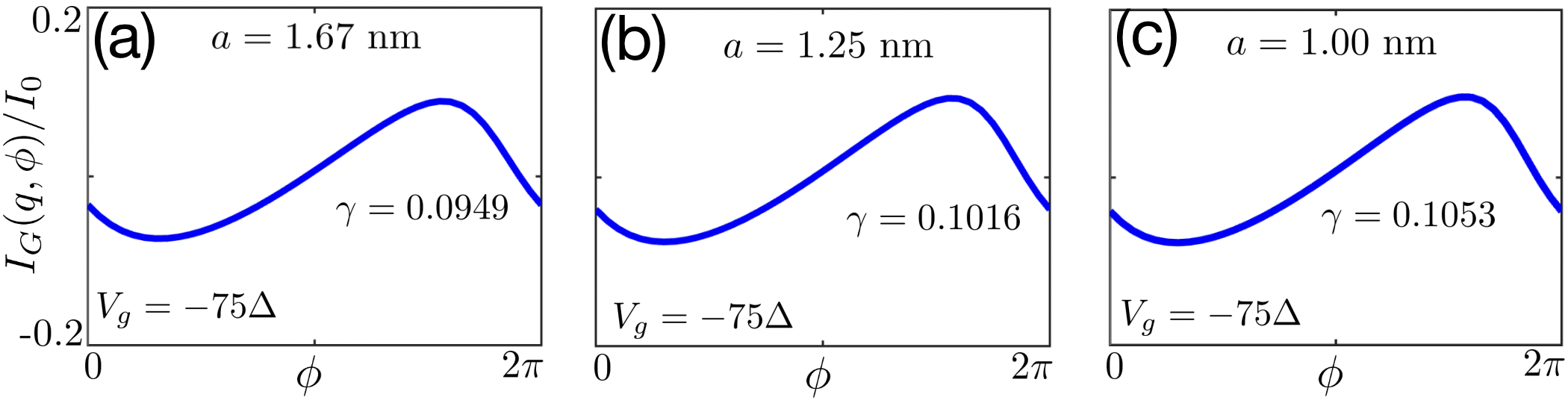}
\end{center}
\caption{Convergent test for $I_G(q,\phi$ vs. $\phi$ for three values of the lattice parameter $a$ 
(``step size'' $\delta x$), panels (a), (b) and (c). The current phase relation is obtained by solving the 
continuum BdG eigenvalue problem via numerical discretization [\eqref{fvmkfk}] (equivalent  
to diagonalizing the corresponding tight-binding Hamiltonian \eqref{maindvmdl}).  Convergence  is clearly 
attained at sufficiently small spacing $a=1$ nm [Panel (c)] for both $I_G(q,\phi)$ and
the corresponding diode efficiency $\gamma$. For the system of $L_S=3.75$ $\mu$m and $L=450$ nm, the total number of sites $N_T=2N+M$ of panels (a), (b), and (c) are $2520$,  $3360$, and $4200$, respectively.}  
\label{FIGConvIs}
\end{figure*}

\subsubsection{Tight-binding dispersions in the normal leads and some estimates} \label{Qestimate}
 
For clarity and definiteness of the parameters used in the main text (e.g., Fermi wave vector and energy), 
next we show the energy bands for normal leads. In this case, the tight binding bands (left and right leads), assuming 
periodic boundary conditions with wave vector $k$, are give by
\begin{equation}\label{dispersion_leads}
\varepsilon(k) = -2t_0+\mu_S + 2t_0\cos(ka),
\end{equation}
where we have straightforwardly Fourier transformed the first two terms of Eq.~\eqref{0maindvmdl}. The 
above expression shows that the bottom of the band (we use $t_0<0$ in the main text) is $\varepsilon(0) = \mu_s$, 
while the Brillouin zone edge is $\varepsilon(\pi/a) = -4t_0 +\mu_s$. In this case, a half-filled band corresponds to a Fermi wave vector 
$k_F=\pi/2a$. In our simulations, we are always away from half filling. Because we are interested in InAs-based 
semiconducting systems, we choose a $k_F$ closer to the bottom of the band, e.g., $k_F= \pi/5a$ in Fig.~\ref{FIGDB}. In this case, we can approximate 
the above dispersion by expanding it to second order in k
\begin{equation}\label{dispesion_leads}
\varepsilon(k) \simeq -2t_0+\mu_S + 2t_0(1 - \frac{(ka)^2}{2}) = \mu_S - t_0a^2k^2= \mu_S + \frac{\hbar^2 k^2}{2m^*}.
\end{equation}
We can then define an effective mass
\begin{equation}
m^*= -\frac{\hbar^2}{2t_0a^2}.
\end{equation}
Assuming $m^* \simeq 0.03 m_0$ for InAs-based systems \cite{dartiailh2021phase}, we find $t_0 \simeq - 50$ meV (this is the value we use in Fig.~\ref{FIGDB}). 
By using a typical electron density $n_e=8.14\times 10^{11}$ cm$^{-2}$ in InAs-based wells \cite{dartiailh2021phase}, we can estimate the magnitude of 
Fermi wave vector for a 2D electron gas, $k_F = \sqrt{2\pi n_e} = 0.226 \text{ nm}^{-1}$. If we use, instead, $k_F = \pi n_e/2$ for a 1D electron gas and consider $n_e^{1D} \sim 10^{6}$ cm$^{-1}$ \cite{degtyarev2017features}, 
we obtain $k_F^{1D} \simeq 0.157$ nm$^{-1}\Rightarrow 40$ nm. 
In our simulations, we take $k_F= \pi/5a$, with lattice parameter $a=5$ nm so that $k_F=0.125$ nm$^{-1}$ (20\% of the Brillouin zone) 
for which the quadratic dispersion is valid. This number is consistent with the above estimates for $k_F$ based on the electron 
density. We note that the $a=5$ nm used here corresponds to a small enough step $\delta x =a$ so that the numerical solution  
is converged; in the next subsection we discuss convergence in more detail.

\textit{Coherence length $\xi$.}  The proximitized InAs/Al  $\xi$ is given by
superconductor is 
\begin{equation}
\xi =\frac{\hbar v_F}{\pi \Delta} = \frac{\hbar^2 k_F}{\pi m^* \Delta} 
\end{equation} 
Since $\hbar^2/2m_0 = 1Ry a_0$ (eV $\AA$), with $1\text{Ry} = 13.606$ eV and $a_0=0.529\AA$, we can write
\begin{equation}
\xi =\frac{2\times 13.606}{\pi \times 0.03 \times 0.22\times 10^{-c}} \times a_0(a_0k_F) =  831 \text{nm}.
\end{equation} 
Note that $k_F\xi = 0.226 \text{ nm$^{-1}$} \times 831 \text{ nm} = 187$
\begin{equation}
\lambda_F \ll \xi, \text{ i.e., 27 nm $<<$ 831 nm}.
\end{equation} 

To estimate the chemical potential $\mu_S$ of the superconductor, we define the Fermi energy as the zero of energy, i.e., 
\begin{equation} \label{fvnafj}
\varepsilon(k_F)=\varepsilon_F = 0=\mu_S + \frac{\hbar^2 k_F^2}{2m^*} \Rightarrow \mu_S = -\frac{\hbar^2k_F^2}{2m^*}<0.
\end{equation}
Using 
$k_F= 0.125$ nm$^{-1}$, which is consistent with the usual electrons densities in InAs-basel well and wires \cite{degtyarev2017features}, 
we find $\mu_S = - 20.1$ meV from Eq. \eqref{fvnafj}, which is similar to $\mu_S = - 19$ meV the numbers we used to generate Fig.~\ref{FIGDB} with energy dispersion \eqref{dispersion_leads}.

\textit{Estimate of the parameter $q$.} As mentioned in the main text, $q$ parametrizes the finite
momentum $\hbar q$ of the Cooper pairs that, in principle, can generate (an anomalous) 
supercurrent $I_G(q,\phi)$ at $\phi=0$. Cooper pairs with 
a finite momentum can arise from, e.g., (i) the interplay of (Rashba) spin-orbit interaction and Zeeman field in InAs-based proximitized 2D electron 
gases \cite{yuan2021topological,dartiailh2021phase}, (ii) screening currents (Meissner effect) due to a weak external magnetic field  \cite{davydova2022universal,zhu2021discovery}, (iii) orbital effects leading to 
the acquision of an inhomogeneous phase \cite{banerjee2023phase}, (v) exchange interaction  \cite{fulde1964superconductivity}, (iv) injected current only \cite{levine1965dependence,hansen1969observation,bardeen1962critical}. In our proximitized InAs/Al nanowire system, Cooper pairs originating from the Al layer can leak into the InAs region, inducing superconductivity in InAs. When an external magnetic field is applied perpendicular to the InAs/Al nanowire, the Meissner effect leads to the formation of unidirectional edge screening currents, similar to the edge current of quantum Hall effect. These screening currents result in the presence of finite-momentum Cooper pairs that also leak into the InAs nanowire. Consequently, we observe the existence of finite-momentum Cooper pairs in the proximitized InAs. The momentum of these Cooper pairs arising from this mechanism can be estimated using the expression  $\hbar q\simeq eB_y\lambda_L$~\cite{davydova2022universal}. For a sizable London penetration depth $\lambda_L = 140$ nm~\cite{fletcher2007penetration},  a magnetic field of approximately $B_y\simeq 20$ mT is required to generate sizable momentum $\hbar qv_F \simeq 2.7 \Delta\sim\Delta$~\cite{zhu2021discovery}.  Importantly, this momentum is much smaller than the Fermi momentum $\hbar k_F$, i.e., $q/k_F \simeq 0.03$. Here, we have utilized experimentally available parameters, including the order parameter $\Delta = 0.5 $ meV, Fermi momentum $k_F=\pi/5a\simeq0.125$ nm$^{-1}$, Fermi velocity $v_F=\hbar k_F/m^{*}\simeq 5*10^5$ m/s,  and effective mass $m^* \simeq 0.03 m_0$(consistent with the parameters employed in all figures).

\subsubsection{Numerical convergence of the continuum model.} 

\begin{figure*}[t!]
\begin{center}
\includegraphics[width=0.9\linewidth]{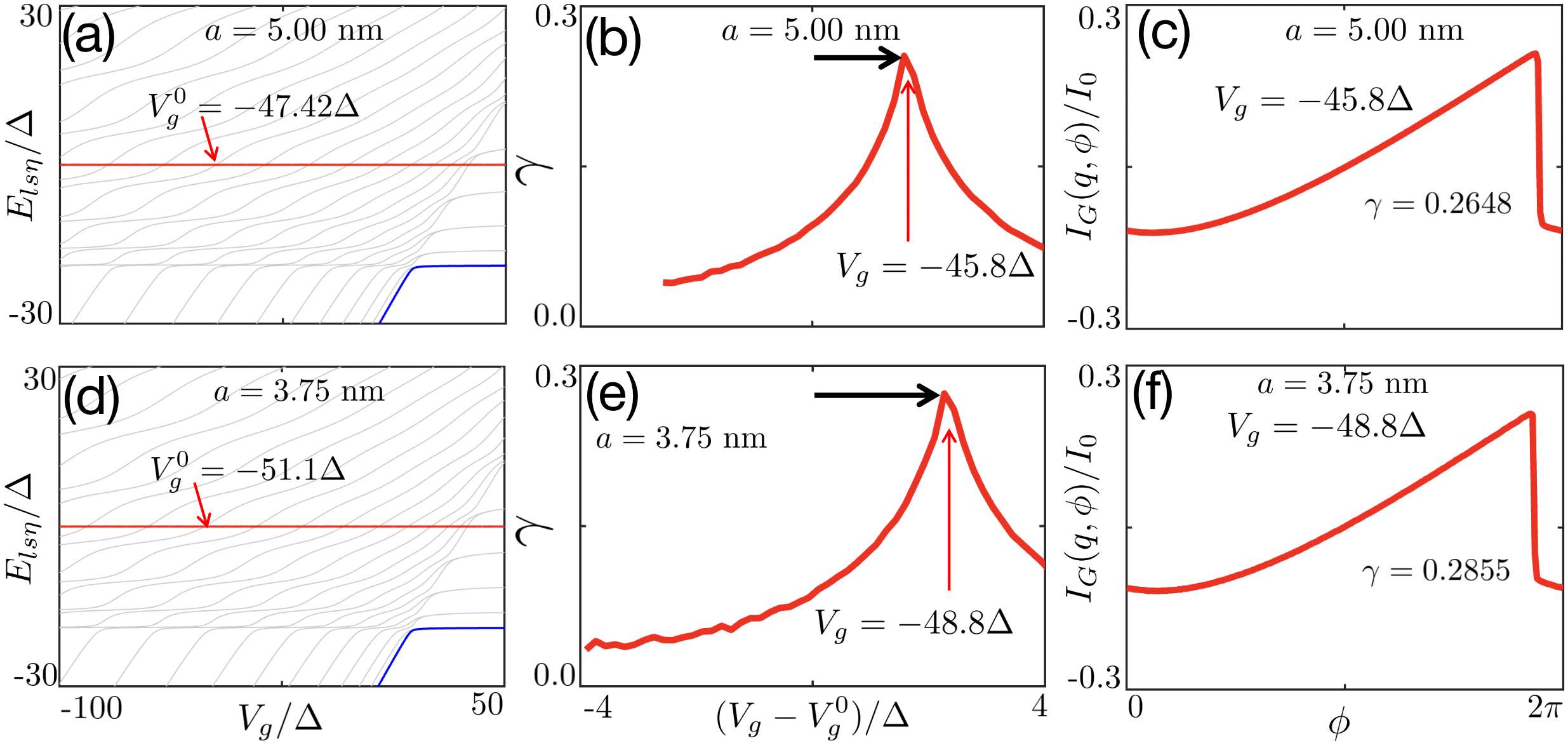}
\end{center}
\caption{Convergent test for two values of the lattice parameter $a$ 
(``step size'' $\delta x$), panels (a) and (b). (a) Dependence on $V_g$ of the non-superconducting well quasi-bound states for $a=5$ nm. Panel (d) corresponds to (a) but for $a=3.75$nm. (b,e)  Diode efficiency $\gamma$ as a function of gate voltage $V_g$. The maximum diode efficiency is shifted to right due to the renormalization of effective bound-state energy from the  lead-dot tunneling coupling, as shown by the black arrows. The current phase relations at the maximum diode efficiency are plotted in panels (c) and (f).   Figs. (a), (b) and (c) here for the same parameters as 
Fig. 1 of the main text.}
\label{FIGConv}
\end{figure*}

Recall that in the continuum model $a=\delta x$ is the discretization step to 
be chosen sufficiently small so that the eigenvalue solution is convergent and, in turn, so are all other calculated 
(physical) quantities. Figures \ref{FIGConvIs} (a), (b), and (c) illustrate the evolution of this convergence (different 
$a$'s) for the ground-state supercurrent [Eq. (6) of main text] vs. the phase difference $\phi=\phi_R-\phi_L$. We 
obtain a convergent result for sufficiently small spacing $a=1$ nm [Fig. \ref{FIGConvIs} (c)]. 

However, the convergent result numerically calculated from the tight-binding model \eqref{0maindvmdl} requires huge system sites (see Fig. \ref{FIGConvIs}) and hence costs long time for numerical calculations. Figure \ref{FIGConv} plots the bare quasi-bound states, diode efficiency, and current-phase relation for smaller spacing. We note that the maximum diode efficiency is shifted to the right-hand side due to the renormalization of effective well bound-state energy from the  lead-dot tunneling coupling, as shown by the black arrows in Figs. \ref{FIGConv} (b) and (e). This renormalization depends on $\tilde{\mu}_S=\mu_S-2t_0$ [Eq. \eqref{fdvndfnk}] and hence varies with $a$ via $t_0$ [Eq. \eqref{fvmfdkk}] when $a$ is not small enough. This shift is well-explained in Sec. \ref{sec-asym}. The current phase relations at the maximum diode efficiency are plotted in panels (c) and (f). Though they are not perfect convergent results, all of them qualitatively show same behaviours, the shifted resonant tunneling and current-phase relation. Therefore, in the main text, we use a larger spacing $a=5$ nm, which can capture the main physics in our model.

\subsection{Gate-tunable Andreev levels} \label{gatetunableAL}

\begin{figure*}[t!]
\begin{center}
\includegraphics[width=0.7\linewidth]{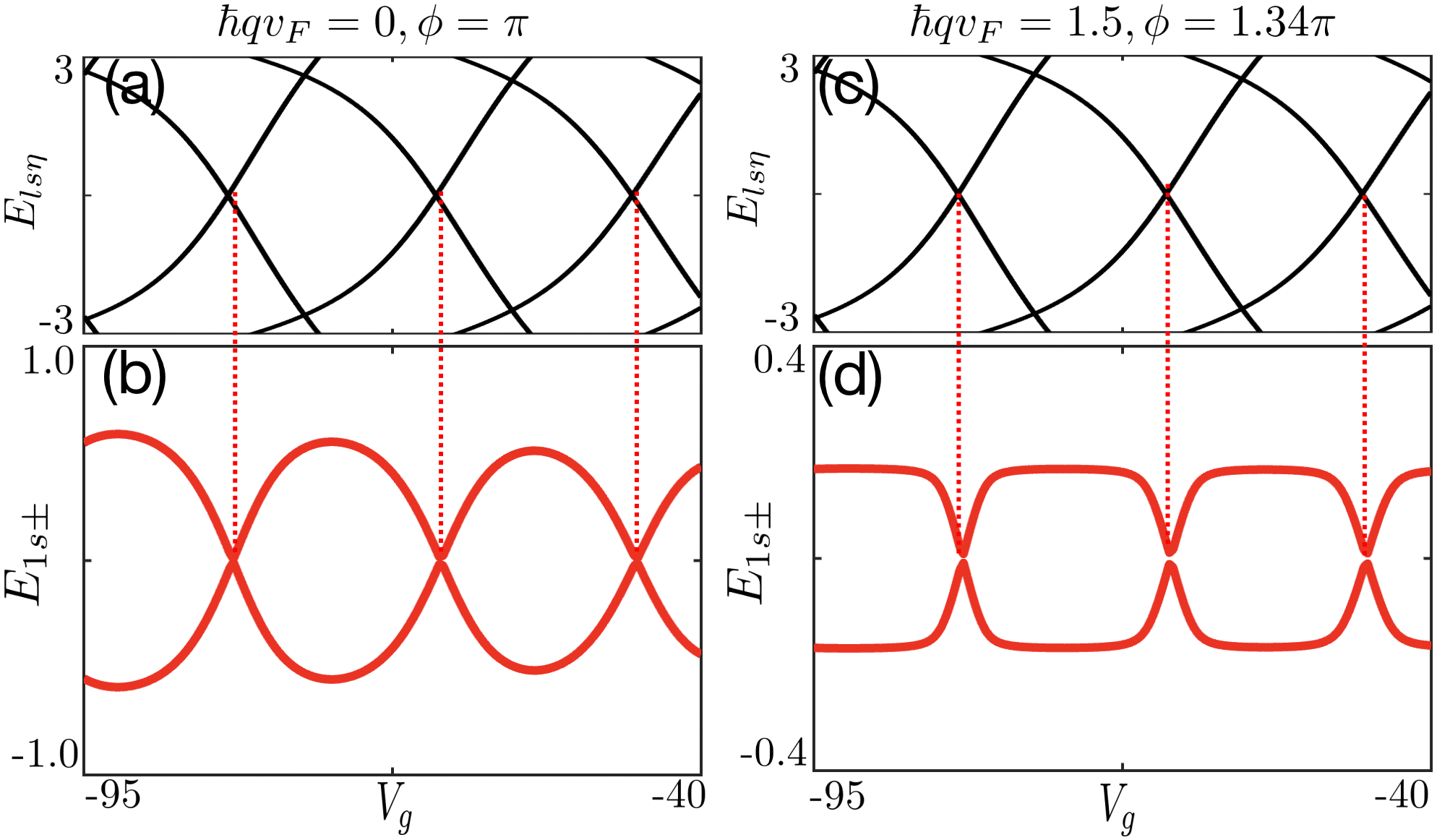}
\end{center}
\caption{Dependence on $V_g$ of the non-superconducting well quasi-bound states and 
corresponding Andreev bound states $E_{1s\pm}$ for the double-barrier resonant-tunneling Josephson junction 
in Fig.~\ref{Effectivemodel}(a), with $\hbar qv_F=0$ and $\phi=\pi$ [left panels (a), (b)] and $\hbar qv_F=1.5\Delta$ and $\phi=1.34\pi$ 
[right panels (c), (d)]. Note that the resonant Andreev bound states (b) and (d) stem from the ordinary quasi-bound 
states of the well (a), (c) [see vertical dotted lines]. As the well depth $V_g$ is increased, the confined 
states in the well [(a) and (c)] successively cross  the zero-energy baseline and so do the corresponding 
Andreev bound states [(b) and (d)]. This induces periodic oscillations in the Andreev levels for both 
$\hbar qv_F=0$ and $\hbar qv_F=1.5\Delta$ cases. Note that for $q=0$ the Andreev levels cross at zero energy for $\phi=\pi$, while 
for $\hbar qv_F=1.5\Delta$ this cross at zero energy occur for $\phi=1.34\pi$ because the spectrum is asymmetric for $q\neq 0$. 
The parameters here are exactly the same parameters as in Fig. 1(f).
}
\label{FIGGTA}
\end{figure*}

Here we explore how the gate voltage $V_g$ controlling the well depth tunes the Andreev levels 
of our hybrid semiconductor-superconductor Josephson junction. Figure~\ref{FIGGTA} shows the $V_g$ dependence of 
(a) the ordinary quasi-bound states of the double-barrier potential well in the absence of superconductivity and (b) the corresponding Andreev 
bound states of our Josephson junction for $q=0$ and $\phi=\pi$. 
Figure~\ref{FIGGTA}(c) (repeated on the right panel for convenience) is identical to Fig.~\ref{FIGGTA}(a), 
while  Fig.~\ref{FIGGTA}(d) is similar to Fig.~\ref{FIGGTA}(b) 
but for finite momentum $\hbar qv_F=1.5\Delta$ and $\phi=1.34 \pi$. For $q=0$ the Andreev levels touch each other as the 
superconducting gap closes at $\phi=\pi$ and oscillate as a function of $V_g$. 
Interestingly, gap closing is now periodic in $V_g$ and happens whenever one of the $V_g$-tunable quasi-bound 
states of the well crosses zero [vertical red dashed lines in (a),(b)]. For $\hbar qv_F=1.5\Delta$ this gap closing takes place    
at $\phi=1.34\pi$ because the Andreev spectrum is asymmetric for $q \neq 0$. Similarly to the $q=0$ case, 
Andreev levels are still highly $V_g$ tunable for $q\neq 0$ and successively cross the zero-energy 
baseline as the well depth is varied [vertical dashed lines (c) and (d)]. 
Unlike the $q=0$ case, the $q \neq 0$ Andreev levels $E_{1s\pm} (V_g$) feature flat 
regions as a function of $V_g$ due the avoided crossing between $V_g$ tunable quasi-bound state of the well and the gap edge of the finite-momentum superconducting lead; cf. solid red lines in (d) and (b).
 
\subsection{From a multi-level double-barrier quantum well to a single-site quantum dot} \label{fvandkfk}

\begin{figure}[h!]
\begin{center}
\includegraphics[width=0.8\linewidth]{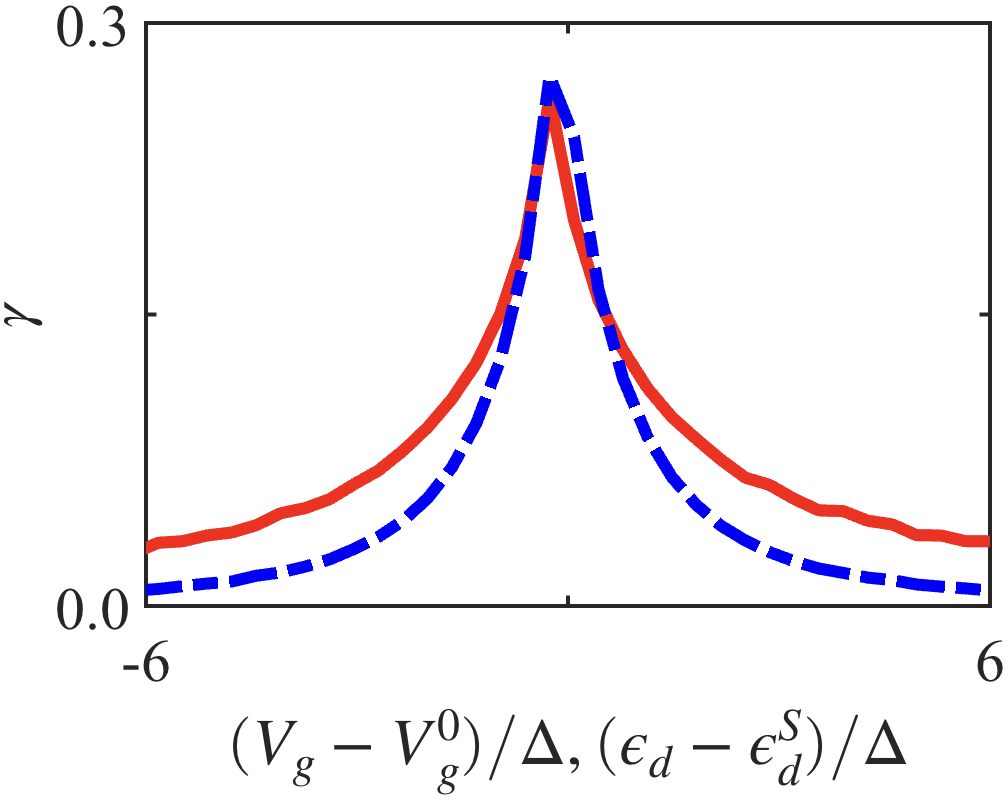}
\end{center}
\caption{Gate-tunable diode efficiency $\gamma$ for the continuum and single-site dot models.  We consider the resonance at 
$V^0_g=-45.8\Delta$ of Fig.~1(f). The red and blue curves correspond to the $\gamma$ resonant peaks for the double-barrier and 
single-site dot models, respectively. The system parameters used to obtain these curves are given in tables \ref{tab:my_label1} 
and \ref{tab:my_label2}. Here, $\epsilon^S_d=2.7\Delta$ is the correction of the dot energy from the lead-dot tunneling coupling, which is discussed in  Fig. \ref{AFIG2} (c). 
}
\label{FIGCP}
\end{figure}

As mentioned in the main text and at the beginning of this section, we would also consider a simpler 1D tight binding model 
describing a single-site quantum dot Josephson junction as a means to further investigate SDE 
considering additional parameters. This is illustrated in Fig.~\ref{Effectivemodel}(c). The single-site dot model retains just one 
of the resonances of the continuum model, which is selected by choosing $\epsilon_d(V_g)$ to coincide with one of the 
double-barrier well quasi-bound levels. Below, for completeness, we reproduce the 1D tight biding model for the 
single-site dot coupled to superconducting leads shown in Eq.~(1) of the main text,
\begin{align} \label{maindvmdl1}
H&=\sum^{N}_{n=1}\sum_{j=L,R}\sum_{s=\uparrow,\downarrow}\epsilon_{jns}c^{\dagger}_{jns}c^{}_{jns}+
\sum_{s=\uparrow,\downarrow}(\epsilon_d+sh_d)d^{\dagger}_{s}d^{}_{s}\notag  \\
&+\sum^{N-1}_{n=1}\sum_{j=L,R}\sum_{s=\uparrow,\downarrow}\left(t_{0}c^{\dagger}_{jns}c^{}_{jn+1s}+h.c.\right)\notag \\
&+\sum^{N}_{n=1}\sum_{j=L,R}\left(\Delta^{j}_{n}c^{\dagger}_{jn\uparrow}c^{\dagger}_{jn\downarrow}+h.c. \right)\notag \\
&+t\sum_{ss'}\left[c^{\dagger}_{LNs} \mathcal{U}_{ss'}(\theta_{so})d_{s'}+
d^{\dagger}_{s} \mathcal{U}_{ss'}(\theta_{so})c_{R1s'} +h.c.\right].
\end{align} 
Physically, the tunnel coupling between the quantum dot and superconducting leads $t$ is related to the the double-barrier 
potential [Eq.~\eqref{fvvdldlv}] of the continuum model. Here we treat $t$ as a fitting parameter to reproduce one of the 
resonances of the continuum model.  Figure~\ref{FIGCP} shows the resonance peak in the diode efficiency $\gamma$ obtained 
via the continuum model (red solid line) and the single-site quantum dot model (blue dotted line). The parameters used in 
both simulations are listed in tables \ref{tab:my_label1} and \ref{tab:my_label2}. The spin rotation matrix $\mathcal{U} (\theta_{\text{so}})=e^{i\theta_{\text{so}} s^{y}/2}$ is responsible for accounting for the spin rotation induced by spin-orbit coupling ($\theta_{\text{so}}$) during the tunneling process between the quantum dot and the last (first) site $c_{LNs}$ ($c_{R1s}$) of the left (right) lead. Spin-orbit coupling exists in both the semiconducting quantum dot and the proximitized InAs/Al leads, creating an effective magnetic field that leads to a momentum-dependent spin rotation of itinerant electrons. In our work, the purpose of the single-site quantum dot Josephson junction is to obtain approximate expressions for the Andreev levels using Green functions. This approach allows us to gain insight into the important role played by the phase shifts $\delta \phi_{L,R}[q,E(q,\phi)]$ (refer to Sec.~\ref{Additionalphaseshifts}). To achieve this objective, we incorporate the effect of spin-orbit coupling on the Andreev levels in a simple manner, treating it as a spin rotation during the tunneling process between the quantum dot and the superconducting leads. It is worth noting that the Andreev levels are derived from the reduced determinant equation [Eq. \eqref{mfdvkdfmv} and Eq.~\eqref{fdvkdfmv}], where the degrees of freedom associated with the superconducting leads are integrated out to generate a self-energy term that captures the overall influence of the superconducting leads on the Andreev levels. Interestingly, the form of the self-energy [Eq. (\ref{mselfenergy})] is quite general and does not depend on the specific details of the superconducting leads. As a result, this spin rotation approach effectively encapsulates the effects of spin-orbit coupling from both the semiconducting quantum dot and the proximitized InAs/Al leads on the Andreev levels.

\begin{widetext}

\begin{table}[]
    \centering
       \caption{Parameter for the continum model used in Fig.~\ref{FIGCP}.}
    \begin{tabular}{|c|c|c|c|c|c|c|c|c|c|c|c|}
    \hline
       $\Delta$ & $t_0$ & $N$ & $\mu_S$ & $a$ & $k_F$& $m^{*}$ & $L_M$ & $L_W$  & $V_B$ & $V_g$    \\\hline
          0.5 meV & -50 meV& 2000  & -19 meV & 5 nm & $\pi/5$& 0.03 $m_0$ & 0.45 $\mu$m & 0.15 $\mu$m& 9 meV & -23 meV  \\\hline
    \end{tabular}
    \label{tab:my_label1}
\end{table}

\begin{table}[]
   \centering
   \caption{Parameter for the single-site dot model used in Fig.~\ref{FIGCP}.}
    \begin{tabular}{|c|c|c|c|c|c|c|c|c|c|}
    \hline
         $\Delta$ & $t_0$ & $N$& $\mu_S$ & $a$ & $k_F$ & $m^{*}$  & $t$  &  $\epsilon_d$  \\ \hline
           0.5 meV & -50 meV & 2000& -19 meV & 5 nm & $\pi/5$& 0.03 $m_0$ & -6.5 meV  & 1.3 meV   \\ \hline
    \end{tabular}
        \label{tab:my_label2}
\end{table}

\end{widetext}

\section{Derivation of the Hamiltonian [Eq.~(7)] in the main text}
\label{sec1}

In this section, we write out the Bogoliubov-de-Gennes (BdG) Hamiltonian, corresponding to $H$ in Eq.~\eqref{maindvmdl}, in Nambu-spin space and detail the  steps to formally derive Eq.~(2) in the main text. 

\subsection{Generalized Nambu field operator and the BdG Hamiltonian}

Let us rewrite the total Hamiltonian [Eq.~(1) in main text] in  the Nambu space of the hybrid quantum-dot/superconducting-lead system described by the field operator 
\begin{equation}
 \Psi = 
    \begin{bmatrix}
      \Psi_d \\
      \Psi_L\\
      \Psi_R
    \end{bmatrix} 
    \equiv \Psi_d  \oplus \Psi_L  \oplus \Psi_R 
 \end{equation}
where $\Psi_d$ and $\Psi_j$, $j=L,R$ denote the field operators for dot and superconducting leads, respectively. More explicitly, we have
\begin{align} \label{tnambu}
    \Psi=
    \begin{bmatrix}
      d^{}_{\uparrow}\\
      d^{}_{\downarrow}\\
      -d^{\dagger}_{\downarrow}\\
      d^{\dagger}_{\uparrow}
    \end{bmatrix} 
   \oplus
    \left(\bigoplus_{n}
    \begin{bmatrix}
      c^{}_{Ln\uparrow}\\
      c^{}_{Ln\downarrow}\\
      -c^{\dagger}_{Ln\downarrow}\\
      c^{\dagger}_{Ln\uparrow}
    \end{bmatrix} \right)
    \oplus
   \left(\bigoplus_{n}
    \begin{bmatrix}
      c^{}_{Rn\uparrow}\\
      c^{}_{Rn\downarrow}\\
      -c^{\dagger}_{Rn\downarrow}\\
      c^{\dagger}_{Rn\uparrow}
    \end{bmatrix} \right),
\end{align}
where $\bigoplus_{n} X_{n}$ concatenates $X_{n}$ vertically.
Using the generalized Nambu field operator $\Psi$ above, we can straightforwardly  recast $H$ [Eq.~(1), main text] in the form
\begin{align} \label{fvdvldv}
    H= \frac{1}{2} \Psi^{\dagger}  \mathcal{H}_{\text{BdG}} \Psi+\mathcal{E},
\end{align}
where  the BdG Hamiltonian matrix is given by 
\begin{align} \label{fdnkvamk}
    \mathcal{H}_{\text{BdG}}=\left[\begin{array}{cccc} 
    \mathcal{H}_{D}  & \mathcal{T}_{L} & \mathcal{T}_{R} \\
   \mathcal{T}^{\dagger}_{L}& \mathcal{H}_{L}  & 0 \\
   \mathcal{T}^{\dagger}_{R}& 0 & \mathcal{H}^{\dagger}_{R}
    \end{array}\right].
\end{align}
and
\begin{equation}
 \mathcal{E}=\sum_{jn}\epsilon_{jn},
 \label{constant-E}
 \end{equation}
is a $\phi$-independent constant (recall that $\epsilon_{jn}$ is the site energy, Eq.~(1) in the main text).
The factor of 1/2 in Eq. \eqref{fvdvldv} arises from the artificial doubling in the BdG formalism \cite{bernevig2013topological}.
 The  quantum dot is described by the non-interacting Hamiltonian 
\begin{align}   \label{dot-H}
   \mathcal{H}_{D}&=\epsilon_d\tau_z\otimes \openone +h_d\openone \otimes s^{z}\\
   &= 
  \begin{bmatrix}
    \epsilon_d +h_d & 0 &  0 & 0\\
    0 & \epsilon_d -h_d&  0 & 0\\
    0 & 0 & -\epsilon_d +h_d  & 0\\
    0 & 0 &0 &  -\epsilon_d - h_d
    \end{bmatrix}_{4\times 4}, \notag
\end{align}
where $\tau_z$ and $s^{z}$ are Pauli matrices acting within the Nambu and spin spaces, respectively. 
The Hamiltonian $\mathcal{H}_j$ describes the superconducting leads $j=L,R$,
\begin{align} \label{tvfnvmk} 
    \mathcal{H}_j=
   \begin{bmatrix}
       \mathcal{H}_{j1} & \mathcal{T}_{t_0} & & & \\
      \mathcal{T}^{\dagger}_{t_0} & \mathcal{H}_{j2} & \mathcal{T}_{t_0} & & & \\
        & \mathcal{T}^{\dagger}_{t_0} & \mathcal{H}_{j3} & \ddots & \\
        &  & \ddots & \ddots & \mathcal{T}_{t_0} \\
        &  & & \mathcal{T}^{\dagger}_{t_0} & \mathcal{H}_{jN}
 \end{bmatrix}_{4N\times 4N},
\end{align}
with
\begin{align} \label{bgbfmk}
   \mathcal{T}_{t_0}= t_0\mathcal{T}_{0},
\end{align}
\begin{align} 
    \mathcal{T}_{0}= \tau_z\otimes \openone =
    \begin{bmatrix}
    1 & 0 &  0 & 0\\
    0 & 1 &  0 & 0\\
    0 & 0 & -1  & 0\\
    0 & 0 &0 &  -1
    \end{bmatrix}_{4\times 4},
\end{align}
$t_0$ denotes the dot-lead tunnel coupling amplitude, and
\begin{align} \label{vnskfm}
    \mathcal{H}_{jn}=
    \begin{bmatrix}
    \epsilon_{jn\uparrow} & 0 &  -\Delta^j_n & 0\\
    0 & \epsilon_{jn\downarrow} &  0 & -\Delta^j_n\\
    -(\Delta^j_n)^* & 0 & -\epsilon_{jn\downarrow}  & 0\\
    0 & -(\Delta^j_n)^* &0 &  -\epsilon_{jn\uparrow}
    \end{bmatrix}_{4\times 4},
\end{align}
where 
\begin{equation}
\epsilon_{jns}=\epsilon_{jn}+sh_{sc},
\label{epsilon-hsc}
\end{equation}
and the finite-$q$ superconducting pairing gap
\begin{equation}
\Delta^{j}_n=\Delta e^{i\phi_j+2iqna}.
\label{phases}
\end{equation}
For simplicity, in what follows,  apart from distinct phases $\phi_L$, $\phi_R$, we assume 
otherwise identical left and right superconducting leads. 
The tunnel-coupling matrix between the quantum dot and the left and right leads are, respectively, 
\begin{align} \label{fvmfkll}
    \mathcal{T}_L&= t\begin{bmatrix}
      0 &  \cdots & 0 &  U(\theta_{so})
 \end{bmatrix}_{4\times 4N},
\end{align}
\begin{align} \label{fvmfklr}
    \mathcal{T}_R&= t\begin{bmatrix}
      U^{\dagger}(\theta_{so}) &  0 & \cdots & 0
 \end{bmatrix}_{4\times 4N},
\end{align}
where  the unitary matrix $U(\theta_{so})$ accounts for the spin-orbit (SO) effect in the tunneling processes (left and right) and is given by
\begin{equation} \label{fmmvfkd}
U(\theta_{so})=\begin{bmatrix}
\mathcal{U}_{}(\theta_{so})& \\ &-\mathcal{U}_{}(\theta_{so})\end{bmatrix}_{4\times 4},
\end{equation}
with $\mathcal{U}_{}(\theta_{so})=e^{i\theta s^{y}/2}$ describing the spin rotation due to the spin-orbit coupling in the 
tunneling between the quantum dot and superconducting leads.

By numerically diagonalizing the BdG matrix \eqref{fdnkvamk} we can construct the $(q,\phi_L,\phi_R)$-dependent Bogoliubov (quasiparticle) operators $\gamma^{}_{l\sigma\eta}(q,\phi_L,\phi_R)\rightarrow \gamma^{}_{l\sigma\eta}$ as unit vectors in the Nambu space \eqref{tnambu}, i.e.,
\begin{equation}
\gamma^{}_{l\sigma\eta}=\sum^{8N+4}_{m=1}u_{l,m}(q,\phi_L,\phi_R)\Psi_{m}, \text{ with } \sum^{8N+4}_{m=1}\vert u_{l,m}(q,\phi_L,\phi_R)\vert^2=1,
\end{equation}
where $\Psi_{m}$ is the $m$th component of $\Psi$ in Eq. \eqref{tnambu} and $[u_{l,1},u_{l,2},\cdots, u_{l,8N+4}]$ is the $l$-th eigenvector 
that diagonalizes the BdG Hamiltonian; the Bogoliubov operators 
obey the conjugation relation
\begin{align} \label{gbmgblf2}
  \gamma^{\dagger}_{l\sigma\eta}= \gamma^{}_{l-\sigma-\eta}.
\end{align}
We can now recast the total Hamiltonian \eqref{fvdvldv}  in the form of Eq.~(2) in the main text,
\begin{align} \label{gdrtfgblrgk}
    H= \frac{1}{2}\sum^{2N+1}_{l=1}\sum_{\sigma=\Uparrow/\Downarrow}\sum_{\eta=\pm} E_{l\sigma\eta}(q,\phi)\gamma_{l\sigma\eta}^{\dagger} \gamma_{l\sigma\eta }^{}+\mathcal{E}.
\end{align}
Note that in the above the quasi-particle eigenenergies $E_{l\sigma\eta}(q,\phi)$ depend only on 
the phase difference $\phi=\phi_R-\phi_L$ 
and satisfy,
\begin{align} \label{gbmgblf1}
    E_{l\sigma\eta}(q,\phi)=-E_{l-\sigma-\eta}(q,\phi),
\end{align}
due to particle-hole symmetry.
To see that the eigenenergies depend on the phase difference $\phi=\phi_R-\phi_L$, let us do the transformation $c_{Lns}\rightarrow e^{-i\phi_L/2}c_{Lns}$, $d_s\rightarrow e^{+i\phi_L/2} d_{s}$ and $c_{Rns}\rightarrow e^{-i\phi_R}c_{Rns}$ in Eq~1 (main text). The pair potential then transforms as $\Delta^j_n\rightarrow \Delta e^{+2iqna}$ while the tunnel coupling between the dot and left (right) superconducting lead transforms as $t\rightarrow t$ ($t \rightarrow te^{i\phi/2}$). Thus, the eigenenergies depend on only the phase difference $\phi$ [note that a $(\phi_R+\phi_L)/2$ term does not appear in the transformed Hamiltonian.] 

In the main text, we use the orthogonal basis set $\{\gamma_{l\sigma+}\}$ as this choice makes it straightforward to determine the ground-state wave function $|G\rangle$ and energy $\mathcal{E}_{G}$. Using Eqs.~\eqref{gbmgblf2} and \eqref{gbmgblf1}, we can replace all $\gamma_{l\sigma-}$ with $\gamma_{l\sigma+}$ in the Hamiltonian \eqref{gdrtfgblrgk} to obtain its form in this basis,
\begin{align} \label{gdrdatfgblrgk}
    H= \sum^{2N+1}_{l=1}\sum_{\sigma=\uparrow/\downarrow} E_{l\sigma+}(q,\phi)\gamma_{l\sigma+}^{\dagger} \gamma_{l\sigma+ }^{}+\mathcal{E}_{+}(q,\phi),
\end{align}
where
\begin{align} \label{fvdmkf}
\mathcal{E}_{+}(q,\phi)=\mathcal{E}+\sum_{l\sigma}\frac{1}{2}E^{}_{l\sigma-}(q,\phi).
\end{align}
is the $\phi$-dependent energy of the vacuum state $\gamma_{l,\sigma,+}|V\rangle_+=0$. The form in Eq.~\eqref{gdrdatfgblrgk} is also convenient for the calculation of the ensemble-averaged supercurrent (Sec.~\eqref{ensemble-avg}). 

\section{Derivation of Eq. (4) (main text) for the supercurrent $I_G(q,\phi)$} \label{supercurrent}

For completeness, in this section, we derive the supercurrent in Eq.~(6) of the main text following Ref.~\cite{pillet2010andreev,pillet2011tunneling}.  

\subsection{Current definition}

Quite generally, in terms of the time evolution of the electron number operator in the left lead $N_L=\sum_{ns}c^{\dagger}_{Lns}c^{}_{Lns}$,  we can define current as
 \begin{align} \label{fdnkvmkL}
     I_{G}=+e \langle \frac{d}{dt}N_L\rangle =+i\frac{e}{\hbar} \langle [H,N_L]\rangle, 
 \end{align}
where $H$ is given in Eq.~(1) in the main text and $H_T$ (see below) is its tunnel coupling part.
The angle bracket $\langle. \rangle$ in Eq.~\eqref{fdnkvmkL} denotes either (i) the expectation value in the 
ground state [e.g., Eq.~(3) in the main text for our problem] at zero temperature ($T=0$) or (ii) the
grand-canonical ensemble average for non-zero $T$, $e<0$ is the charge of the 
electron and $\hbar$ is the reduced Plank constant. Here we use the subindex ``$G$' in $I_G$ as our focus 
in the main text is the ground-state ($T=0$) supercurrent.
Equivalently, we can define current from the electron number operator in the right lead  $N_R=\sum_{ns}c^{\dagger}_{Rns}c^{}_{Rns}$,
\begin{align} \label{fdnkvmkR}
     I_{G}=-e \langle \frac{d}{dt}N_R\rangle =-i\frac{e}{\hbar} \langle [H,N_R]\rangle.
 \end{align}
Using 
  \begin{align}
     \left\langle \left[\left(\Delta^{j}_n c^{\dagger}_{jn\uparrow}c^{\dagger}_{jn\downarrow}+h.c. \right) ,\sum_{s}c^{\dagger}_{jns}c^{}_{jns}\right]\right\rangle \\
     =-2\left(\Delta^{j}_n \langle c^{\dagger}_{jn\uparrow}c^{\dagger}_{jn\downarrow} \rangle-\Delta^{j*}_n \langle c^{}_{jn\uparrow}c^{}_{jn\downarrow} \rangle\right)\notag =0, \notag 
 \end{align}
which follows from the self-consistency condition of the order parameter, Eqs.~\eqref{fdnkvmkL} and \eqref{fdnkvmkR} become
 \begin{align} \label{fdnkvmkLR}
     I_{G}=+i\frac{e}{\hbar} \langle [H_T,N_L]\rangle=-i\frac{e}{\hbar} \langle [H_T,N_R]\rangle, 
 \end{align}
To calculate the commutators in Eq.~\eqref{fdnkvmkLR}, we first do the transformation $c_{Lns}\rightarrow e^{-i\phi/4}c_{Lns}$ and $c_{Rns}\rightarrow e^{+i\phi/4}c_{Rns}$ in Eq.~(1), main text, which leads to
\begin{align} \label{fdnvkdk}
    H_T=t\sum_{ss'}\left[e^{+i\phi/4}c^{\dagger}_{LNs} \mathcal{U}_{ss'}(\theta_{so})d_{s'} 
    \right.\\
    +\left.e^{-i\phi/4}d^{\dagger}_{s} \mathcal{U}_{ss'}(\theta_{so})c_{R1s'} +h.c. \right], \notag  
\end{align}
and
\begin{align} \label{commutL}
    [H_T,N_L]&=+t\sum_{ss'}\left[e^{+i\phi/4}c^{\dagger}_{LNs} \mathcal{U}_{ss'}(\theta_{so})d_{s'}\right.\\
    &-\left.e^{-i\phi/4}d^{\dagger}_{s'}\mathcal{U}^{*}_{ss'}(\theta_{so})c^{}_{LNs}  \right],\notag
\end{align}
\begin{align} 
    [H_T,N_R]&=-t\sum_{ss'}\left[e^{+i\phi/4}c^{\dagger}_{R1s} \mathcal{U}^{*}_{ss'}(\theta_{so})d_{s'}\right.\\
    &-\left.e^{-i\phi/4}d^{\dagger}_{s}\mathcal{U}^{}_{ss'}(\theta_{so})c^{}_{R1s'}  \right].
    \label{commutR}
\end{align}
Combining Eqs.~\eqref{commutL}, \eqref{commutR} and 
\begin{align}
    \frac{\partial}{\partial \phi} H_T = \frac{it}{4}\sum_{ss'}\left[e^{+i\phi/4}c^{\dagger}_{LNs} \mathcal{U}_{ss'}(\theta_{so})d_{s'} \right.\\
    -\left.e^{-i\phi/4}d^{\dagger}_{s} \mathcal{U}_{ss'}(\theta_{so})c_{R1s'} -h.c. \right],\notag 
\end{align}
we find the identity
\begin{align}
    \langle \frac{\partial}{\partial \phi} H_T\rangle= \frac{i}{4}\left(\langle [H_T,N_L]\rangle-\langle [H_T,N_R]\rangle\right).
\end{align}
Using $\langle [H_T,N_R]\rangle=-\langle [H_T,N_L]\rangle$ [Eq.~\eqref{fdnkvmkLR}], we can write the above as
\begin{align}
    \langle [H_T,N_L]\rangle=-2i\left\langle \left(\frac{\partial}{\partial \phi} H_T\right)\right\rangle.
\end{align}
Hence the current \eqref{fdnkvmkL} becomes
\begin{align} \label{fvkdvldl}
  I_G(q,\phi)=\frac{2e}{\hbar} \left\langle  \left(\frac{\partial}{\partial \phi} H_T\right) \right\rangle,
\end{align}
where we have explicitly written out the $q$ and $\phi$ dependencies of $I_G$ [there is also, in principle, a temperature dependence (not indicated) in the case we are considering a thermal average (Sec.~\ref{ensemble-avg})]. Since we are dealing with a superconducting system with no voltage applied, we will hereafter refer to $I_G$ in \eqref{fvkdvldl} as supercurrent.

\subsection{Thermal average of the supercurrent in the grand-canonical ensemble}
\label{ensemble-avg}

Even though we are interested in the zero-temperature limit ($T\rightarrow 0$) in the main text, here 
we calculate the supercurrent $I_G$ in Eq.~\eqref{fvkdvldl} by performing an ensemble average in 
the grand-canonical ensemble. As we see below, besides being instructive, this provides, as a bonus, an alternative way 
to obtain the ground-state energy $\mathcal{E}_{G}(q,\phi)$ of our system.

The grand potential function $\Omega(T,q,\phi)$ is defined as
\begin{equation}
\Omega(T,q,\phi)=-(1/\beta) \ln (\mathcal{Z})
\end{equation}
 with 
$\mathcal{Z} \equiv\operatorname{tr}\{e^{-\beta H}\}$ being the grand partition function. More explicitly, using $H$ in Eq.~\eqref{gdrdatfgblrgk}, we have 
\begin{align}
    \mathcal{Z}&=\operatorname{tr}\left\{e^{-\beta \mathcal{E}_{+}(q,\phi)} \prod_{l\sigma} e^{-\beta E_{l\sigma+}(q,\phi)\gamma^{\dagger}_{l\sigma+}\gamma_{l\sigma+}}\right \}\notag\\
    &=e^{-\beta \mathcal{E}_{+}(q,\phi)}\prod_{l\sigma}\operatorname{tr}\left\{  e^{-\beta E_{l\sigma+}(q,\phi)\gamma^{\dagger}_{l\sigma+}\gamma_{l\sigma+}}\right \}\notag\\
    &=e^{-\beta \mathcal{E}_{+}(q,\phi)}\prod_{l\sigma}\left( 1+ e^{-\beta E_{l\sigma+}(q,\phi)}\right ).
\end{align}
The ensemble-averaged supercurrent [Eq~\eqref{fvkdvldl}] 
can be expressed as
\begin{align} \label{fdvnkdv}
I_G(T,q,\phi) & =\frac{2e}{\hbar} \frac{1}{\mathcal{Z}} \operatorname{tr}\left\{e^{-\beta H} \frac{\partial}{\partial \phi} H_{T}\right\} \notag \\ 
& =\frac{2e}{\hbar}\frac{1}{\mathcal{Z}} \operatorname{tr}\left\{\frac{\partial}{\partial \phi} e^{-\beta H}\right\} \left(\frac{1}{-\beta}\right)\notag \\
&=\frac{2e}{\hbar} \frac{1}{-\beta} \frac{1}{\mathcal{Z}} \frac{\partial}{\partial \phi} \mathcal{Z}\notag \\
&=\frac{2e}{\hbar} \frac{-1}{\beta} \frac{\partial}{\partial \phi} \ln (\mathcal{Z}) \notag \\
& =\frac{2e}{\hbar} \frac{\partial}{\partial \phi} \Omega(T,q,\phi),  
\end{align}
with 
\begin{align}
    \Omega(T,q,\phi)=\mathcal{E}_{+}(q,\phi)-\frac{1}{\beta} \sum_{l\sigma} \ln \left( 1+ e^{-\beta E_{l\sigma+}(q,\phi)}\right ).
    \label{grand-pot}
\end{align}
Note that the pre-factor $2e/\hbar$ appears naturally in the  supercurrent \eqref{fdvnkdv}, obtained quite generally from  \eqref{fdnkvmkL} or \eqref{fdnkvmkR}.

\subsubsection{$T\rightarrow 0$ limit of $\Omega(T,q,\phi$): ground-state energy}

In the $T\rightarrow 0$ ($\beta\rightarrow \infty$) we have
\begin{align}
    \frac{1}{\beta}\ln \left( 1+ e^{-\beta E_{l\sigma+}(q,\phi)}\right )\rightarrow 
    \left\{
    \begin{matrix*}
        -E_{l\sigma+}(q,\phi), & E_{l\sigma+}(q,\phi)<0;\\
        0, & E_{l\sigma+}(q,\phi)>0.
    \end{matrix*}\right.
\end{align}
Interestingly, in this limit the grand potential $\Omega(T,q,\phi)$ [Eq~\eqref{grand-pot}] reduces to the ground-state energy 
$\mathcal{E}_{G}(q,\phi)$ [Eq.~(5) in the main text], i.e.,
\begin{align} \label{fdnvkdfk}
  \mathcal{E}_{G}(q,\phi)=\Omega(0,q,\phi)=\mathcal{E}_{+}(q,\phi)+ \sum_{E_{l\sigma+}(q,\phi)<0}E_{l\sigma+}(q,\phi),
\end{align}
where  $\mathcal{E}_{+}(q,\phi)$ is defined in Eq.~\eqref{fvdmkf}. Note that the above result, more than just yielding the correct limit, provides an alternative way to derive the ground-state energy $\mathcal{E}_{G}(q,\phi)$. In the main text, we derived $\mathcal{E}_{G}(q,\phi)$ from the \textit{exact} $T=0$ ground-state wave function $|G\rangle$ [Eq.~(3)] via $H|G\rangle= \mathcal{E}_{G}(q,\phi)|G\rangle$. As mentioned in the main text, since $E_{l\sigma\eta}(q,\phi)=-E_{l-\sigma-\eta}(q,\phi)$, we can replace all positive energies in \eqref{fdnvkdfk} [and in \eqref{fvdmkf}] with \textit{negative} eigenenergies $E_{l\sigma\eta}<0$ to obtain
\begin{align} \label{fvkdfmvka}
    \mathcal{E}_{G}(q,\phi)=\mathcal{E}+ \frac{1}{2}\sum_{E_{l\sigma\eta}(q,\phi)<0}E_{l\sigma\eta}(q,\phi),
\end{align}
where $\mathcal{E}$ is the constant defined in Eq.~\eqref{constant-E}, being independent of $q$ and $\phi$. Note that the form of $\mathcal{E}_{G}(q,\phi)$ in \eqref{fvkdfmvka} is particularly convenient to calculate the ground-state supercurrent (see below) as its $\phi$-dependence is solely included in the second term, which involves only negative quasi-particle eigenenergies.

\subsubsection{$T\rightarrow 0$ limit: ground-state supercurrent $I_G(q,\phi)$}

From the general expression for the supercurrent in Eq.~\eqref{fdvnkdv}, we can straightforwardly [using Eq.~\eqref{fdnvkdfk}] obtain the usual ground-state (i.e., $T=0$) supercurrent 
\begin{equation}
I_G(q,\phi) = I_G(0,q,\phi)  \frac{2e}{\hbar} \frac{\partial}{\partial \phi} \mathcal{E}_{G}(q,\phi), 
\label{zeroT-Ig}
\end{equation}
which, upon using Eq.~\eqref{fvkdfmvk}, yields Eq.~(6) in the main text
\begin{align} \label{fddbdjvkdfr}
I_G(q,\phi)=\frac{I_0}{2}\sum_{E_{l\sigma\eta}(q,\phi)<0}\partial_{\phi}E_{l\sigma\eta}(q,\phi),
\end{align}
with $I_0=2e/\hbar$.

\section{Additional phase shifts in finite-$q$ superconductors}\label{AndreevrReflection}

In this section, we define the additional phase shifts $\delta\phi_j(q,\phi)$, central to our discussion and plotted in Figs.~1(b) 
and 1(c) of the main text, in terms of the matrix elements of the anomalous Green functions of the finite-momentum 
superconducting leads. To this end, below we focus on an effective 
description of the quantum dot in which the effect of the leads are 
accounted for via a self energy. We derive a general expression for the self energy that incorporates the 
phase shifts $\delta\phi_j(q,\phi)$ via the superconducting lead (anomalous) Green functions. As we will see later on, these phase shifts (via the self energy) provide a simple mechanism to understand (i) the finite-$q$-induced asymmetry of the Andreev dispersions [Sec.~\eqref{sec-asym}], crucial to the SDE, and (ii) the fermion parity changes in the ground state [Sec.~\eqref{sec-parity}]

\subsection{Defining an effective dot Hamiltonian and self energy} \label{fvdkvoav}

A simpler starting point for studying Andreev bound states in our hybrid  
quantum-dot/superconducting-lead system is to consider an effective quantum dot Hamiltonian in which 
the leads have been integrated out thus giving rise to a self energy, which, as we will see below, 
emulates a pair potential in the quantum 
dot \cite{meng2009self,kurilovich2021microwave,rozhkov1999josephson,bauer2007spectral,meden2019anderson,fatemi2021microwave}. 
To this end, we note that the diagonalization of the BdG matrix \eqref{fdnkvamk} results in the determinantal equation,
\begin{align} \label{fvmldfvdl}
    \det\left[\begin{array}{cccc} 
    \mathcal{H}_{D}-E  & \mathcal{T}_{L} & \mathcal{T}_{R} \\
   \mathcal{T}^{\dagger}_{L}& \mathcal{H}_{L} -E & 0 \\
   \mathcal{T}^{\dagger}_{R}& 0 & \mathcal{H}_{R}-E
    \end{array}\right]=0.
\end{align}
To find the  Andreev (subgap)  levels we can use the identity for determinants of block matrices comprised of four matrices $A,B,C$, and $D$ \cite{silvester2000determinants},
\begin{align} \label{fvnvkdk}
    \det {\begin{pmatrix}A&B\\C&D\end{pmatrix}}=\det(D)\det \left(A-BD^{-1}C\right),
\end{align}
where $D$ is assumed to be invertible. By comparing Eqs.~\eqref{fvmldfvdl} and \eqref{fvnvkdk} we can make 
make the identifications: $A \leftrightarrow [H_D-E]$, $B  \leftrightarrow  [\mathcal{T}_{L} \quad \mathcal{T}_{R}]$, $C  \leftrightarrow B^\dagger$, and
\begin{align} \label{D}
   D \leftrightarrow \left[\begin{array}{cc} 
   \mathcal{H}_{L} -E & 0 \\
   0 & \mathcal{H}_{R}-E
    \end{array}\right].
\end{align}
Now we have to guarantee that the blocks 
$\mathcal{H}_j-E$ (i.e., block $D$) are invertible. As it turns out, this is the case for Andreev eigenenergies (i.e., for in-gap $E$'s). In this case, Eq.~\eqref{fvmldfvdl} [via Eq.~\eqref{fvnvkdk}] reduces to 
\begin{align} \label{fdvkdfmv}
   \det\left\{\mathcal{H}_{D}-E-\Sigma(q,E,\phi_L,\phi_R)\right\}
    =0,
\end{align}
where we have introduced the self energy
\begin{align} \label{fvmkfllgf}
    \Sigma(q,E,\phi_L,\phi_R)&=\begin{bmatrix}
        \mathcal{T}_{L} & \mathcal{T}_{R}
    \end{bmatrix}
    \begin{bmatrix}
        \mathcal{H}_{L} -E & 0 \\
   0 & \mathcal{H}_{R}-E
    \end{bmatrix}^{-1}
    \begin{bmatrix}
        \mathcal{T}^{\dagger}_{L}\\ \mathcal{T}^{\dagger}_{R}
    \end{bmatrix} \notag \\
   & =\sum_{j=L,R}  [\mathcal{T}_{j}]_{4\times 4N}  \left[\frac{1}{\mathcal{H}_{j}-E} \right]_{4N\times 4N} [\mathcal{T}^{\dagger}_{j}]_{4N\times 4}.  
\end{align}
Note that the determinantal Eq.~\eqref{fdvkdfmv} can be viewed as stemming from the effective quantum-dot 4x4 Hamiltonian, 
\begin{equation}
\mathcal{H}_D^{eff}=\mathcal{H}_{D} - \Sigma(q,E,\phi_L,\phi_R),
\label{eff-D}
\end{equation}
in which the superconducting leads have been eliminated thus giving rise to the self energy \eqref{fvmkfllgf}. Interestingly, this   
self-energy emulates a pair potential in the quantum dot [via $\mathcal{H}_{j}$, see Eqs.~ \eqref{tvfnvmk}and \eqref{vnskfm}].

Here we derive an expression for the dot self energy \eqref{fvmkfllgf} making more explicit the contributions from the left and right leads.
By using Eqs. \eqref{fvmfkll} and \eqref{fvmfklr},  Eq. \eqref{fvmkfllgf} becomes
\begin{widetext}
\begin{align} \label{fvmvldl}
       \Sigma(q,E,\phi_L,\phi_R) & =t^2\begin{bmatrix}
      0 &  \cdots & 0 &  U(\theta_{so})
 \end{bmatrix}_{4\times 4N}  \left[\frac{1}{\mathcal{H}_{L}-E} \right]_{4N\times 4N} \begin{bmatrix}0\\
 \vdots\\
 0 \notag \\ 
 U^{+}(\theta_{so})
 \end{bmatrix}_{4N\times 4}\\
 &+t^2\begin{bmatrix}
      U^{+}(\theta_{so})& 0 &  \cdots & 0 
 \end{bmatrix}_{4\times 4N}  \left[\frac{1}{\mathcal{H}_{R}-E} \right]_{4N\times 4N} \begin{bmatrix}   U(\theta_{so})\\
 0\\
 \vdots\\
 0\end{bmatrix}_{4N\times 4}.
\end{align}
\end{widetext}
The quantum dot is coupled to the last site of the left superconducting lead and the first site of the right superconducting lead, as shown by the first and second lines of Eq.  \eqref{fvmvldl}, respectively, and hence only the last  $4\times 4$ block of $[1/(\mathcal{H}_L-E)]_{4N\times 4N}$ and  the first $4\times 4$ block of $[1/(\mathcal{H}_R-E)]_{4N\times 4N}$ participate in the calculation of the self-energy \eqref{fvmvldl}. More explicitly, below we specify the lowermost and uppermost 4x4 diagonal blocks of $G_{L,R}$, denoted by $[G_{L}(q,E,\phi_L)]_{4\times 4}$ and $[G_R(q,E,\phi_R)]_{4\times 4}$,
\begin{align} \label{fdvmdllL}
   G_{L}(q,E,\phi_L)&= \frac{1}{\mathcal{H}_L-E}\\
   &\equiv
    \begin{bmatrix}
        \square & \square & \square & \square \\
        \square & \ddots & \vdots & \vdots \\
        \square & \cdots & \square &  \square\\
        \square & \cdots & \square & [G_{L}(q,E,\phi_L)]_{4\times 4}
 \end{bmatrix}_{4N\times 4N},\notag
\end{align}
\begin{align} \label{fdvmdllR}
       G_{R}(q,E,\phi_R) &=  \frac{1}{\mathcal{H}_R-E}\\
       &\equiv
    \begin{bmatrix}
       [G_{R}(q,E,\phi_R)]_{4\times 4} & \square & \cdots & \square \\
        \square & \square & \cdots & \square  \\
        \vdots & \vdots & \ddots & \square  \\
        \square & \square & \square  & \square  
 \end{bmatrix}_{4N\times 4N}. \notag 
\end{align}
Each empty square in Eqs.~\eqref{fdvmdllL} and \eqref{fdvmdllR} is the $4\times 4$ matrix 
\begin{align} \label{fffdmvmld}
    \square =\begin{bmatrix}
    \vartriangleright & 0 & \vartriangleright & 0 \\
    0 & \vartriangleleft & 0 & \vartriangleleft\\
     \vartriangleright & 0 & \vartriangleright & 0\\
    0 & \vartriangleleft & 0 & \vartriangleleft
    \end{bmatrix}.
\end{align}
In the upper and lower diagonal 2x2 blocks of $\square$ above, the triangles denote couplings between only same spins, while 
in its off-diagonal 2x2 blocks Cooper pairing couples only opposite spins, see Eq. \eqref{vnskfm}. 
Substituting Eqs. \eqref{fdvmdllL} and \eqref{fdvmdllR} into \eqref{fvmvldl}, we obtain the general expression for the self energy
\begin{align} \label{selfenergy}
   \Sigma(q,E,\phi_L,\phi_R) &= t^2U_{}(\theta_{so})[G_{L}(q,E,\phi_L)]_{4\times 4} U^{+}(\theta_{so})\notag \\
   &+t^2U^{+}(\theta_{so})[G_{R}(q,E,\phi_R)]_{4\times 4} U^{}(\theta_{so}).
\end{align}
The first and second terms on the right-hand side of \eqref{selfenergy} correspond to the contributions from the left and right leads, respectively.

\begin{figure*}[t]
\begin{center}
\includegraphics[width=0.8\textwidth]{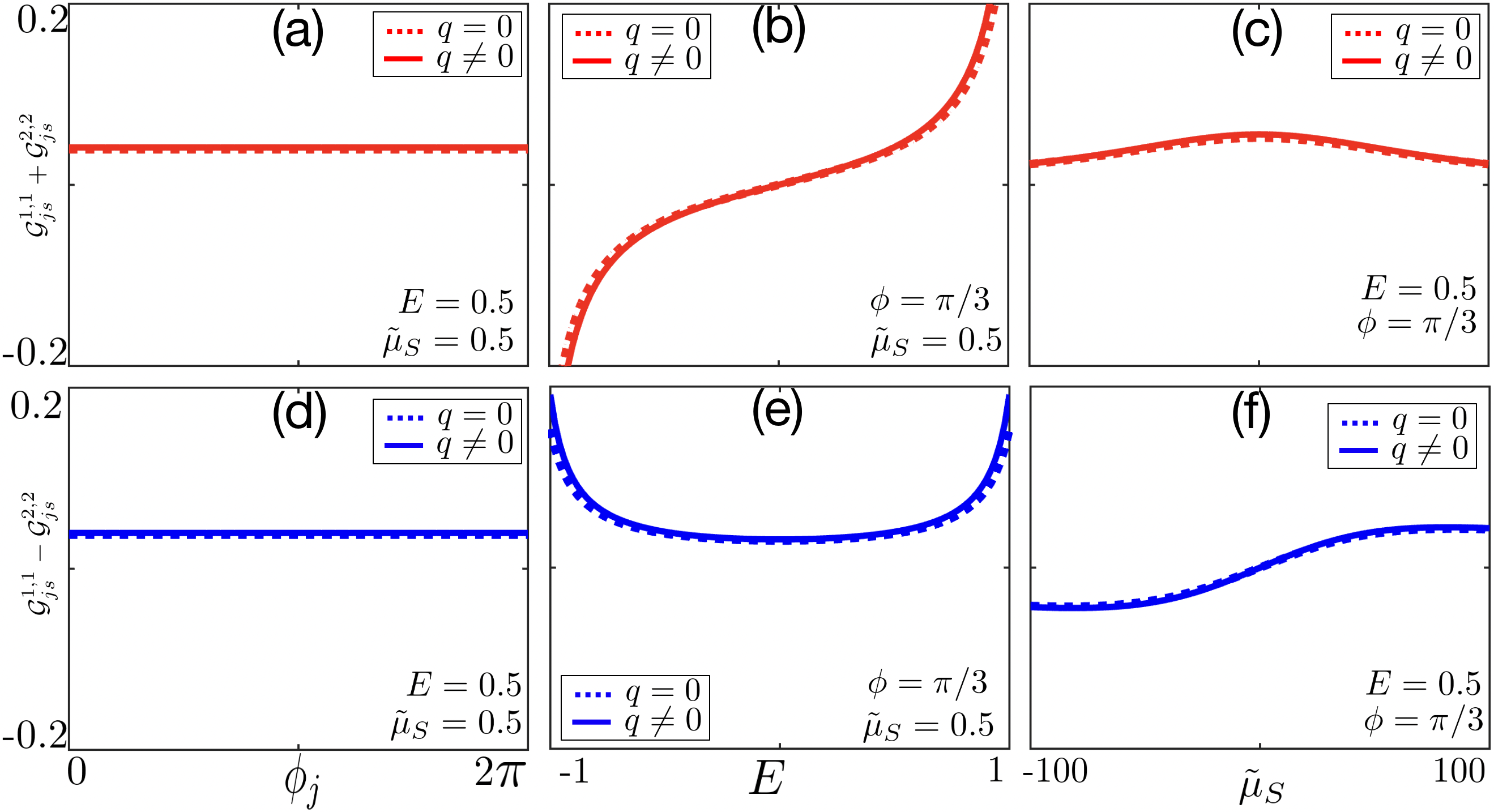}
\end{center}
\caption{Phase $\phi_j$, energy $E$,  and potential $\tilde{\mu}_S$ dependencies of the diagonal Green functions Eq~(\ref{fdvmdkfmkv}) $\mathcal{G}^{l,l}_{js}$ ($l=1,2$, $j=L,R$, $s=\uparrow,\downarrow$).  Panels (a), (b), and (c) 
exhibit $\mathcal{G}^{1,1}_{js}+\mathcal{G}^{2,2}_{js}$ as functions of $\phi_j$, $E$, and $\tilde{\mu}_S$, 
respectively, for different values of 
$q$. Panels (d),  (e), and (f)  correspond to (a), (b), and (c) but for  $\mathcal{G}^{1,1}_{js}-\mathcal{G}^{2,2}_{js}$. From panels (a) and (d) we see that $\mathcal{G}^{1,1}_{js}+\mathcal{G}^{2,2}_{js}$ and $\mathcal{G}^{1,1}_{js}-\mathcal{G}^{2,2}_{js}$ (i.e., $\mathcal{G}^{l,l}_{js}$)  
are independent of $\phi_j$ for both $q=0$ and $q\neq 0$. While $\mathcal{G}^{1,1}_{js}+\mathcal{G}^{2,2}_{js}$ is an odd function of $E$ [panel (b)], $\mathcal{G}^{1,1}_{js}-\mathcal{G}^{2,2}_{js}$ is an even function of $E$ [panel (e)]. However, they show opposite $\tilde{\mu}_S$ dependence. $\mathcal{G}^{1,1}_{js}+\mathcal{G}^{2,2}_{js}$ is an even function of $\tilde{\mu}_S$ [panel (c)] but $\mathcal{G}^{1,1}_{js}-\mathcal{G}^{2,2}_{js}$ is an odd function of $\tilde{\mu}_S$ [panel (f)].
Here, we use the Cooper-pair momentum $\hbar qv_F/\Delta= 1.0$. Other parameters are given by $\Delta=1$, $\epsilon_d=2.6$, $h_{sc}=0$, $t=-13$ and $t_0=-100$, that are the same as Fig. 2 in main text.} 
\label{AFIGG}
\end{figure*}

\begin{figure*}[t]
\begin{center}
\includegraphics[width=0.8\textwidth]{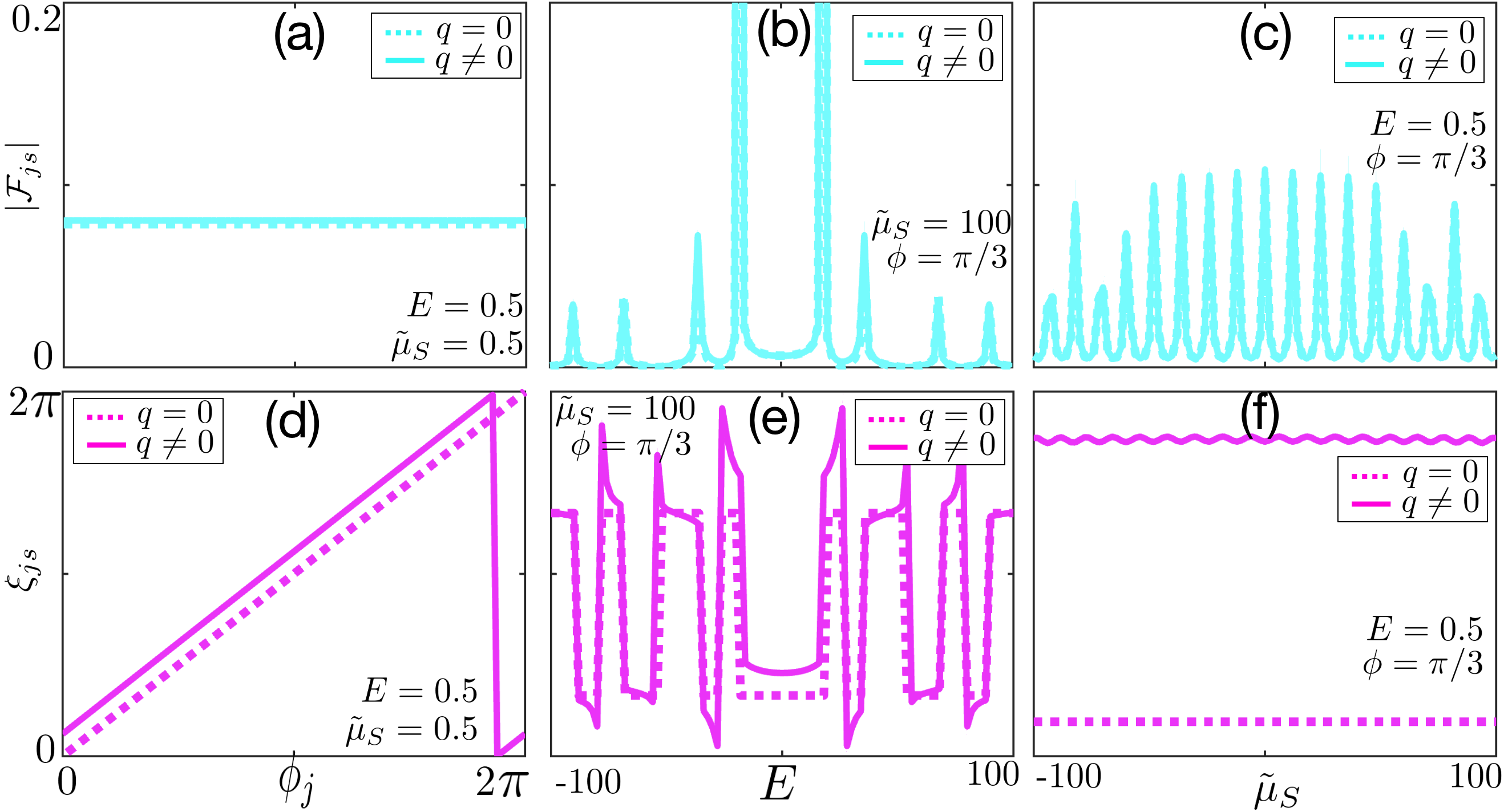}
\end{center}
\caption{Phase $\phi_j$, energy $E$,  and potential $\tilde{\mu}_S$ dependencies of the off-diagonal Green functions Eq~(\ref{fdvmdkfmkv}) $\mathcal{F}_{js}$ ($j=L,R$, $s=\uparrow,\downarrow$).  Panels (a), (b), and (c) 
exhibit $\left| \mathcal{F}_{js}\right|$ as functions of $\phi_j$, $E$,  and $\tilde{\mu}_S$
respectively, for different values of 
$q$. Panels (d), (e), and (f)  correspond to (a), (b), and (c) but for  $\xi_{js}$. Both $\left| \mathcal{F}_{js}\right|$ and $\xi_{js}$ are even function of $E$ and $\tilde{\mu}_S$ [panels (b), (c), (e), and (f)]. While the moduli $\left| \mathcal{F}_{js}\right|$ is independent of $\phi_j$, 
the phases $\xi_{js}$ vary linearly with $\phi_j$ for both $q=0$ and $q\neq 0$; however, for $q\neq 0$ the phases $\xi_{js}$ 
show an abrupt transition from $2\pi$ to $0$, when plotted in the range $[0,2\pi]$. The linear dependence on $\phi_j$  together with this  
transition imply the existence of an additional phase shift $\delta \phi_j(q,E)$ for non-zero $q$, see panel (d). This additional 
phase shifts arise from the Andreev reflections in the finite-momentum superconducting leads. In fact, for $q=0$,  
$\xi_{js}=\phi_j$ mod($2\pi$)  as  $\delta \phi_j(q,E)=0$, see dotted line in (d), and  
$\xi_{js}(q,E,\tilde{\mu}_S,\phi_j)=\phi_j + \delta \phi_j(q,E,\tilde{\mu}_S)$ for finite $q$, see solid line (magenta) in (d). We emphasize 
that the phase shifts $\delta \phi_j(q,E)$ do not depend on $\phi_j$. Other parameters are the same as Fig. \ref{AFIGG}.}
\label{AFIGF}
\end{figure*}

\subsection{Additional phase shifts $\delta\phi_j(q,\phi)$}

Here we write out the $4\times 4$ blocks $[G_{j}(q,E,\phi_j)]_{4\times 4}$,  $j=L,R$ [Eqs.~\eqref{fdvmdllL} and \eqref{fdvmdllR}], explicitly making use  of their hermitian properties. By introducing diagonal $\mathcal{G}^{}_{js}$ and off-diagonal $\mathcal{F}^{}_{js}$ 
matrix elements we can write
\begin{align} \label{fdvmdkfmkv}
    [G_{j}&(q,E,\phi_j)]_{4\times 4}\equiv\begin{bmatrix}
    \mathcal{G}^{1,1}_{j\uparrow} & 0 & -\mathcal{F}^{}_{j\uparrow} & 0 \\
    0 & \mathcal{G}^{1,1}_{j\downarrow} & 0 & -\mathcal{F}^{}_{j\downarrow}\\
     -\mathcal{F}^{*}_{j\uparrow} & 0 & \mathcal{G}^{2,2}_{j\uparrow} & 0\\
    0 & -\mathcal{F}^{*}_{j\downarrow} & 0 & \mathcal{G}^{2,2}_{j\downarrow}
    \end{bmatrix},
\end{align}
where we have omitted the $q$, $E$, and $\phi_j$ dependencies of the matrix elements 
for simplicity. The off-diagonal functions $\mathcal{F}^{}_{js}$ correspond to \text{anomalous} $G_{L,R}$ matrix elements of the 
superconducting leads. Note that all Green functions in
Eq.~\eqref{fdvmdkfmkv}  [also in Eqs.~\eqref{fdvmdllL} and \eqref{fdvmdllR}] have a magnetic field  
dependence $h_{sc}$ [see, Eqs.~\eqref{tvfnvmk},  \eqref{vnskfm} and \eqref{epsilon-hsc}]; whenever discussing non-zero $h_{sc}$ effects, we 
should make the substitution $E \rightarrow E - sh_{sc}$ in Eq.~\eqref{fdvmdkfmkv} to make the $h_{sc}$ dependence explicit.
As usual, hermiticity requires that the diagonal Green functions 
$\mathcal{G}^{l,l}_{js}(q,E,\phi_j)$ ($l=1,2$) be real numbers.

The anomalous Green functions $\mathcal{F}_{js}(q,E,\tilde{\mu}_S,\phi_j)$, on the other hand, can in principle be 
complex, where $\tilde{\mu}_S=\mu_S-2t_0$. To extract the moduli and phases of $\mathcal{F}_{js}(q,E,\tilde{\mu}_S,\phi_j)$,  we use their polar form
\begin{align} \label{fdavkkfl}
    \mathcal{F}_{js}(q,E,\tilde{\mu}_S,\phi_j) \equiv \left| \mathcal{F}_{js}(q,E,\tilde{\mu}_S,\phi_j)\right| e^{i\xi_{js}(q,E,\tilde{\mu}_S,\phi_j)},
\end{align}
where $\xi_{js}(q,E,\tilde{\mu}_S,\phi_j)$ denote their phases. By comparing the numerically determined $\mathcal{F}_{js}(q,E,\tilde{\mu}_S,\phi_j)$ 
with Eq.~\eqref{fdavkkfl}, we can 
determine the moduli $\left| \mathcal{F}_{js}(q,E,\tilde{\mu}_S,\phi_j)\right|$ and phases $\xi_{js}(q,E,\tilde{\mu}_S,\phi_j)$. The minus sign in front of $\mathcal{F}_{js}$ in Eq.~(\ref{fdvmdkfmkv}) is added so that $\xi_{js}(q,E,\tilde{\mu}_S,\phi_j)=\phi_j$ when $q=0$.

Figure~\ref{AFIGG} (a) shows $\mathcal{G}^{1,1}_{js}(q,E,\tilde{\mu}_S,\phi_j)+\mathcal{G}^{2,2}_{js}(q,E,\tilde{\mu}_S,\phi_j)$ as a function of phase $\phi_j$ for different values of 
$q$. Figures \ref{AFIGG} (b) and (c) are similar to Fig.~\ref{AFIGG} (a) but as a function of the energy $E$ and $\tilde{\mu}_S$, respectively. Panels (d),  (e), and (f)  correspond to (a), (b), and (c) but for  $\mathcal{G}^{1,1}_{js}(q,E,\tilde{\mu}_S,\phi_j)-\mathcal{G}^{2,2}_{js}(q,E,\tilde{\mu}_S,\phi_j)$. Note that the Green $\mathcal{G}^{1,1}_{js}(q,E,\tilde{\mu}_S,\phi_j)+\mathcal{G}^{2,2}_{js}(q,E,\tilde{\mu}_S,\phi_j)$ and $\mathcal{G}^{1,1}_{js}(q,E,\tilde{\mu}_S,\phi_j)-\mathcal{G}^{2,2}_{js}(q,E,\tilde{\mu}_S,\phi_j)$ [i.e., $\mathcal{G}^{l,l}_{js}(q,E,\tilde{\mu}_S,\phi_j)$]  
are independent of $\phi_j$ for both $q=0$ and $q\neq 0$, as shown in  Fig.~\ref{AFIGG}(a) and Fig.~\ref{AFIGG}(d), 
respectively. From Figs.~ref{AFIGG} (b) and (e) we see that  $\mathcal{G}^{1,1}_{js}(q,E,\tilde{\mu}_S,\phi_j)+\mathcal{G}^{2,2}_{js}(q,E,\tilde{\mu}_S,\phi_j)$ is an odd function of $E$ but $\mathcal{G}^{1,1}_{js}(q,E,\tilde{\mu}_S,\phi_j)-\mathcal{G}^{2,2}_{js}(q,E,\tilde{\mu}_S,\phi_j)$ is an even function of $E$. While $\mathcal{G}^{1,1}_{js}(q,E,\tilde{\mu}_S,\phi_j)+\mathcal{G}^{2,2}_{js}(q,E,\tilde{\mu}_S,\phi_j)$ is an even function of $\tilde{\mu}_S$ [Figs.~\ref{AFIGG}  (c)], $\mathcal{G}^{1,1}_{js}(q,E,\tilde{\mu}_S,\phi_j)-\mathcal{G}^{2,2}_{js}(q,E,\tilde{\mu}_S,\phi_j)$ is an odd function of $\tilde{\mu}_S$ [Figs.~\ref{AFIGG}  (f)]. Figs.~\ref{AFIGF} (a), (b), and (c)  
correspond to Figs.~\ref{AFIGG} (a), (b), and (c) but for $\left| \mathcal{F}_{js}(q,E,\tilde{\mu}_S,\phi_j)\right|$. Figures \ref{AFIGF} (d), (e), and (f) correspond to (a), (b) 
and (c) but for $\xi_{js}(q,E,\tilde{\mu}_S, \phi_j)$. From Fig.~\ref{AFIGF}(a), we see that the moduli 
$\left| \mathcal{F}_{js}(q,E,\tilde{\mu}_S,\phi_j)\right| $ of the anomalous Green functions  are independent  of $\phi_j$. It becomes even functions of $E$ and $\tilde{\mu}_S$,  as shown in Figs.~\ref{AFIGF}(b) and (c). Their  phases  $\xi_{js}(q,E,\tilde{\mu}_S,\phi_j)$ 
are  linear in $\phi_j$ [Figs.~\ref{AFIGF}(d)] for $q=0$ and $q\neq 0$. These phases are independent of $E$ for $q=0$, while strongly 
energy dependent for  $q\neq 0$, as shown
in Fig.~\ref{AFIGF}(e). Thus, we suggest the ans\"atz 
\begin{equation}
\xi_{js}(q,E,\tilde{\mu}_S,\phi_j)=\phi_j+\delta\phi_j(q,E,\tilde{\mu}_S),
\label{phase-shifts}
\end{equation}
where $\delta\phi_j(q,E,\tilde{\mu}_S)$ are additional phase shifts due to the Andreev reflections in 
the finite-momentum superconducting leads. These phase shifts, as it turns out, do not explicitly 
depend on $\phi_j$. As we will see in the next section, the additional phase shifts 
\begin{equation}
\delta\phi_j(q,E,\tilde{\mu}_S) = \xi_{js}(q,E,\tilde{\mu}_S,\phi_j) - \phi_j.
\label{dphase-shifts}
\end{equation}
play an important role in giving rise to asymmetric Andreev dispersions, which, in turn, underlie the SDE in our Andreev QD setup.

To obtain a more compact and convenient form for further calculations involving $[G_{j}(q,E,\tilde{\mu}_S,\phi_j)]_{4\times 4}$, let us define the 
operation
\begin{align}
\bigotimes_{s=\uparrow,\downarrow} [X^{s}]_{2\times 2}\equiv
\begin{bmatrix}
    X^{\uparrow}_{1,1} & 0 & X^{\uparrow}_{1,2} & 0 \\
    0 & X^{\downarrow}_{1,1} & 0 & X^{\downarrow}_{1,2}\\
     X^{\uparrow}_{2,1} & 0 & X^{\uparrow}_{2,2} & 0\\
    0 & X^{\downarrow}_{2,1} & 0 & X^{\downarrow}_{2,2}
    \end{bmatrix},
\end{align}
where  $[X^{s}]_{2\times 2}$ denotes a generic $2\times 2$ matrix. 
Using this notation and Eq.~\eqref{phase-shifts}, we can now recast the $4\times4$ blocks $[G_{j}(q,E,\tilde{\mu}_S,\phi_j)]_{4\times 4}$ in Eq.~(\ref{fdvmdkfmkv}) as 
\begin{widetext}
\begin{align} \label{fdvadavfv}
    [G_{j}(q,E,\tilde{\mu}_S,\phi_j)]_{4\times 4}&=\bigotimes_{s=\uparrow,\downarrow}(E-sh_{sc})\begin{bmatrix}
    \mathcal{K}(q,E-sh_{sc},\tilde{\mu}_S) &  \\
    0  & \mathcal{K}(q,E-sh_{sc},\tilde{\mu}_S)
    \end{bmatrix} \\
 &+\bigotimes_{s=\uparrow,\downarrow}\tilde{\mu}_S\begin{bmatrix}
    +\mathcal{D}(q,E-sh_{sc},\tilde{\mu}_S)  & 0 \\
    0  &  -\mathcal{D}(q,E-sh_{sc},\tilde{\mu}_S)
    \end{bmatrix}\notag\\
    &+\bigotimes_{s=\uparrow,\downarrow}\mathcal{R}(q,E-sh_{sc},\tilde{\mu}_S)\begin{bmatrix}
    0 & - e^{+i\phi_j+i\delta\phi_j(q,E-sh_{sc},\tilde{\mu}_S}) \\
    - e^{-i\phi_j-i\delta\phi_j(q,E-sh_{sc},\tilde{\mu}_S})  &  0
    \end{bmatrix}\notag 
\end{align}
where  we have defined the functions 
\begin{align} \label{fvmslmvG}
           \mathcal{K}(q,E-sh_{sc},\tilde{\mu}_S) \equiv \frac{1}{ 2(E -sh_{sc})}[\mathcal{G}^{1,1}_{js}(q,E-sh_{sc},\tilde{\mu}_S,\phi_j)+\mathcal{G}^{2,2}_{js}(q,E-sh_{sc},\tilde{\mu}_S,\phi_j)],
\end{align}
\begin{align} \label{fvmslmvGa}
           \mathcal{D}(q,E-sh_{sc},\tilde{\mu}_S) \equiv \frac{1}{ 2\tilde{\mu}_S}[\mathcal{G}^{1,1}_{js}(q,E-sh_{sc},\tilde{\mu}_S,\phi_j)-\mathcal{G}^{2,2}_{js}(q,E-sh_{sc},\tilde{\mu}_S,\phi_j)],
\end{align}
\end{widetext}
\begin{align} \label{fvmslmvF}
      \mathcal{R}(q,E-sh_{sc},\tilde{\mu}_S)  \equiv \left| \mathcal{F}_{js}(q_j,E-sh_{sc},\tilde{\mu}_S,\phi_j)\right\vert .
\end{align}
In the above we have explicitly put in the dependence on the magnetic field $h_{sc}$ in the arguments 
of the functions $\mathcal{D}(q,E-sh_{sc},\tilde{\mu}_S)$,  $\mathcal{K}(q,E-sh_{sc},\tilde{\mu}_S)$ and $\mathcal{R}(q,E-sh_{sc},\tilde{\mu}_S)$; note that these are real functions,   
being independent of $\phi_{L,R}$, see Fig.~\ref{AFIGG} and ~Fig.~\ref{AFIGF}.

In the next two sections we use the form \eqref{fdvadavfv} to determine the self energy \eqref{selfenergy}, from which 
we can obtain the relevant Andreev dispersions.




\section{Origin of the asymmetric Andreev dispersions for finite $q$}
\label{sec-asym}

To gain some insight into how the finite-momentum $q$ of the Cooper pairs affects the Andreev reflections, and,  
in particular, how it leads to asymmetric dispersions, below we look in some detail at the simpler case  
with no Zeeman fields ($h_d=h_{sc}=0$) and no spin-orbit coupling ($\theta_{so}=0$).

\subsection{Self energy for $h_d=h_{sc}=\theta_{so}=0$}


First of all let us write out the self energy for this case in terms of the matrix elements of the superconducting lead Green functions.
Substituting Eq. \eqref{fdvadavfv} into the self-energy \eqref{selfenergy}, we obtain
\begin{widetext}
\begin{align} \label{selfenergy1}
    \Sigma(q,E,\tilde{\mu}_S,\phi_L,\phi_R)=2t^2\bigotimes_{s=\uparrow,\downarrow}\begin{bmatrix}
    E\mathcal{K}(q,E,\tilde{\mu}_S)+\tilde{\mu}_S\mathcal{D}(q,E,\mu_S)  & \mathcal{\bar R}(q,E,\tilde{\mu}_S,\phi_{L,R},\delta\phi_{L,R}(q,E,\tilde{\mu}_S))  \\
    \mathcal{\bar R}^{*}(q,E,\tilde{\mu}_S,\phi_{L,R},\delta\phi_{L,R}(q,E,\tilde{\mu}_S)) &  E\mathcal{K}(q,E,\mu_S)-\tilde{\mu}_S\mathcal{D}(q,E,\mu_S)
    \end{bmatrix},
\end{align}
\end{widetext}
where we have defined,
\begin{align} \label{fvjdfjvf}
    &\mathcal{\bar R}_{}(q,E,\tilde{\mu}_S,\phi_{L,R},\delta\phi_{L,R}(q,E,\tilde{\mu}_S))\\
    &=\frac{1}{2}\sum_{j=L,R} \mathcal{R}(q,E)  e^{i\phi_j+i\delta\phi_j(q,E,\tilde{\mu}_S}).\notag
\end{align}
The above describes the interference between the Andreev reflections in the left-lead and right-lead paths 
[Fig.~1(b), main text] and can be rewritten as
\begin{align} \label{fvjdfjv}
    &\mathcal{\bar R}_{}(q,E,\tilde{\mu}_S,\phi,\delta\phi(q,E,\tilde{\mu}_S),\phi_\text{eff})\\
    &=\mathcal{R}(q,E,\tilde{\mu}_S)\cos\left[\frac{\phi+\delta\phi(q,E,\tilde{\mu}_S)}{2}\right]e^{i\phi_{\text{eff}}},\notag 
\end{align}
with $\phi_{\text{eff}}=[\delta\phi_R(q,E,\tilde{\mu}_S)+\delta\phi_L(q,E,\tilde{\mu}_S)]/2$, where $\phi=\phi_R-\phi_L$ is the flux-tunable phase difference 
between the left and right superconducting leads and 
$\delta\phi(q,E,\tilde{\mu}_S)=\delta\phi_R(q,E,\tilde{\mu}_S)-\delta\phi_L(q,E,\tilde{\mu}_S)$ is the additional \textit{phase-shift difference} between 
the left and right phase shifts arising from the Andreev reflections in the left and right finite-momentum 
superconductors. Remarkably, the form of the off-diagonal component of the self 
energy [i.e., Eq. \eqref{fvjdfjv}] is quite general. Irrespective of the details of the 
superconducting leads, their effect on the Andreev reflections are  captured by the additional phase-shift difference 
$\delta\phi(q,E,\tilde{\mu}_S)$ that can be numerically calculated.


\subsection{Andreev dispersions for $h_d=h_{sc}=\theta_{so}=0$}
\label{ssec-andreev-dispersions}

By substituting the self energy \eqref{selfenergy1} into Eq.~\eqref{fdvkdfmv}, we can determine the Andreev in-gap states 
from the secular equation
\begin{align}  \label{fdvmdfkvm}
    &E^2\left[1+2 t^2\mathcal{K}(q,E,\tilde{\mu}_S ) \right]^2 -[\epsilon_d-2\mu_St^2\mathcal{D}(q,E,\tilde{\mu}_S)]^2\notag\\
    &= 4t^4\mathcal{R}^2(q,E,\tilde{\mu}_S)\cos^2\left[\frac{\phi+\delta\phi(q,E,\tilde{\mu}_S)}{2}\right],
\end{align}
where $\phi_{\text{eff}}$ dropped out. Equation~\eqref{fdvmdfkvm} can be formally solved as an implicit solution  
for the Andreev eigenenergies $E_{\eta}$,
\begin{widetext}
\begin{align} \label{tgogbokee} 
  E_{\eta}&=\eta \frac{\sqrt{[\epsilon_d-2\tilde{\mu}_St^2\mathcal{D}(q,E_{\eta},\tilde{\mu}_S)]^2}+4t^4\mathcal{R}^2(q,E_{\eta},\tilde{\mu}_S)\cos^2\left[\frac{\phi+\delta\phi(q,E_{\eta},\tilde{\mu}_S)}{2}\right]}{1+2 t^2\mathcal{K}(q,E_{\eta},\tilde{\mu}_S)},
\end{align}
\end{widetext}
with $\eta=\pm$. 
From the compact implicit equation above we can obtain the Andreev levels, e.g, iteratively.  
Note that $E_{\eta} = E_{\eta}(q, \phi, \delta\phi)$. Here, the phase shifts $\delta\phi_j(q,\phi,E,\tilde{\mu}_S)$ obey particle-hole symmetry, i.e., $\delta\phi_j(q,\phi,-E,\tilde{\mu}_S)=\delta\phi_j(q,\phi,E,\tilde{\mu}_S)$, as shown by Fig. \ref{AFIGF} (e).

Note that $\mathcal{R}(q,E_{\eta},\tilde{\mu}_S) $,  $\mathcal{D}(q,E_{\eta},\tilde{\mu}_S) $ and $\mathcal{K}(q,E_{\eta},\tilde{\mu}_S)$ are complicated  functions and can only be  calculated numerically. To have some analytical result, let us perform the transformation of the fermionic lead operators
\begin{align} \label{transf-leads}
    c_{jns}=\frac{1}{\sqrt{N}}\sum^{N}_{l=1} e^{ik_l na} c_{jk_ls}, \text{ with } k_l=-\frac{\pi}{a}+\frac{2\pi l}{Na} \text{  and }  l \in Z,
\end{align} 
with
\begin{align} \label{fvmdfv}
    \sum^{N}_{n=1}e^{i(k_l+k_{l'}) na}=N \delta_{ll'}.
\end{align}
The above corresponds to a change of basis in the leads, i.e., from the original site (position) representation 
to the k-space representation. Substituting Eq. \eqref{transf-leads} into the superconducting lead Hamiltonian (c.f., Eq.~(1) 
in the main text)
\begin{align} \label{maindvmdla}
    H_j&=\sum^{N}_{n=1}\sum_{s=\uparrow,\downarrow}\epsilon_{jns}c^{\dagger}_{jns}c^{}_{jns}+\sum^{N}_{n=1}\left(\Delta^j_n c^{\dagger}_{jn\uparrow}c^{\dagger}_{jn\downarrow}+h.c. \right)\\
   &+\sum^{N-1}_{n=1}\sum_{s=\uparrow,\downarrow}\left(t_{0}c^{\dagger}_{jns}c^{}_{jn+1s}+h.c.\right)\notag ,
\end{align}
we obtain 
\begin{align} \label{2maindvmdl}
   H_j&=\sum^{N}_{l=1} \sum_{s=\uparrow,\downarrow}\epsilon_{j}c^{\dagger}_{jk_{l}s}c^{}_{jk_{l}s}+\sum^{N}_{l=1}\left(\Delta e^{i\phi_j} c^{\dagger}_{jk_{l}+q\uparrow}c^{\dagger}_{j-k_{l}+q\downarrow}+h.c. \right)\notag \\
   &+\sum^{N}_{l=1}\sum_{s=\uparrow,\downarrow}2t_0\cos(k_la) c^{\dagger}_{jk_{l}\uparrow}c^{}_{jk_{l}\downarrow}\notag\\
   &+\frac{1}{N}\sum_{s=\uparrow,\downarrow}\left(t_{0}c^{\dagger}_{jk_{l}s}c^{}_{jk_{l'}s} e^{-ik_{l} Na} e^{ik_{l'} (N+1)a}+h.c.\right).
\end{align}
where we have use the identity  \eqref{fvmdfv}. In the limit of $N\rightarrow \infty$, we can omit the third line of Eq. \eqref{2maindvmdl} for simplicity and therefore the lead Hamiltonian \eqref{2maindvmdl} becomes diagonal in $k_l$ space. 
We denote the $k_l$-space version of the Nambu space of the superconducting leads by adding a superscript to $\Psi_j$,
\begin{align} \label{tnamabu}
    \Psi_j^{k}=
    \bigoplus_{l}
    \begin{bmatrix}
      c^{}_{jk_l+q\uparrow}\\
      c^{}_{jk_l+q\downarrow}\\
      -c^{\dagger}_{j-k_l+q\downarrow}\\
      c^{\dagger}_{j-k_l+q\uparrow}
    \end{bmatrix}.
\end{align}
The transformation \eqref{transf-leads} does not change 
the many-body Hamiltonian of our system [\eqref{fvdvldv}] but it does change the representation of the Bogoliubov 
Hamiltonian $\mathcal{H}_{\text{BDG}}$, \eqref{fdnkvamk}. As a matter of fact, only $\mathcal{H}_{D}$ does note change; all the other 
blocks of $\mathcal{H}_{\text{BDG}}$, i.e., $\mathcal{H}_L$,  $\mathcal{H}_R$, $\mathcal{T}_L$, and $\mathcal{T}_R$ do change. Below we add a $k$ superscript in all these matrices so as to emphasize the k-space representation used in the leads.
\begin{align}  \label{fvdkmvk}
    \mathcal{H}^{(k)}_j=
    \begin{bmatrix}
       \mathcal{H}_{jk_1} &  & & & \\
       & \mathcal{H}_{jk_2} &  & & & \\
        &  & \mathcal{H}_{jk_3} &  & \\
        &  &  & \ddots &  \\
        &  & & & \mathcal{H}_{jk_N}
 \end{bmatrix}_{4N\times 4N},
\end{align}
with
\begin{align} \label{vnsdfkfm}
    \mathcal{H}_{jk_l}=
    \begin{bmatrix}
    +\epsilon_{jk_l+q\uparrow} & 0 &  -\Delta e^{+i\phi_j} & 0\\
    0 & +\epsilon_{jk_l+q\downarrow} &  0 & -\Delta e^{+i\phi_j}\\
    -\Delta e^{-i\phi_j} & 0 & -\epsilon_{j-k_l+q\downarrow}  & 0\\
    0 & -\Delta e^{-i\phi_j} &0 &  -\epsilon_{j-k_l+q\uparrow}
    \end{bmatrix}_{4\times 4},
\end{align}
where 
\begin{align}
\epsilon_{jk_l\downarrow}=-2t_0+\mu_S+2t_0\cos(k_la)
\end{align}
is the energy spectrum of  the lead $j$. 
The tunnel-coupling matrix between the quantum dot and the left and right leads in the Nambu basis \eqref{tnamabu} are, respectively, 
\begin{align} \label{fvmfkll1}
    \mathcal{T}^{(k)}_L&= t \begin{bmatrix}
      \mathcal{T}_0 &  \cdots & \mathcal{T}_0 & 
 \end{bmatrix}_{4\times 4N},
\end{align}
\begin{align} \label{fvmfklr1}
    \mathcal{T}^{(k)}_R&= t\begin{bmatrix}
      \mathcal{T}_0 & \cdots & \mathcal{T}_0
 \end{bmatrix}_{4\times 4N},
\end{align}
where the $4\times 4$ tunneling matrix $\mathcal{T}_0$ is given by Eq. \eqref{bgbfmk}. Note that the real-space coupling between the dot and the last (first) site of 
the left (right) lead, translates into a 
$k_l$-independent tunnel coupling matrix in the $k_l$-space representation of the leads. 
Following the same procedure as in Eqs. (\ref{fvmldfvdl}-\ref{fdvkdfmv}), we obtain the self-energy of the Andreev quantum dot 
\begin{align} \label{fdnkngn}
&\Sigma(q,E,\tilde{\mu}_S,\phi_L,\phi_R)=\begin{bmatrix}
        \mathcal{T}^{(k)}_{L} & \mathcal{T}^{(k)}_{R}
    \end{bmatrix}\\
    &\times 
    \begin{bmatrix}
        \mathcal{H}^{(k)}_{L} -E & 0 \\
   0 & \mathcal{H}^{(k)}_{R}-E
    \end{bmatrix}^{-1}
    \begin{bmatrix}
        (\mathcal{T}^{(k)}_{L})^{\dagger}\\ (\mathcal{T}^{(k)}_{R})^{\dagger}
    \end{bmatrix} \notag \\
   & =\sum_{j=L,R}  [ \mathcal{T}^{(k)}_{j}]_{4\times 4N}  \left[\frac{1}{\mathcal{H}^{(k)}_{j}-E} \right]_{4N\times 4N} [(\mathcal{T}^{(k)}_{j})^{\dagger}]_{4N\times 4}.   \notag 
\end{align}
We emphasize that the above self energy is exacly the same as the one calculated in the real-space site representation, Eq.~(\ref{fvmkfllgf}). This is so because we performed the change of basis (k-space) only on the leads.
Substituting Eqs. \eqref{fvdkmvk} and \eqref{vnsdfkfm} into the self-energy \eqref{fdnkngn}, we obtain 
\begin{align} \label{tgbvmdakvm}
&\Sigma(q,E,\tilde{\mu}_S,\phi_L,\phi_R)  =t^2\begin{bmatrix}
      \mathcal{T}_0 &  \cdots &   \mathcal{T}_0
 \end{bmatrix}_{4\times 4N}  \\
& \times\left[\frac{1}{\mathcal{H}^{(k)}_{L}}-E \right]_{4N\times 4N} \begin{bmatrix}\mathcal{T}_0\\
 \vdots\\
 \mathcal{T}_0
 \end{bmatrix}_{4N\times 4}\notag \\
 &+t^2\begin{bmatrix}
      \mathcal{T}_0 &  \cdots & \mathcal{T}_0 
 \end{bmatrix}_{4\times 4N}  \left[\frac{1}{\mathcal{H}^{(k)}_{R}-E} \right]_{4N\times 4N} \begin{bmatrix}  \mathcal{T}_0\\
 \vdots\\
 \mathcal{T}_0\end{bmatrix}_{4N\times 4}.\notag
\end{align}
Note that $[1/(\mathcal{H}^{(k)}_{j}-E)]_{4N\times4N}$ is diagonal in $k_l$ space [see Eq. \eqref{fvdkmvk}]. The  huge matrix calculation of the above self-energy can be written as a summation over $k_l$ 
\begin{align} \label{fadavmdakvm}
&\Sigma(q,E,\tilde{\mu}_S,\phi_L,\phi_R)\\
&=t^2\bigotimes_{s=\uparrow,\downarrow}\sum_{jk_l}\tau_z 
\left[\begin{array}{cc}
 \mathcal \epsilon_{jk_l+q}-E & -\Delta e^{+i\phi_j}\\
-\Delta e^{-i\phi_j} &-\epsilon_{j-k_l+q}-E
\end{array}\right]^{-1}\tau_z,\notag 
\end{align}
where $\tau_z$ is the Pauli matrix in Nambu space. Let us define the Green functions as follows
\begin{align}
     G_{js}(E,\tilde{\mu}_S,&\phi_j,\epsilon_{jk_l})\equiv \left[\begin{array}{cc}
 \epsilon_{jk_l+q}-E & -\Delta e^{+i\phi_j}\\
-\Delta e^{-i\phi_j} &-\epsilon_{j-k_l+q}-E
\end{array}\right]^{-1} \notag \\
&=\frac{1}{\text{det}_{jk_ls}}
     \begin{bmatrix}
      \epsilon_{j-k_l+q}+E & -\Delta e^{+i\phi_j}\\
     -\Delta e^{-i\phi_j}  & -\epsilon_{jk_l+q}+E
 \end{bmatrix},
\end{align}
where $\text{det}_{jk_ls}=\Delta^2-(E-\epsilon_{j-k_l+q})(E+\epsilon_{jk_l+q})$. The self-energy  \eqref{fadavmdakvm} then becomes 
\begin{equation} \label{tfadavmdakvm}
\Sigma(q,E,\tilde{\mu}_S,\phi_L,\phi_R)=t^2\bigotimes_{s=\uparrow,\downarrow}\sum_{jk_l}\tau_z 
G_{js}(E,\phi_j,\epsilon_{jk_l})\tau_z.
\end{equation}

\subsubsection{Zero Cooper pair momentum: symmetric dispersions}

Here, for simplicity, we first consider the case with zero Cooper pair momentum $q=0$. 

\begin{figure*}[t]
\begin{center}
\includegraphics[width=1.0\textwidth]{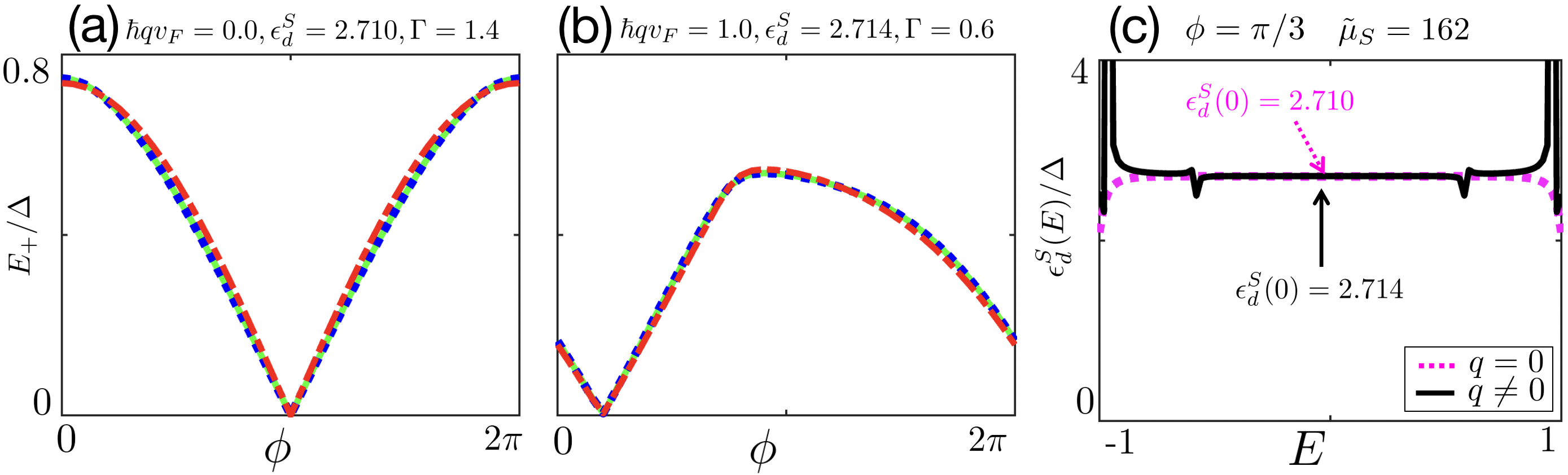}
\end{center}
\caption{(a,b) Andreev levels $E_{+}$ as a function of phase difference $\phi$ for $\hbar qv_F=0$ with $\Gamma=1.4$ and (b) $\hbar qv_F=1.0$ with $\Gamma=0.6$. The latter corresponds to the results of Fig. 2 in the main text. Panel (c) plots the additional dot energy $\epsilon_d$ [Eq. \eqref{fdvndfnk}] induced by the lead-dot tunneling coupling. For $q=0$, we have $\epsilon^S_d\simeq 2.7$. Though it shows showing some poles in the range ($-\Delta$, $\Delta$) for $q\neq 0$, we also obtain $\epsilon^S_d\simeq 2.7$ for $E$ away from poles. Therefore, it is reasonable to use $\epsilon^S_d\simeq 2.7$ during the fitting [red lines of panels (a) and (b)]. In panels (a) and (b), the blue line is calculated from the numerical diagonalization of the BDG Hamiltonian \eqref{fdnkvamk}, the green line is obtained from Eq. \eqref{tgogbokee} after the substitutions of $\mathcal{R}(q,E_{+},\tilde{\mu}_S)$, $\mathcal{D}(q,E_{+},\tilde{\mu}_S)$,$\mathcal{K}(q,E_{+},\tilde{\mu}_S)$, and $\delta\phi(q,E_{+},\tilde{\mu}_S)$, and the red lines correspond to $E_{+}= \sqrt{(\epsilon^{}_d-\epsilon^S_d)^2+4\Gamma^2\cos^2\left(\frac{\phi}{2}\right)}/(1+2\Gamma/\Delta)$ and $E_{+}=\eta \sqrt{(\epsilon^{}_d-\epsilon^S_d)^2+4\Gamma^2\cos^2\left[\frac{\phi+\delta\phi[q,E_{\eta}(q,\phi)]}{2}\right]}/(1+2\Gamma/\Delta)$ for $q=0$ and $q\neq 0$, respectively. The blue lines coincide with green ones, which verifies our methodology of the additional phase shift and asymmetric Andreev dispersion for finite-$q$ superconductor in Secs.~\ref{AndreevrReflection} and \ref{sec-asym}. We can see clear asymmetric dispersion for $q\neq 0$, crucial to the superconducting diode effect. Other parameters: $\Delta=1$, $\epsilon_d=2.6$, $h_{sc}=0$, $h_d=0$, $t=-13$, and $t_0=-100$.}
\label{AFIG2}
\end{figure*}

The summation over momenta $k_f$ in each element $\Sigma^{a,b}(q,E,\tilde{\mu}_S,\phi_L,\phi_R)$ of the self-energy \eqref{tfadavmdakvm} can be replaced by integration as follows
\begin{align} \label{fdnvkn}
    \sum_{k_f}G^{a,b}_{js}(E,\tilde{\mu}_S,\tilde{\mu}_S,\phi_j,\epsilon_{jk_f})&=\int \frac{d k_f}{Na}G^{a,b}_{js}(E,\tilde{\mu}_S,\tilde{\mu}_S,\phi_j,\epsilon_{jk_f})\notag \\
    &=\int d\epsilon \nu_{j}(\epsilon)G^{a,b}_{js}(E,\tilde{\mu}_S,\tilde{\mu}_S,\phi_j,\epsilon),
\end{align}
where $\nu_{j}(\epsilon)$ is the density of states of superconductor $j$ per spin. Then, Eq.~\eqref{tfadavmdakvm} reduces to
\begin{widetext}
\begin{align} \label{selfenergy2}
    \Sigma(q,E,\tilde{\mu}_S,\phi_L,\phi_R)=2t^2\bigotimes_{s=\uparrow,\downarrow}\begin{bmatrix}
    E\mathcal{K}(q,E,\tilde{\mu}_S)+\tilde{\mu}_S\mathcal{D}(q,E,\tilde{\mu}_S)  & \mathcal{R}(q,E,\tilde{\mu}_S)\cos\left(\frac{\phi}{2}\right) \\
    \mathcal{R}(q,E,\tilde{\mu}_S)\cos\left(\frac{\phi}{2}\right) &  E\mathcal{K}(q,E,\tilde{\mu}_S)-\tilde{\mu}_S\mathcal{D}(q,E,\tilde{\mu}_S)
    \end{bmatrix},
\end{align}
\end{widetext}
with
\begin{align} \label{fdvndfnk}
    t^2\tilde{\mu}_S\mathcal{D}(q,E,\tilde{\mu}_S) \equiv \epsilon^{S}_d (E).
\end{align}
We have $t^2E\mathcal{K}(q,E,\tilde{\mu}_S) \simeq \Gamma \frac{E}{\sqrt{\Delta^2-E^2}} $ and  $t^2\mathcal{R}(q,E,\tilde{\mu}_S) \simeq \Gamma\frac{\Delta}{\sqrt{\Delta^2-E^2}} $, where $\Gamma=\pi\nu_F t^2$ and  $\nu_F=\nu_L=\nu_R$ is the density of states of superconducting leads at the Fermi energy for each spin species and is assumed to be $\epsilon$-independent for simplicity. Then, the Andreev levels \eqref{tgogbokee} becomes
\begin{align}
\label{tgogdvbokee} 
  E_{\eta}&=\eta \frac{\sqrt{\left[\epsilon_d-\epsilon^{S}_d (E_{\eta})\right]^2+\Delta^2\left(\frac{2\Gamma}{\sqrt{\Delta^2-E^2_{\eta}}}\right)^2\cos^2\left(\frac{\phi}{2}\right)}}{1+\left(\frac{2\Gamma}{\sqrt{\Delta^2-E^2_{\eta}}}\right)}.
\end{align}
Obviously, Andreev levels at $q=0$ [Eq. \eqref{tgogdvbokee}] is  a symmetric dispersion, i.e., $E_{\eta} (q=0,\pi - \phi) = E_{\eta}(q=0,\pi+\phi)$. As a result, if we find a critical forward supercurrent $I^c_{+}$ at $\phi=\phi_{+}$, we can always obtain a critical reverse suppercurrent $I^c_{-}$ at $\phi=\phi_{-}=2\pi-\phi_{+}$ with the same magnitude, i.e., $\vert I^c_{-}\vert=I^c_{+}$. Note that the superconducting lead-quantum dot tunneling coupling renormalizes  effective dot energy $\epsilon^{\text{eff}}_d=\epsilon_d- \epsilon^{S}_d $, superconducting proximity effect with effective pair potential  $\Delta^{\text{eff}}=\left(\frac{2\Gamma}{\sqrt{\Delta^2-E^2_{\eta}}}\right)\Delta\cos\left(\frac{\phi}{2}\right)$, as well as Andreev levels [it appears in the  denominator of the Andreev levels \eqref{tgogbokee}]. In weak tunneling and low energy limit ($\Gamma,E\ll \Delta$), the renormalization effects from  $2t^2\mathcal{K}(q,E_{\eta}) \simeq  \frac{2\Gamma}{\Delta} $ become independent of $E$, and the above Andreev level  reduces to  $E_{\eta}=\eta \sqrt{(\epsilon^{}_d-\epsilon^S_d)^2+4\Gamma^2\cos^2\left(\frac{\phi}{2}\right)}/(1+2\Gamma/\Delta)$. Thus, we obtain better agreement for smaller $t$ as shown by the red lines in Figs. \ref{AFIG2} (a).

\subsubsection{Finite Cooper-pair momentum $q$: asymmetric dispersions} \label{asymmetricdispersions}

For $q\neq 0$,  the $\phi$-dependent additional phase difference $\delta\phi[q,E(q,\phi)]$ results in the asymmetric dispersion $E_{\eta} (q\neq 0,\pi - \phi)\neq E_{\eta}(q\neq 0,\pi+\phi)$ [Fig.~2 (a) in the main text] thus giving rise to the superconducting diode effect [see Fig.~2 (c) in the main text]. To see how the asymmetric $\phi$-dependence of the additional phase difference $\delta\phi(q,E,\tilde{\mu}_S)$ gives rise to asymmetric dispersions, let us consider the weak tunneling and low energy case ($\Gamma,E\ll \Delta$) for simplicity  in which we assume i) $\epsilon^S_d (\simeq 2.7$  ) is independent of the $E$ inside the finite-momentum superconductor gap [Fig. \ref{AFIG2} (c)], ii) $2t^2\mathcal{R}(q,E_{\eta},\tilde{\mu}_S)\rightarrow 2\Gamma$, and iii) $2 t^2\mathcal{K}(q,E_{\eta},\tilde{\mu}_S)\rightarrow \frac{2\Gamma}{\Delta}$ that eliminate the $E$-dependence of the renormalization effects in both effective pair potential and Andreev level. Equation~\eqref{tgogbokee} then reduces to $E_{\eta}(q,\phi)=\eta \sqrt{(\epsilon_d-\epsilon^S_d)^2+4\Gamma^2\cos^2\left[\frac{\phi+\delta\phi[q,E_{\eta}(q,\phi)]}{2}\right]}/(1+2\Gamma/\Delta)$, which is plotted by the red lines of Fig. \ref{AFIG2} (b). Numerically, we can obtain the exact $\phi$-dependence of $E_{\eta}(q,\phi)$ via the diagonalization of the  BdG Hamiltonian \eqref{fdnkvamk}. To calculate the additional phase difference $\delta\phi[q,E_{\eta}(q,\phi)]$ for each phase difference $\phi=\phi_L-\phi_R$, we have to substitute $\phi_L=-\phi/2$, $\phi_R=+\phi/2$, and $E=E_{\eta}(q,\phi)$ into Eqs.~\eqref{fdvmdllL} and \eqref{fdvmdllR} and then determine the additional phase shifts of the Andreev reflections in the left and right  superconductors, i.e., $\delta\phi_L[q,E_{\eta}(q,\phi)]$ and $\delta\phi_R[q,E_{\eta}(q,\phi)]$.   We find that the additional phase difference  $\delta\phi[q,E_{\eta}(q,\phi)]$ is an asymmetric function of $\phi$ for $q\neq 0$ [see Fig.~1 (c) in the main text] and therefore we obtain asymmetric dispersions with respect to $\phi$, i.e., $E_{\eta} (q\neq 0, \pi - \phi)\neq E_{\eta}(q\neq0,\pi+\phi)$ (see Fig. \ref{AFIG2}). Our fitting agrees with the results obtained from the numerical diagonalization of the BDG Hamiltonian \eqref{fdnkvamk}.

\section{Ground state Fermion parity changes}
\label{sec-parity}

In this section, we derive a condition for the ground-state fermion parity changes in our system in the presence of both Zeeman 
fields $h_d$ and $h_{sc}$. We discuss the cases with and without SO interaction. 

\subsection{Zero SO interaction ($\theta_{so}=0$) and $h_d\neq 0$, $h_{sc}\neq 0$} \label{zeroSO}
 
Here the secular equation \eqref{fdvkdfmv} for the Andreev states in the dot reads
\begin{align} \label{fvnkfkg}
    &\left[sh_d-E^s-2t^2(E-sh_{sc})\mathcal{K}_{}(q,E-sh_{sc})\right]^2 \\
    &-[\epsilon_d-2\tilde{\mu}_St^2\mathcal{D}(q,E,\tilde{\mu}_S)]^2 \notag \\
    &=4t^4\mathcal{R}^2(q,E-sh_{sc})\cos^2\left[\frac{\phi+\delta\phi(q,E)}{2}\right]. \notag 
\end{align}
The above equation can be obtained from Eq. \eqref{fdvmdfkvm} by a replacement $E\rightarrow E-sh_{d}$ ($E\rightarrow E-sh_{sc}$) in the quantum dot (superconducting leads).
The $\mathcal{K}(q,E-sh_{sc})$ function renormalizes both Andreev levels and its spin splitting. Thus, Eq. \eqref{fvnkfkg} reduces to 
\begin{align} \label{yefvnkfkg}
    &\left\{sh^r_d-E\left[1+2t^2\mathcal{K}(q,E-sh_{sc})\right]\right\}^2-(\epsilon^r_d)^2\\
&=4t^4\mathcal{R}^2(q,E-sh_{sc})\cos^2\left[\frac{\phi+\delta\phi(q,E-sh_{sc})}{2}\right],\notag
\end{align}
with
\begin{align}
    \epsilon^r_d=\epsilon_d-2\tilde{\mu}_St^2\mathcal{D}(q,E,\tilde{\mu}_S),
\end{align}
\begin{align}
    h^r_d=h_d+2t^2\mathcal{K}(q,E-sh_{sc},\tilde{\mu}_S)h_{sc}.
\end{align}
Therefore, we have obtained a compact implicit equation from which we can easily obtain the Andreev levels explicitly by iterations
\begin{widetext}
\begin{align} \label{gogbokee}
     E_{s\eta}&=s\frac{h_d+2t^2\mathcal{K}(q,E_{s\eta}-sh_{sc})h_{sc}}{1+2t^2\mathcal{K}(q,E_{s\eta}-sh_{sc})}+\eta \frac{\sqrt{[\epsilon_d-2\tilde{\mu}_St^2\mathcal{D}(q,E_{s\eta}-sh_{sc},\tilde{\mu}_S)]^2+4t^4\mathcal{K}^2(q,E_{s\eta}-sh_{sc})\cos^2\left[\frac{\phi+\delta\phi(q,E_{s\eta}-sh_{sc})}{2}\right] }}{1+2t^2\mathcal{K}(q,E_{s\eta}-sh_{sc})}.
\end{align}
The fermion parity changes happen at $E_{s\eta}=0$ and hence we obtain the condition for it to occur
\begin{align} \label{fvndkdvk}
\phi_{\pm}&=-\delta\phi(q,h_{sc})\pm2\arccos\left[\frac{[h_d+2 t^2h_{sc}\mathcal{K}(q,h_{sc})]^2- [\epsilon_d-2\tilde{\mu}_St^2\mathcal{D}(q,h_{sc},\tilde{\mu}_S)]^2}{4t^4\mathcal{R}^2(q,h_{sc})}\right]^{1/2}.
\end{align}
\end{widetext}
where we have $\delta\phi(q,h_{sc})=\left.\delta\phi(q,E-sh_{sc})\right\vert_{E=0}$, $\mathcal{K}(q,h_{sc})\equiv\left.\mathcal{K}(q,E-sh_{sc})\right\vert_{E=0}$, $\mathcal{D}(q,h_{sc})\equiv\left.\mathcal{D}(q,E-sh_{sc})\right\vert_{E=0}$, and $\mathcal{R}(q,h_{sc})\equiv\left.\mathcal{R}(q,E-sh_{sc})\right\vert_{E=0}$, where we remove the $s$ dependence because $\delta\phi(q,E)$, $\mathcal{K}(E)$, $\mathcal{D}(E)$, and $\mathcal{R}(E)$ are even function of $E$ (Figs.~\ref{AFIGG} and \ref{AFIGF}).
Equation \eqref{fvndkdvk}  helps explain how fermion parity changes can modify the SDE in the absence of spin-orbit coupling, Fig.~2 in the main text.

\subsection{Nonzero SO interaction ($\theta_{so} \neq 0$) and $h_d\neq 0$, $h_{sc}\neq 0$}

In the presence of spin-orbit interaction, spin-up and spin-down electrons mix via tunneling matrix [Eq. \eqref{fmmvfkd}]. Substituting the Green function \eqref{fdvadavfv} and the tunneling matrix \eqref{fmmvfkd} into the self-energy \eqref{selfenergy}, we obtain 
\begin{align}
    \hat{\Sigma}(E)=t^2 
    \begin{bmatrix}
    \hat{\mathcal{G}}^{1,1}(E) &  \hat{\mathcal{F}}(E)  \\
    \hat{\mathcal{F}}^{+}(E) & \hat{\mathcal{G}}^{2,2} (E)
    \end{bmatrix}.
\end{align}
The normal contributions are given by
\begin{widetext}
\begin{align} \label{fdjqvdk11}
    \hat{\mathcal{G}}^{1,1}(E)&=\mathcal{U}(\theta_{so})
    \begin{bmatrix}
    (E-h_{sc})\mathcal{K}(q,E-h_{sc},\tilde{\mu}_S) & 0  \\
    0 & (E+h_{sc})\mathcal{K} (q,E+h_{sc},\tilde{\mu}_S)
    \end{bmatrix}
    \mathcal{U}^{+}(\theta_{so})\notag \\
    &+\mathcal{U}(\theta_{so})
    \begin{bmatrix}
    +\tilde{\mu}_S\mathcal{D}(q,E-h_{sc},\tilde{\mu}_S) & 0  \\
    0 & +\tilde{\mu}_S\mathcal{D}(q,E+h_{sc},\tilde{\mu}_S)
    \end{bmatrix}
    \mathcal{U}^{+}(\theta_{so})\notag \\
    &+\mathcal{U}^{+}(\theta_{so})
    \begin{bmatrix}
    (E-h_{sc})\mathcal{K}(q,E-h_{sc},\tilde{\mu}_S) & 0  \\
    0 & (E+h_{sc})\mathcal{K} (q,E+h_{sc},\tilde{\mu}_S)
    \end{bmatrix}
    \mathcal{U}^{}(\theta_{so})\notag \\
    &+\mathcal{U}^{+}(\theta_{so})
    \begin{bmatrix}
    +\tilde{\mu}_S\mathcal{D}(q,E-h_{sc},\tilde{\mu}_S) & 0  \\
    0 & +\tilde{\mu}_S\mathcal{D}(q,E+h_{sc},\tilde{\mu}_S)
    \end{bmatrix}
    \mathcal{U}^{}(\theta_{so}),
\end{align}
\begin{align} \label{fdjqvdk22}
    \hat{\mathcal{G}}^{2,2}(E)&=\mathcal{U}(\theta_{so})
    \begin{bmatrix}
    (E-h_{sc})\mathcal{K}(q,E-h_{sc},\tilde{\mu}_S) & 0  \\
    0 & (E+h_{sc})\mathcal{K} (q,E+h_{sc},\tilde{\mu}_S)
    \end{bmatrix}
    \mathcal{U}^{+}(\theta_{so})\notag \\
    &+\mathcal{U}(\theta_{so})
    \begin{bmatrix}
    -\tilde{\mu}_S\mathcal{D}(q,E-h_{sc},\tilde{\mu}_S) & 0  \\
    0 & -\tilde{\mu}_S\mathcal{D}(q,E+h_{sc},\tilde{\mu}_S)
    \end{bmatrix}
    \mathcal{U}^{+}(\theta_{so})\notag \\
    &+\mathcal{U}^{+}(\theta_{so})
    \begin{bmatrix}
    (E-h_{sc})\mathcal{K}(q,E-h_{sc},\tilde{\mu}_S) & 0  \\
    0 & (E+h_{sc})\mathcal{K} (q,E+h_{sc},\tilde{\mu}_S)
    \end{bmatrix}
    \mathcal{U}^{}(\theta_{so})\notag \\
    &+\mathcal{U}^{+}(\theta_{so})
    \begin{bmatrix}
    -\tilde{\mu}_S\mathcal{D}(q,E-h_{sc},\tilde{\mu}_S) & 0  \\
    0 & -\tilde{\mu}_S\mathcal{D}(q,E+h_{sc},\tilde{\mu}_S)
    \end{bmatrix}
    \mathcal{U}^{}(\theta_{so}).
\end{align}
Then, the  anomalous contributions are given by 
 \begin{align} \label{fdjvdkf}
    \hat{\mathcal{F}}(E)&=\mathcal{U}(\theta_{so})
    \begin{bmatrix}
    \mathcal{R}(q,E-h_{sc},\tilde{\mu}_S)e^{-i\phi/2+i\delta\phi^L(q,E-h_{sc},\tilde{\mu}_S)} & 0  \\
    0 & \mathcal{R} (q,E+h_{sc},\tilde{\mu}_S)e^{-i\phi/2+i\delta\phi^L(q,E+h_{sc},\tilde{\mu}_S)}
    \end{bmatrix}
    \mathcal{U}^{+}(\theta_{so})\notag \\
    &+\mathcal{U}^{+}(\theta_{so})
    \begin{bmatrix}
    \mathcal{R}(q,E-h_{sc},\tilde{\mu}_S)e^{+i\phi/2+i\delta\phi^R(q,E-h_{sc},\tilde{\mu}_S)} & 0  \\
    0 & \mathcal{R}(q,E+h_{sc},\tilde{\mu}_S)e^{+i\phi/2+i\delta\phi^R(q,E+h_{sc},\tilde{\mu}_S)}
    \end{bmatrix}
    \mathcal{U}^{}(\theta_{so}),
\end{align}
Here we are interested in the zero energy Andreev bound state, $E=0$. 
Then, Eqs. (\ref{fdjqvdk11}-\ref{fdjvdkf}) reduce to 
\begin{align} \label{0fdjqvdk11}
    \hat{\mathcal{G}}^{1,1}(0)&=-h_{sc}\mathcal{K}(q,h_{sc},\tilde{\mu}_S)\mathcal{U}(\theta_{so})
    s^z
    \mathcal{U}^{+}(\theta_{so})+\tilde{\mu}_S\mathcal{D}(q,h_{sc},\tilde{\mu}_S)\mathcal{U}(\theta_{so})
    s^o
    \mathcal{U}^{+}(\theta_{so}) \\
    &-h_{sc}\mathcal{K}(q,h_{sc},\tilde{\mu}_S)\mathcal{U}^{+}(\theta_{so})
    s^z
    \mathcal{U}^{}(\theta_{so})+\tilde{\mu}_S\mathcal{D}(q,h_{sc},\tilde{\mu}_S)\mathcal{U}^{+}(\theta_{so})
    s^o
    \mathcal{U}^{}(\theta_{so})\notag,
\end{align}
\begin{align} \label{0fdjqvdk22}
    \hat{\mathcal{G}}^{2,2}(0)&=-h_{sc}\mathcal{K}(q,h_{sc},\tilde{\mu}_S)\mathcal{U}(\theta_{so})
    s^z
    \mathcal{U}^{+}(\theta_{so})+\tilde{\mu}_S\mathcal{D}(q,h_{sc},\tilde{\mu}_S)\mathcal{U}(\theta_{so})
    s^o
    \mathcal{U}^{+}(\theta_{so}) \\
    &-h_{sc}\mathcal{K}(q,h_{sc},\tilde{\mu}_S)\mathcal{U}^{+}(\theta_{so})
    s^z
    \mathcal{U}^{}(\theta_{so})-\tilde{\mu}_S\mathcal{D}(q,h_{sc},\tilde{\mu}_S)\mathcal{U}^{+}(\theta_{so})
    s^o
    \mathcal{U}^{}(\theta_{so})\notag,
\end{align}
\begin{align} \label{0fdjvdkf}
    \hat{\mathcal{F}}(0)&=e^{-i\phi/2+i\delta\phi^L(q,h_{sc},\tilde{\mu}_S)}\mathcal{R}(q,h_{sc},\tilde{\mu}_S)\mathcal{U}(\theta_{so})s^o
    \mathcal{U}^{+}(\theta_{so})\\
    &+ e^{+i\phi/2+i\delta\phi^R(q,h_{sc},\tilde{\mu}_S)}\mathcal{R}(q,h_{sc},\tilde{\mu}_S)\mathcal{U}^{+}(\theta_{so})
    s^o
    \mathcal{U}^{}(\theta_{so}),\notag 
\end{align}
where we have used the fact that $\delta\phi^R(q,E,\tilde{\mu}_S)$, $\mathcal{K}(q,E,\tilde{\mu}_S)$, $\mathcal{D}(q,E,\tilde{\mu}_S)$, and $\mathcal{R}(q,E,\tilde{\mu}_S)$  are even function of $E$. Then, we reach 
\begin{align}
    \hat{\mathcal{G}}^{1,1}(0)=-2\mathcal{K}(q,h_{sc},\tilde{\mu}_S)h_{sc}\cos\theta_{so} s^z+\tilde{\mu}_S\mathcal{D}(q,h_{sc},\tilde{\mu}_S),
\end{align}
\begin{align}
    \hat{\mathcal{G}}^{2,2}(0)=-2\mathcal{K}(q,h_{sc},\tilde{\mu}_S)h_{sc}\cos\theta_{so} s^z-\tilde{\mu}_S\mathcal{D}(q,h_{sc},\tilde{\mu}_S),
\end{align}
\begin{align} \label{tr3fffkvmfg}
    \hat{\mathcal{F}}^{}(0)&= 2\mathcal{R}(q,h_{sc},\tilde{\mu}_S)\cos\left[\frac{\phi+\delta\phi(q,h_{sc},\tilde{\mu}_S)}{2}\right]e^{+i\phi_{\text{eff}}}.
\end{align} 
Solving the reduced determinantal equation \eqref{fdvkdfmv} at $E=0$, we find
\begin{align}
\phi_{\pm}&=-\delta\phi(q,h_{sc},\tilde{\mu}_S)\pm2\arccos\left\{\frac{[h_d+2\cos\theta_{\text{so}} t^2h_{sc}\mathcal{K}(q,h_{sc},\tilde{\mu}_S)]^2- [\epsilon_d-\tilde{\mu}_S\mathcal{D}(q,h_{sc},\tilde{\mu}_S)]^{2}}{4t^4\mathcal{R}^2(q,h_{sc},\tilde{\mu}_S)}\right\}^{1/2}.
\end{align}
These are the phase values at which fermion parity changes occur for the case with spin-orbit interaction and Zeeman magnetic fields, Fig.~3 of the main text.

\end{widetext}

\end{document}